\pdfoutput=1 
\documentclass[traditabstract]{aa} 
\usepackage{graphicx,caption}
\usepackage{marginnote}
\usepackage{txfonts}
\usepackage{subfigure}
\usepackage{natbib}
\usepackage{amstext}
\usepackage[verbose]{placeins}
\usepackage{tikz}
\usepackage{array}
\usepackage{amssymb}
\usepackage{xcolor}
\newcommand{\bd}{\begin{displaymath}}
\newcommand{\ed}{\end{displaymath}}
\newcommand{\be}{\begin{equation}}
\newcommand{\ee}{\end{equation}}
\newcommand{\beaa}{\begin{eqnarray*}}
\newcommand{\eeaa}{\end{eqnarray*}}
\newcommand{\bea}{\begin{eqnarray}}
\newcommand{\eea}{\end{eqnarray}}

\def\ourlens{J1148+1930} 
\def\HST{{\it HST}{}}
\def\CH{{Cosmic Horseshoe}{}}
\def\GLEE{\textsc{Glee}}

\begin{document}
   \title{The inner dark matter distribution of the Cosmic Horseshoe (J1148+1930) with gravitational lensing and dynamics}

  \titlerunning{The inner dark matter distribution of the Cosmic Horseshoe}

   \author{S. Schuldt\inst{1}\inst{,2}
           \and
          G. Chiriv\`{\i}\inst{1} 
          \and
          S. H. Suyu\inst{1}\inst{,2}\inst{,3}
          \and
          A. Y{\i}ld{\i}r{\i}m\inst{1}\inst{,4}
          \and
          A. Sonnenfeld\inst{5}
          \and
          A. Halkola\inst{6}
          \and
          G. F. Lewis\inst{7}
          }

   \institute{Max-Planck-Institut f\"ur Astrophysik, Karl-Schwarzschild Str.~1, 85741 Garching, Germany \\
              \email{schuldt@mpa-garching.mpg.de}
         \and  
             Physik Department, Technische Universit\"at M\"unchen, James-Franck Str. 1, 85741 Garching, Germany
         \and
             Institute of Astronomy and Astrophysics, Academia Sinica, P.O. Box 23-141, Taipei 10617, Taiwan
         \and
              Max Planck Institute for Astronomy, K\"onigstuhl~17, 69117 Heidelberg, Germany
         \and
             Kavli Institute for the Physics and Mathematics of the Universe, The University of Tokyo, 5-1-5 Kashiwanoha; Kashiwa, 277-8583
         \and
             Py\"orrekuja 5 A, 04300 Tuusula, Finland
         \and
            Sydney Institute for Astronomy, School of Physics, A28, The University of Sydney, NSW 2006, Australia
             }

   \date{Received --; accepted --}


  \abstract
  {We present a detailed analysis of the inner mass structure of the \CH\ (\ourlens) strong gravitational lens system observed with the \textit{Hubble Space Telescope} (\HST) Wide Field Camera 3 (WFC3). In addition to the spectacular Einstein ring, this systems shows a radial arc. We obtained the redshift of the radial arc counter image $z_\text{s,r} = 1.961 \pm 0.001$ from Gemini observations. To disentangle the dark and luminous matter, we consider three different profiles for the dark matter distribution: a power-law profile, the NFW, and a generalized version of the NFW profile. For the luminous matter distribution, we base it on the observed light distribution that is fitted with three components: a point mass for the central light component resembling an active galactic nucleus, and the remaining two extended light components scaled by a constant M/L. To constrain the model further, we include published velocity dispersion measurements of the lens galaxy and perform a self-consistent lensing and axisymmetric Jeans dynamical modeling. Our model fits well to the observations including the radial arc, independent of the dark matter profile. Depending on the dark matter profile, we get a dark matter fraction between 60\% and 70\%. With our composite mass model we find that the radial arc helps to constrain the inner dark matter distribution of the \CH\ independently of the dark matter profile.
}
   {}
   {}
   {}
   {}

   \maketitle
%
\section{Introduction}
\label{sec:introduction}

In the standard cold dark matter (CDM) model, the structure of
dark matter halos is well understood through large numerical
simulations based only on gravity \citep[e.g.,][]{dubinski91,
  navarro96b, navarro96a, ghigna00, diemand05, graham06b, gao12}. From
these N-body dark matter only simulations it appears that halos are
well described by the NFW profile \citep*{navarro97}. This profile has
characteristics slopes; it falls at large radii as $\rho_{\text{r} \gg
  \text{r}_\text{s}} \propto r^{-3}$, while, for small radii, it goes
as $\rho_{\text{r} \ll \text{r}_\text{s}} \propto r^{-1} $ and thus
forms a central density cusp. The so-called scale radius $r_\text{s}$
is the radius where the slope changes. Nowadays, simulations with
higher resolution predict shallower behavior for the density slope at
very small radii and thus a deviation from this simple profile
\citep[e.g.,][]{golse02, graham06a, navarro10, gao12}. Thus, the
distribution is more cored than cuspy \citep[e.g.,][]{collett17,
  dekel17}. These simulations are also showing that DM halos are not
strictly self-similar as first expected for a CDM universe
\citep[e.g.,][]{ryden91, moutarde95, chuzhoy06, lapi11}.

In realistic models for halos one has to include the baryonic
component, and that modifies the distribution and the amount of dark
matter. The distribution of stars, dark matter, and gas depends on
processes such as gas cooling, which allows baryons to condense
towards the center \citep[e.g.,][]{blumenthal86, gnedin04, sellwood05,
  gustafsson06, pedrosa09, abadi10, sommer-larsen10}, active galactic
nuclei (AGNs) feedback \citep[e.g.,][]{peirani08, martizzi13,
  peirani17, li17}, dynamical heating in the central cuspy region due
to infalling satellites and mergers
\citep[e.g.,][]{el-zant01,el-zant04, nipoti04, romano-diaz08,
  tonini06, laporte15} and thermal and mechanical feedback from
supernovae \citep[e.g.,][]{navarro96a, governato10, pontzen12}.

Therefore, detailed observations of the mass distribution include
important information of these complex baryonic processes. Of
particular interest is the radial density profile of DM on small
scales. In addition, at small radii we expect to have the densest
regions of the DM particles, therefore these regions are ideal to
learn more about their interactions and nature \citep{spergel00,
  abazajian01, kaplinghat05, peter10}.

Strong gravitational lensing has arisen as a good technique to obtain
the mass distribution for a wide range of systems. Gravitational
lensing provides a measurement of the total mass within the Einstein
ring since the gravitational force is independent of the mass nature
\citep[e.g.,][]{treu10, treu15}. \citet{dye05} showed that strong lens
systems with a nearly-full Einstein ring are better than those
observations where the source is lensed into multiple point-like
images if one wants to construct a composite profile of baryons and
dark matter. With such observations, one can very well fit the profile
near the region of the Einstein ring, but the inner part cannot be
well constrained due to the typical absence of lensing data in the
inner region. The presence of a radial arc, even though seldom
observed in galaxy-scale lenses, can help break the lensing
degeneracies and put constraints on the inner mass
distribution. Another possibility is to combine lensing and dynamics,
which is now a well established probe to get for instance the density
profile for early-type galaxies \citep[ETGs; e.g.,][]{mortlock00,
  treu02, treu04, gavazzi07, barnabe09, auger10, ven10, barnabe11,
  grillo13}.
  
In this paper, we present a detailed study of the inner mass structure
of the Cosmic Horseshoe lens through lensing and combine these
information with those coming from dynamical modeling. The Cosmic
Horseshoe, discovered by \citet{belokurov07}, is ideal for such a
study: the deflector galaxy's huge amount of mass results in a
spectacular and large Einstein ring, and near the center of the lens
exists a radial arc, which helps to constrain the mass distribution in
the inner part of the Einstein ring. To include the radial arc and our
association for its counter image in the models, we have spectroscopy
measurements for the counter image to get its redshift.

The outline of the paper is as follows. In Sec.~\ref{sec:horseshoe} we
introduce the imaging and spectroscopic observations with their
characteristics and describe the data reduction and redshift
measurement for the radial arc counter image. Then we revisit briefly
in Sec.~\ref{sec:multiplane} the multiple-lens-plane theory. In
Sec.~\ref{sec:compositeModel} we present our results of the composite
mass model of baryons and dark matter using lensing-only, while in
Sec.~\ref{sec:Dynamics} we present the results of our models based on
dynamics-only. In Sec.~\ref{sec:LensingDynamics} we combine lensing
and dynamics and present our final models. Sec.~\ref{sec:conclusion}
summarizes and concludes our results.

Throughout this work, we assume a flat $\Lambda$CDM cosmology with
Hubble constant $H_0 = 72\, \text{km}\, \text{s}^{-1}\,
\text{Mpc}^{-1}$ \citep{bonvin17} and $\Omega_\text{M} =1 -
\Omega_\Lambda = 0.32 $ whose values correspond to the updated Planck
data \citep{planck16}. Unless specified otherwise, each quoted
parameter estimate is the median of its one-dimensional marginalized
posterior probability density function, and the quoted uncertainties
show the 16$^{\text{th}}$ and 84$^\text{th}$ percentiles (that is, the bounds of a 68\%
credible interval).

\FloatBarrier 
\section{The Cosmic Horseshoe (\ourlens)}
\label{sec:horseshoe}
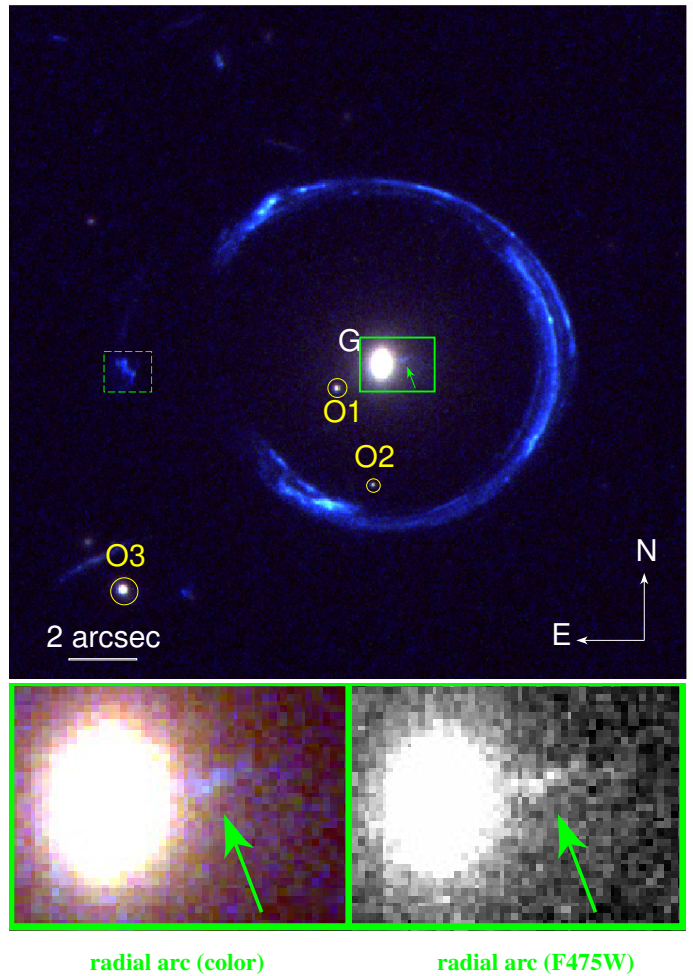
\begin{figure}[ht]
\centering
\begin{picture}(250,390)
\put(0,110){\includegraphics[trim=100 110 90 30, clip, width=1.0\columnwidth]{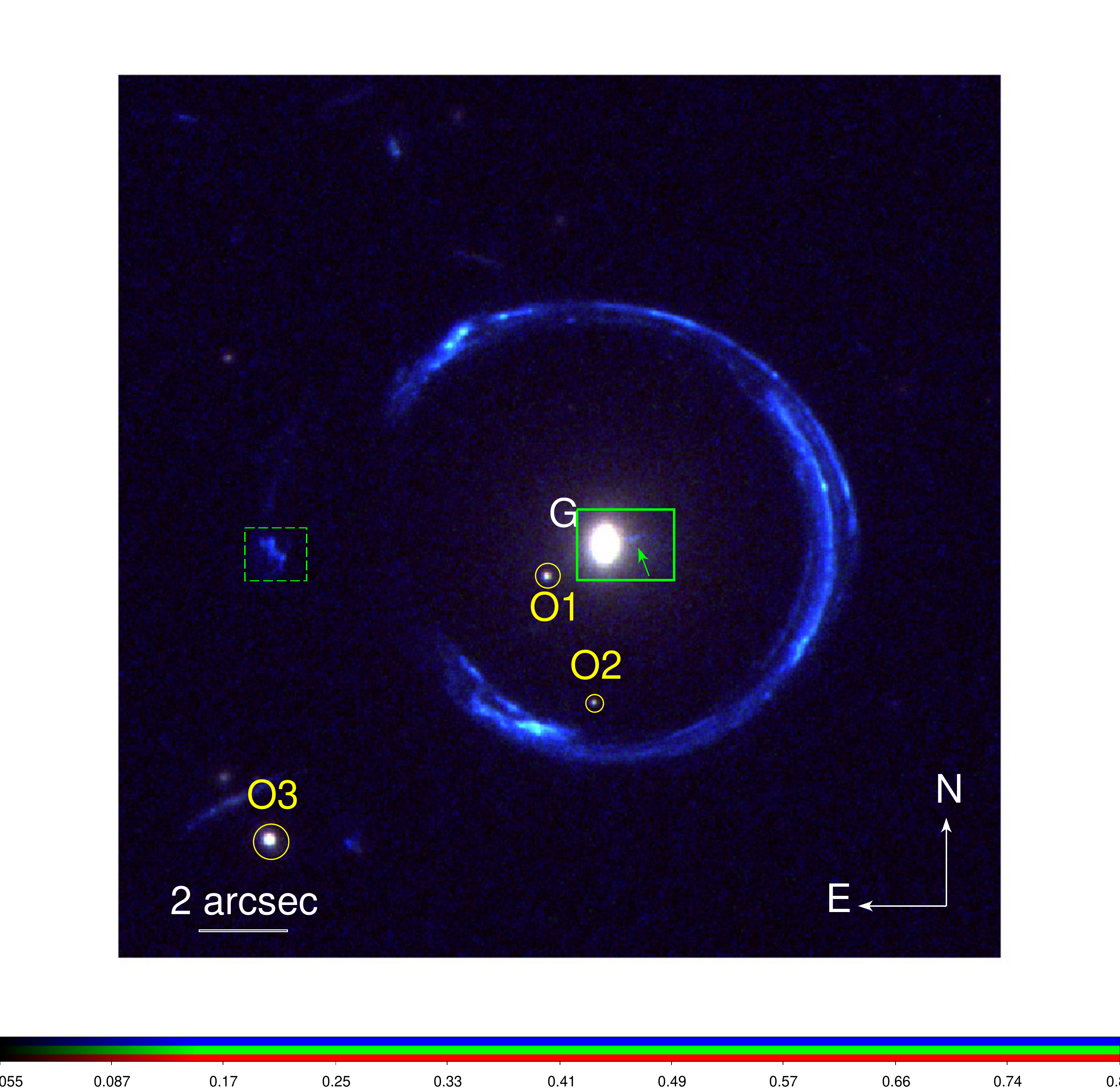}}
\put(125,15){\includegraphics[trim=482 435 432 433, clip, width=0.5\columnwidth]{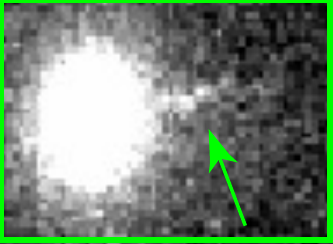}}
\put(0,15){\includegraphics[trim=482 435 432 433, clip, width=0.5\columnwidth]{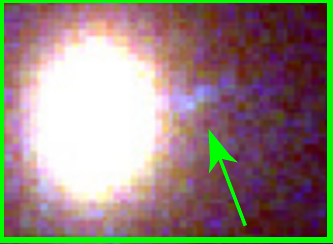}}
\linethickness{0.3mm}
\put(160,0){\color{green} \textbf{radial arc (F475W)}}
\put(30,0){\color{green} \textbf{radial arc (color)}}
\put(50,375){\color{blue} \textbf{\Large{The \CH}}} 
\end{picture}
\caption{Color image of the \CH\ obtained through a combination of the
  F475W, F606W, and F814W filter images from the \HST\ WFC3. The size
  of this image is $20 \arcsec \times 20 \arcsec$. One can see the
  $\approx 300^\circ$ wide blue Einstein ring of the \CH. In addition,
  the \CH\ observation includes a radial arc which is marked with a
  green solid box. This is shown in detail in in the bottom panel, in
  color (left) and from the F475W filter (right). We associate this
  radial arc to its counter image, marked in the main figure with a
  dashed green box and located around $8 \arcsec$ on the east side of
  the lens galaxy G. Both the radial arc and its counter image
  correspond to a source at redshift $z_\text{s,r} = 1.961 $ (see
  Sec.~\ref{sec:horseshoe:arc}). The three star-like objects in the field of view,
  which we include in our light model, are circled in yellow. The figures
  are oriented such that North is up and East is left.}
\label{fig:CH475}
\end{figure}

The Cosmic Horseshoe, also known as SDSS \ourlens, was discovered by
\citet{belokurov07} within the \textit{Sloan Digital Sky Survey}
(\textit{SDSS}). A color image of this gravitational lensed image is
shown in Fig.~\ref{fig:CH475}. The center of the lens galaxy G, at a
redshift of $z_\text{d} = 0.444$, lies at ($11^h 48^m 33^s.15;
19^\circ 30^{\prime} 3^{\prime\prime}.5$) of the epoch J2000 \citep{belokurov07}.  The
tangential arc is a star-forming galaxy at redshift $z_\text{s,t}
=2.381$ \citep{quider09} which is strongly lensed into a nearly full
Einstein ring ($\approx 300^\circ$), whose radius is around
$5^{\prime\prime}$ and thus one of the largest Einstein rings observed
up to now. This large size shows that this lens galaxy must be very
massive. A first estimate of the enclosed mass within the Einstein
ring is $ \approx 5 \times 10^{12} M_\odot$ \citep{dye08} and thus the
lens galaxy, a luminous red galaxy (LRG), is one of the most massive
galaxies ever observed.  Apart from the nearly full Einstein ring and
the huge amount of mass within the Einstein ring, which makes this
observation already unique, the \CH\ observations reveal a radial
arc. This radial arc is in the west of the lens, as marked in the
green solid box in Fig.~\ref{fig:CH475}. We include this radial arc in
our models as well as our association of its counter image, marked
with a green dashed box in Fig.~\ref{fig:CH475}. For this counter
image we have Gemini measurements (see Sec.~\ref{sec:horseshoe:arc})
to yield a redshift of $z_\text{s,r}= 1.961 \pm 0.001$.  A summary of
various properties about the \CH\ is given in Table
\ref{tab:CHproperties}.

\begin{table}[ht]
  \begin{center}
\caption{Properties of the \CH\ (\ourlens)}             
\label{tab:CHproperties}      
\renewcommand{\arraystretch}{1.4}  
\begin{tabular}{lll}        
\hline                 
Component & Properties & Value \\    
\hline                        
Lens	& Right ascension$^a$		& $11^h 48^m 33^s$				\\
	& Declination$^a$		& $19^\circ 30^{\prime} 3^{\prime\prime}.5$	\\
	& Redshift, $z_{\text{d}}$ $^a$	& $0.444$					\\
tangential arc source	& Redshift, $z_\text{s,t}$ $^b$	& $2.381$			\\
	& Star forming rate$^b$		& $\approx 100 \, M_\odot \, \text{yr}^{-1}$	\\
Ring	& Diameter$^a$			& $10.2\arcsec$					\\
	& Length$^a$			& $ \approx 300^\circ$				\\
	& Enclosed mass$^{c,d}$ 		& $ \approx 5 \times 10^{12} \, M_\odot$	\\
radial arc source& Redshift, $z_\text{s,r}$ $^d$ & $ 1.961 $				\\
\hline                                   
\end{tabular}
\end{center}
References:\\
~$^a$ \citet{belokurov07} \\
~$^b$ \citet{quider09} \\
~$^c$ \citet{dye08} \\
~$^d$ result presented in this paper
\end{table}

\subsection{Hubble Space Telescope imaging}
\label{sec:horseshoe:hst}

The data we analyse in this work come from the \textit{Hubble Space
  Telescope} (\HST) Wide Field Camera 3 (WFC3) and can be downloaded
from the Mikulski Archive for Space
Telescopes\footnote{http://archive.stsci.edu/hst/search.php}. The
observations with filters F475W, F606W, F814W, F110W, and F160W were
obtained in May 2010 (PI: Sahar Allam) and the observations with the
F275W filter in November 2011 (PI: Anna Quider).

For the data reduction we use \HST\ DrizzlePac\footnote{DrizzlePac is
  a product of the Space Telescope Science Institute, which is
  operated by AURA for NASA.}.  The size of a pixel after reduction is
0.04\arcsec for WFC3 UVIS (i.e. the F275W F475W, F814W and F606W band)
and 0.13\arcsec for the WFC3 IR (i.e. the F160W and F110W band),
respectively.  The software includes a sky background subtraction. In
our case the subtracted background appears to be overestimated since
many of the pixels have negative value, possibly due to the presence
of a very bright and saturated star in the lower-right corner of the
WFC3 field of view ($\approx 160 \arcsec \times 160 \arcsec$). Since
negative intensity is unphysical and we fit the surface brightness of
the pixels, we subtract the median of an empty patch of sky that we
pick to be around 25\arcsec N-E to the Cosmic
Horseshoe 
from all pixels of the reduced F160W-band image.  After our background
correction, around 300 pixels ($\approx$ 1.3\% of the full cutout) of
the corrected image still have negative values, which is consistent
with the number given by background noise fluctuations. We proceed in
a similar way with the F475W band, where the number of negative pixel
is still high but in the range of background fluctuations.

To align the images of the different filters we are using in this
paper, we model the light distribution of the star-like objects O2 and O3
(see Fig. \ref{fig:CH475}) in the F475W band, masking out all the
remaining light components (such as arc, lens and object O1). We do
not include object O1 in the alignment since we do not model the
light distribution of the lens in this band and the lens has
significant flux in the region of O1 that could affect the light
distribution of O1. From this model and our lens light model in
the F160W band, which we present in
Sec.~\ref{sec:compositeModel:lensLight}, we get the coordinates of the
centers of both objects in the two considered bands. Under the
assumption these coordinates should match, we are able to align the
F475W and the F160W images.

\subsection{Spectroscopy: redshift of the counter image of radial arc}
\label{sec:horseshoe:arc}

We obtained a spectrum of the counter-image to the radial arc using
the Gemini Near-InfraRed Spectrograph \citep[GNIRS;][]{GNIRS} on the
Gemini North Telescope (Program ID: GN-2012B-Q-42, PI Sonnenfeld). We
used GNIRS in cross-dispersed mode, with the 32 l/mm grating, the SXD
cross-dispersing prism, short blue camera ($0.15''$/pix) and a
$7''\times0.675''$ slit. This configuration allowed us to achieve
continuous spectral coverage in the range $9,000-25,000\AA$ with a
spectral resolution $R\sim900$. We obtained $18\times300s$ exposures,
nodding along the slit with an ABBA template.

\begin{figure}
\begin{tabular}{c}
\includegraphics[width=\columnwidth]{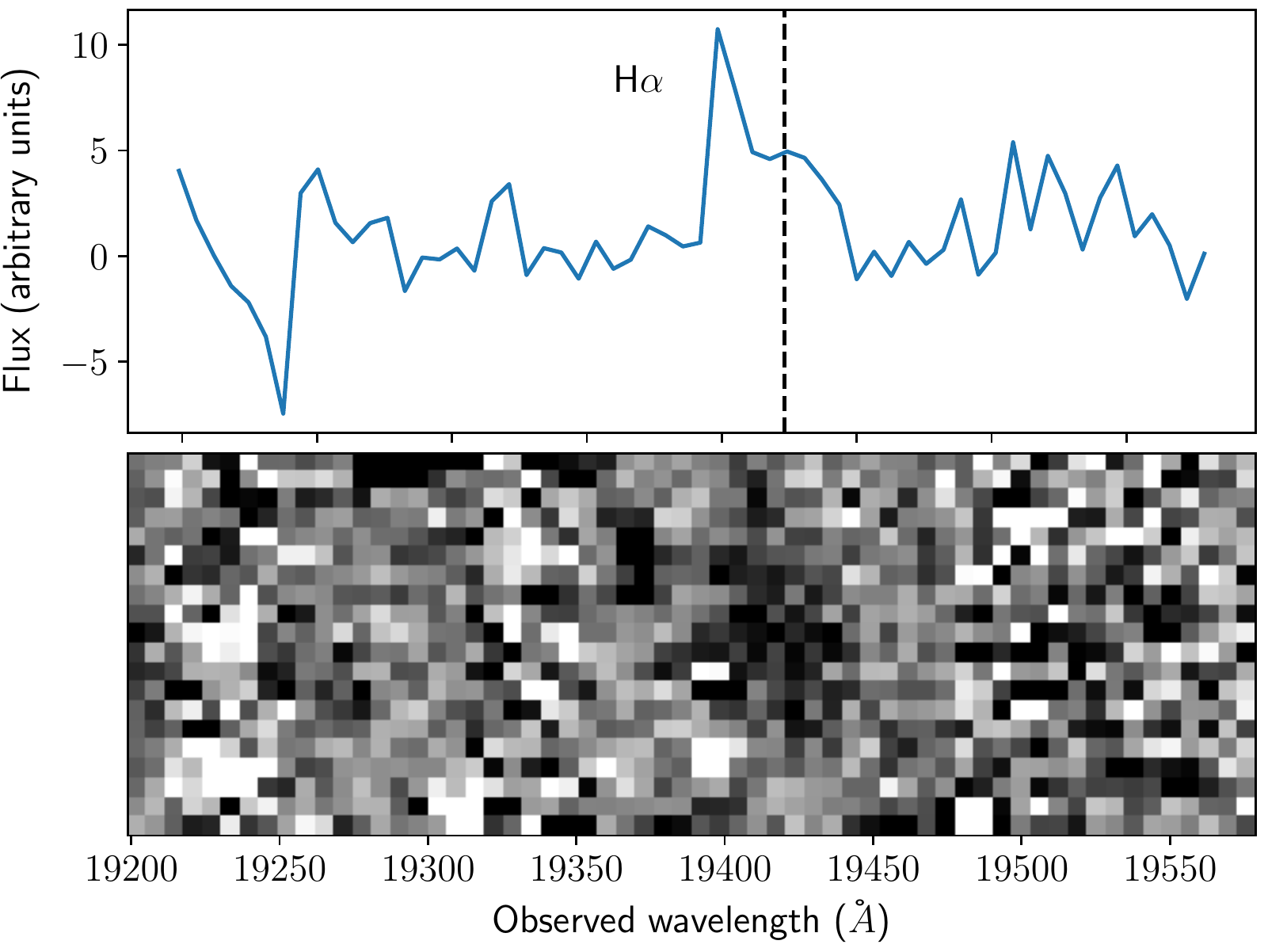} \\
\includegraphics[width=\columnwidth]{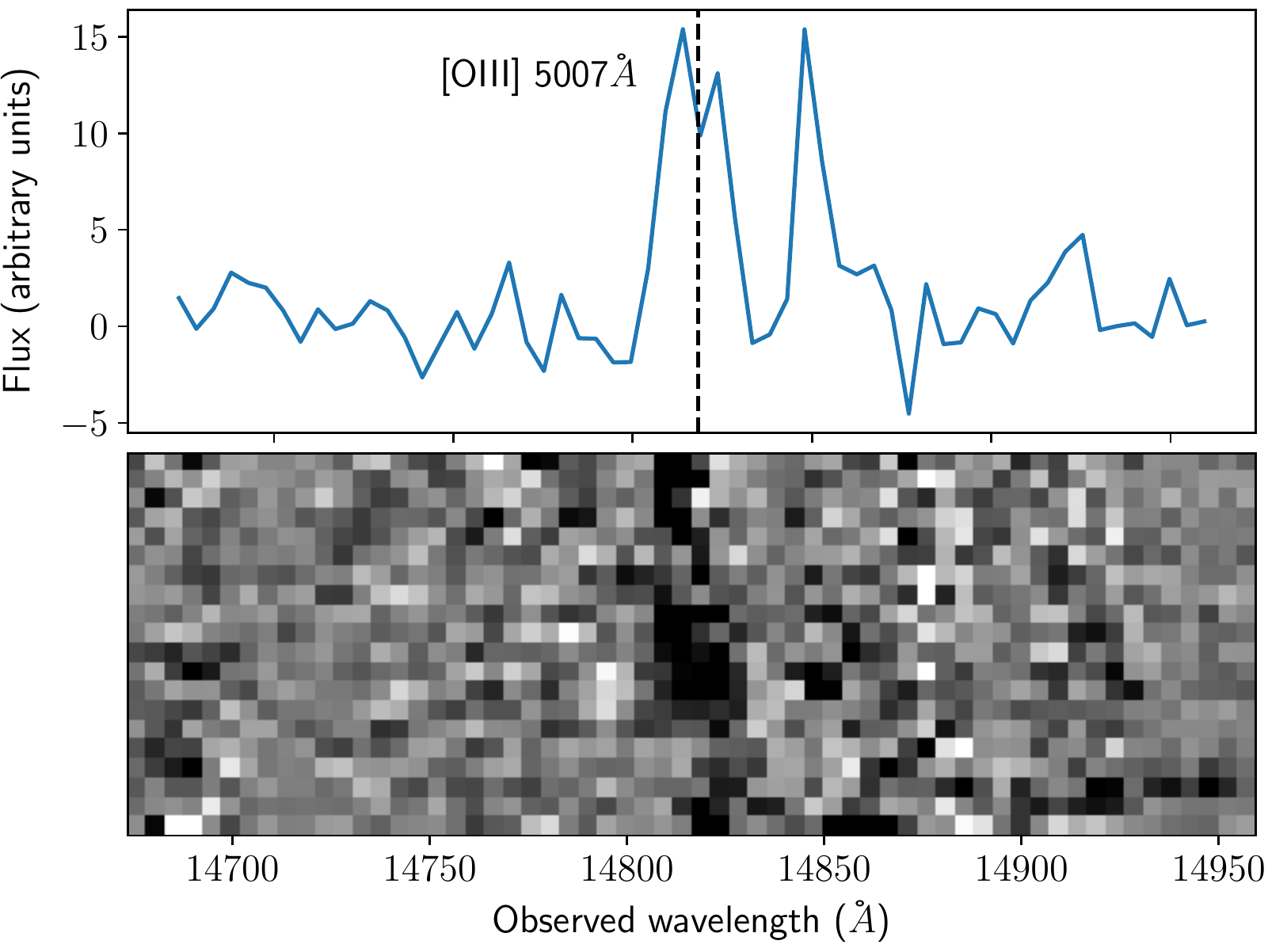}
\end{tabular}
\caption{{\em Top (Bottom)}: 2d and 1d spectrum around the H$\alpha$ ([OIII]~5007$\AA$) emission line from the counter-image to the radial arc, obtained from GNIRS observations. The 1D spectrum is extracted from a 5~pixel, corresponding to $0.75''$, aperture around each line.The secondary peak redward of [OII] visible in the 1D spectrum is due to a cosmic ray that was not properly removed in the data reduction process.}
\label{fig:2d_spec}
\end{figure}

We reduced the data using the Gemini IRAF package. We identified two
emission lines in the 2D spectrum, plotted in Fig. \ref{fig:2d_spec}:
these are H$\alpha$ and [OIII]~5007$\AA$, at a redshift $z_\text{s,r}
= 1.961\pm0.001$. From here on, we take this to be the redshift of the
radial arc and its counter-image.  

\FloatBarrier
\section{Multi-plane Lensing}
\label{sec:multiplane}
In this work we employ multi-plane gravitational lensing, given the
presence of two sources at different redshifts (corresponding to the
tangential and radial arcs, respectively). We therefore briefly
revisit in this section the single plane and generalized multi-plane
gravitational lens formalism. In the single plane formalism a light
ray of a background source is deflected by one single lens whereas, in
the multi-plane case, the same light ray is deflected several times by
different deflectors at different redshifts
\citep[e.g.,][]{blandford86, schneider06, gavazzi08}. The lens
equation of the multi-plane lens theory, which gives the relation
between the angular position $\vec{\theta}_j$ of a light ray in the
$j$-th lens plane and the angular position in the $j=1$ plane, which
is the observed image plane, is given by
\begin{equation}
\vec{\theta_j} (\vec{\theta}_1) = \vec{\theta}_1 - \sum^{j-1}_{k=1} \frac{D_{kj}}{D_j} \vec{\hat{\alpha}} ( \vec{\theta_k})~,
\end{equation}
where $\vec{\theta}_\text{N} = \vec{\beta}$ corresponds to the source
plane if $N$ is the number of planes, $\vec{\theta}_k$ is the image
position on the $k$-th plane, $\vec{\hat{\alpha}} ( \vec{\theta}_k )$
is the deflection angle on the $k$-th plane, $D_{kj}$ is the angular
diameter distance between the $k$-th and $j$-th plane, and $D_j$ is
the angular diameter distance between us and the $j$-th
plane. The total deflection angle $\vec{\alpha}_\text{tot}$ is then the sum
over all deflection angles on all planes \be \vec{\alpha}_\text{tot} =
\sum_{k=1}^{N-1} \frac{D_{kN}}{D_N} \vec{\hat{\alpha}}
(\vec{\theta}_k)~.
\label{eq:alphatot}
\ee
In the case of $N=2$ the general formula reduces to the well known lens equation for the single plane formalism, namely
\be
\vec{\beta} = \vec{\theta} - \frac{D_\text{ds}}{D_\text{s}} \vec{\hat{\alpha}} (D_\text{d}\vec{\theta})~.
\ee
Here the only lens is at $\vec{\theta} = \vec{\theta}_1$, the source
at $\vec{\beta} = \vec{\theta}_2$, $\vec{\hat{\alpha}}$ is the (total)
deflection angle, and $D_\text{ds}$, $D_\text{s}$ and $D_\text{d}$ the
distances between deflector (lens) and source, observer and source,
and observer and deflector, respectively \citep[e.g.][]{schneider06}.

The magnification $\mu$ is in the multi-plane formalism defined in the
same way as in the single plane formalism, namely
\be
\mu = \frac{1}{\det{\vec{A}}}
\ee
with the Jacobian matrix
\be
\vec{A} = \frac{\partial \vec{\beta}}{\partial \vec{\theta}} = \frac{\partial \vec{\theta}_N}{\partial \vec{\theta}_1}~.
\label{eq:jacobianmatrix}
\ee

For the surface mass density $\Sigma (R)$ one needs the convergence
$\kappa$, sometimes also called the dimensionless surface mass
density. In the single-lens plane case, the convergence is
\be
2\kappa = \frac{\partial \alpha_1}{\partial \theta_1} + \frac{\partial \alpha_2}{\partial \theta_2}= \vec{\nabla}_\theta \cdot \vec{\alpha} ~,
\label{eq:kappa}
\ee
where $\vec{\alpha} = (D_{\rm
  ds}/D_{s})\vec{\hat{\alpha}}$. This can then be
multiplied with
\be
\Sigma_\text{crit} = \frac{c^2}{4 \pi G} \frac{D_\text{s}}{D_\text{d}D_\text{ds}}
\ee
to derive $\Sigma (R)$ using the definition of convergence
\be
\kappa = \frac{\Sigma (R)}{\Sigma_\text{crit}}~.
\ee
We can then compute the average surface mass density with the formula
\be
\Sigma (<R) = \frac{\int_0^R \Sigma(R')~ 2\pi R'  \, \mathrm{d} R'}{\pi R^2} ~.
\ee
These general equations hold in the single plane case, but for the
multi-plane case one defines similar, so-called effective,
quantities. For calculating the effective convergence
$\kappa_\text{eff}$ one replaces in Eq. \ref{eq:kappa} the deflection
angle $\vec{\alpha}$ with the total deflection angle
$\vec{\alpha}_\text{tot}$ from Eq. \ref{eq:alphatot}. In analogy to
the case above one computes the effective average surface mass density
$\Sigma_\text{eff}(<R)$, now using $\kappa_\text{eff}$ instead of
$\kappa$. The consequence is that this quantity
$\Sigma_\text{eff}(<R)$ is the gradient of the total deflection angle
$\alpha_\text{tot}$ instead of a physical surface density.

\FloatBarrier
\section{Lens mass models}
\label{sec:compositeModel}

Since the position of an observed gravitationally lensed image depends
on both baryonic and dark matter, one can use gravitational lensing as
a probe for the total mass, i.e. baryonic and dark matter together. We
start with a model of the lensed source positions, i.e. surface
brightness peaks in the observed Einstein ring, with a single
power-law plus external shear for the total mass. In addition to the
main arc, which is the tangential arc, this model includes the radial
arc and its counter image and is presented in
Sec.~\ref{sec:compositeModel:LensMass}. Based on this, we construct a
composite mass model to describe the total mass. To disentangle the
visible (baryonic) matter from the dark matter, we model the lens
light distribution (see Sec.~\ref{sec:compositeModel:lensLight}) which
is then scaled by a constant mass-to-light ratio $M/L$, for the
baryonic mass. Combining the total mass and the baryonic mass, we
construct in Sec.~\ref{sec:compositeModel:DMmass} a composite mass
model of baryons and dark matter assuming a power-law
\citep{barkana98}, a NFW profile \citep*{navarro97}, or a generalized
NFW profile for the dark matter distribution. We use then a model
based on the full \HST\ images (Sec.~\ref{sec:compositeModel:esr}) to
refine our image positions (Sec.~\ref{sec:compositeModel:redef}). In
these models we always include the radial arc and our assumption for
its counter image. Only in the last section with the redefined image
positions we treat explicitly models with and without the radial arc
as constraints. This would allow us to quantify the additional
constraint on the inner dark matter distribution of the \CH\ from the
radial arc, which is the primary goal of this paper.

For the modeling, we use \GLEE\ (Gravitational Lens Efficient
Explorer), a gravitational lensing software developed by S.~H.~Suyu
and A.~Halkola \citep{suyu10, suyu12}. This software contains several
types of lens and light profiles and uses Bayesian analysis such as
simulated annealing and Markov Chain Monte Carlo (MCMC) to infer the
parameter values of the profiles.  The software also employs the {\sc
  Emcee} package developed by \citet{ForemanMackey2013} for sampling
the model parameters.

\subsection{Power-law model for total mass distribution}
\label{sec:compositeModel:LensMass}

In this section, we consider a simple power-law model for the total
lens mass distribution, which has been shown by previous studies to
describe well the observed tangential arc \citep[e.g.,][]{belokurov07,
  dye08, quider09, bellagamba17}. This will allow us to compare our
model, that includes the radial arc, with previous models.  We
visually identify and use as constraints six sets of multiple image
positions, where each set comes from a distinct source component. For
modeling the lensed source positions we choose the image of the F475W
band, since one can distinguish better between the different parts of
the Einstein ring and since the arc is bluer than the lens
galaxy. This is an indicator that the lens galaxy is fainter and
therefore one can better identify multiple images in
F475W. Here we use a singular power-law elliptical mass distribution
\citep[SPEMD;][]{barkana98} with slope $\gamma' = 2g+1$ for the lens (where the convergence $\kappa(\vec{\theta}) \propto \vec{\theta}^{\gamma'-1})$ with an external shear.
We infer the best-fit parameters of this model by minimizing
\be
\label{eq:chi2pt}
 \chi^2 = \sum_{j=1}^{N_{\rm pt}} \frac{\left(\vec{\theta}_j^\text{obs}-\vec{\theta}_j^\text{pred}\right)^2}{\sigma_j^2}
\ee
with \GLEE.
Here $N_{\rm pt}$ is the number of data points,
$\vec{\theta}_j^\text{pred}$ the predicted and
$\vec{\theta}_j^\text{obs}$ the observed image position, with
$\sigma_j$ the corresponding uncertainty of point $j$.

This model contains six sets of multiple images in addition to the
radial arc and its counter image (see
Fig. \ref{fig:sourcePositionModel_redef} with refined identifications
that will be described in Sec.~\ref{sec:compositeModel:redef}). This model has a $\chi^2$ of 12.6 for the image positions and the
best-fit parameter and median values with 1-$\sigma$ uncertainties are
given in Table \ref{tab:SPEMDparameters}. The obtained marginalized and best-fit values for the total mass model are in agreement with models from previous studies \citep[e.g.,][]{dye08, spiniello11}.
 
\begin{table}[ht!]			
\renewcommand{\arraystretch}{1.4} 
 \caption{Best-fit and marginalized parameter values for the model assuming a power-law profile plus external shear.
          }
 \begin{center}
 \begin{tabular}[width=\textwidth]{c|c|c|c}
  component & parameter & best-fit value & marginalized value\\ \hline 
            & $x \ [\arcsec]$       & 10.86 &$10.92^{+0.05}_{-0.05} $   \\ 
            & $y \ [\arcsec]$       & 9.60 &$9.61^{+0.04}_{-0.04} $   \\
            & $q$       & 0.76 & $ 0.78^{+0.04}_{-0.04} $ \\
  power-law & $\theta$ [rad]  & 0.58 &$0.51^{+0.07}_{-0.08} $ \\
            & $\theta_\text{E}$  & 8.06 &$7.6^{+0.5}_{-0.5} $ \\
            & $r_\text{c} \ [\arcsec]$ & 0.01 &$0.29^{+0.3}_{-0.3} $ \\
            & $\gamma' $ & 1.7 & $ 2.0^{+0.4}_{-0.2}$ \\ \hline
  shear     & $\gamma_\text{ext}$ & 0.08 & $ 0.07^{+0.02}_{-0.02}
$ \\
            & $\phi_\text{ext} \ [\text{rad}]$ & 3.5 & $ 3.2^{+0.2}_{-0.3}$ \\
 \end{tabular}
 \end{center}
 \label{tab:SPEMDparameters}
  Note. The parameters $x$ and $y$ are centroid coordinates with
  respect to the bottom-left corner of our cutout, $q$ is the axis
  ratio, $\theta$ is the position angle measured counterclockwise from
  the $x$-axis, $\theta_\text{E}$ is the Einstein radius, $r_\text{c}$
  is the core radius, $\gamma'$ is the slope, $\gamma_\text{ext}$ is
  the external shear magnitude, and $\phi_\text{ext}$ is the external
  shear orientation. The constraints for this model are the selected
  multiple image systems. The best-fit model has an image position
  $\chi^2$ of 12.6.
\end{table}

\subsection{Components for composite mass model}

Since the light deflection depends on both the baryonic and the dark
matter, we can construct a composite mass model. For the baryonic
component, we need a model of the lens light to scale it by a
mass-to-light ratio (Sec.~\ref{sec:compositeModel:lensLight}). Since
we do not have other information to infer the dark matter component,
we fit to the data using different types of mass profiles
(Sec.~\ref{sec:compositeModel:DMmass}).

\subsubsection{Lens light distribution for baryonic mass}
\label{sec:compositeModel:lensLight}

To disentangle the baryonic matter from the dark matter, we need a
model of the lens light distribution. For this we mask out all flux
from other components such as stars and the Einstein ring in the image
of the F160W filter. We then fit the parameters to the observed
intensity value by minimizing the $\chi^2_\text{lens}$, which is
defined as
\begin{equation}
\chi^2_\text{lens} = \sum_{j=1}^{N_{\rm p}} \frac{ \left( I^{\text{obs}}_j - \text{PSF} \otimes I^{\text{sersic}}_j \right)^2}{\sigma_{\text{tot},j}^2}~.
 \label{eq:chi2}
\end{equation}

Here $N_\text{p}$ is the number of pixels, $\sigma_{\text{tot},j}$ the
total noise, i.e. background and Poisson noise (see below for
details), of pixel $j$, and $\otimes$ represents the convolution of
the point spread function (PSF) and the predicted intensity. It is
necessary to take the convolution with the PSF into account due to
telescope effects. Here we use a normalized bright star $\approx 40
\arcsec$ S-W of the \CH\ lens as the PSF. We subtract also from the PSF a constant to counterbalance the background coming from a very bright object in the field of view which scatters light over the image.

We approximate the background noise $\sigma_\text{bkgd}$ as a constant
that is set to the standard deviation computed from an empty
region. We also include the contribution of the astrophysical Poisson
noise \citep{hasinoff12}, which is expressed as a count rate for pixel
$i$
\begin{equation}
\sigma^2_{\text{poisson}, i} = \left( \frac{\sigma'_{\text{tot},i}}{t_i} \right)^2 = \left( \frac{\sqrt{d_i t_i}}{t_i} \right)^2 = \left| \frac{d_i}{t_i} \right|~,
\end{equation}
where $t_i$ is the exposure time, $d_i$ the observed intensity of
pixel $i$ (in $e^- $-counts per second) and $\sigma'_{\text{tot},i}$
is the total Poisson noise (labeled with an apostrophe as it is not a
rate like $\sigma_{\text{poisson,}i}$). We include the contribution of
the astrophysical Poisson noise only if it is larger than the
background noise. We sum the background noise and astrophysical noise
in quadrature, such that $\sigma_{\text{tot},j}^2$ in
Eq.~(\ref{eq:chi2}) is
\be
\sigma_{\text{tot},j}^2 = \sigma_{\text{bkgd},j}^2 + \sigma_{\text{poisson},j}^2 ~.
\ee

\subsubsection*{Sersic}

\label{sec:compositeModel:lensLight:sersic}

To describe the surface brightness of the Cosmic Horseshoe lens galaxy, we use the commonly adopted Sersic profile  \citep{sersic63}, which is the generalization of the de~Vaucouleurs law \citep[also called $r^{1/4}$ profile,][]{DeVaucouleurs48}. For modeling the lens light distribution we choose the observation in
the F160W band, since the lens is brighter in F160W than in the other
bands, and infrared bands trace better the stellar mass of the lens
galaxy.

The best-fit model obtained by using two Sersic profiles and two
stellar profiles (in this model we include two star-like objects, labelled object O1
and object O2 in Fig. \ref{fig:CH475}) has $\chi^2 = 2.73 \times 10^4$
(corresponding to a reduced $\chi^2$ of 1.74).

\subsubsection*{Chameleon}
\label{sec:compositeModel:lensLight:cham}

In addition to our lens light distribution model with the Sersic
profile, we also model with another type of profile which mimics the
Sersic profile well and allows analytic computations of lensing
quantities \citep[e.g.,][]{maller00, dutton11, suyu14}.  It is often
called chameleon and composed by a difference of two isothermal
profiles:
\bea
\label{eq:isothermal}
\displaystyle	
{\Large L(x,y)}  = & {\Large \frac{L_0}{1+q_\text{L}} \left( \frac{1}{\sqrt{x^2 + y^2 / q_\text{L}^2 + 4 w_\text{c}^2 / (1+q_\text{L})^2}} \right.} \nonumber  \\
& { \Large - \left. \frac{1}{\sqrt{x^2 + y^2 / q_\text{L}^2 + 4 w_\text{t}^2 / (1+q_\text{L})^2}} \right)~.}
\eea
In this equation, $q_\text{L}$ is the axis ratio, and $w_\text{t}$ and $w_\text{c}$ are parameters of the profile with $w_\text{t}>w_\text{c}$ to keep $L>0$.

By modeling with the chameleon profile we assume the same background
noise as using the Sersic profile (see
Sec.~\ref{sec:compositeModel:lensLight:sersic}). Since the model
including two isothermal profile sets and two stellar profiles for the
two objects, as used above with the Sersic profile, has a $\chi^2$ of
around two times the Sersic-$\chi^2$, we add a third chameleon profile
and get a $\chi^2$ of $2.89\times 10^4$ which corresponds to a reduced
$\chi^2$ of 1.85. In this model we include also objects O1, O2 and O3 
(numbering follows Fig. \ref{fig:CH475}), since we want to use
the coordinates for the alignment of the two considered bands,
F160W and F475W.

\begin{figure*}[ht!]
\centering
\subfigure[observed]{\includegraphics[trim=50 48 50 0, clip, width=0.65\columnwidth]{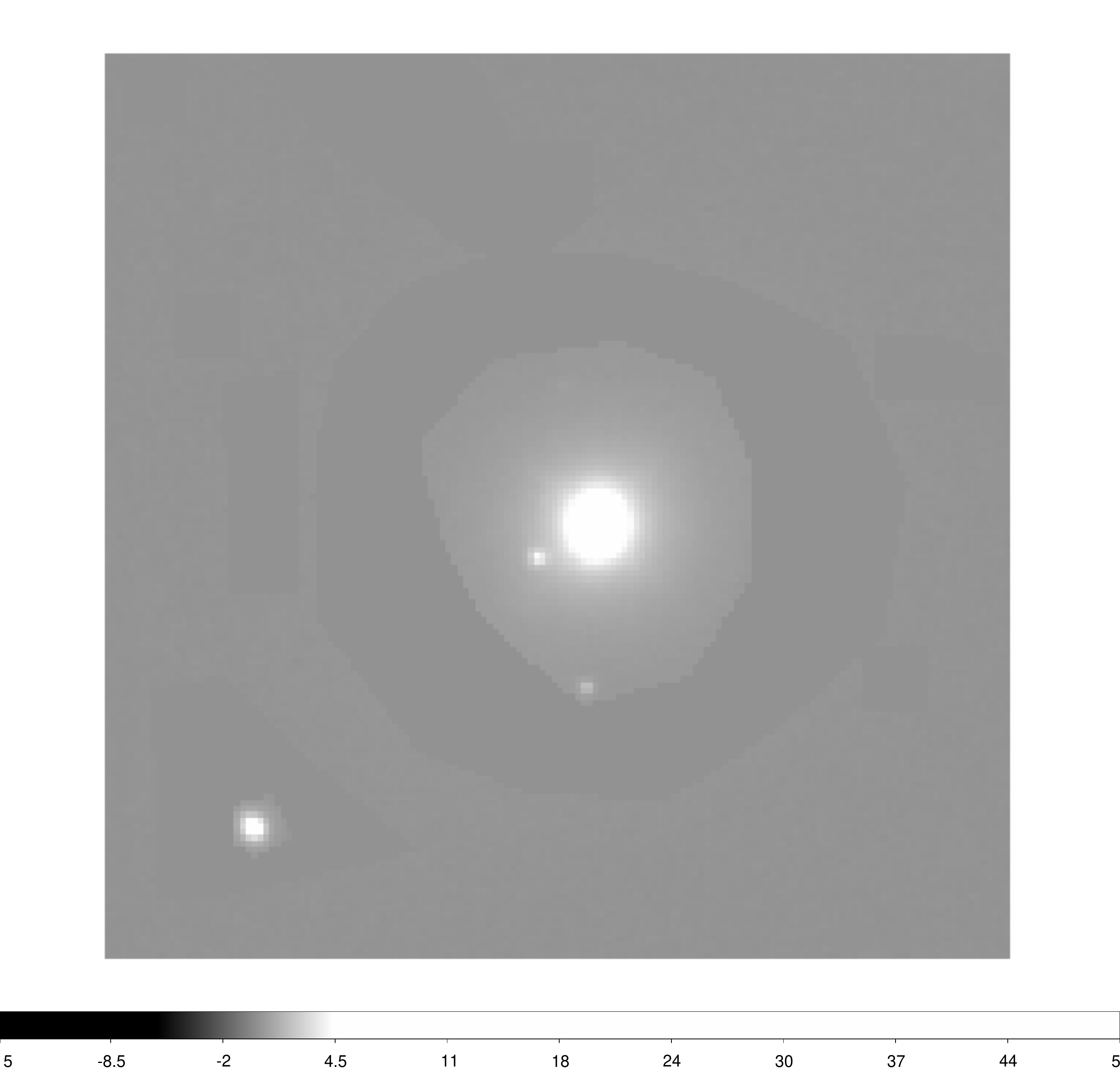} \label{fig:bestfit_cham:obs}}\hfill
\subfigure[model]{\includegraphics[trim=50 48 50 0, clip, width=0.65\columnwidth,]{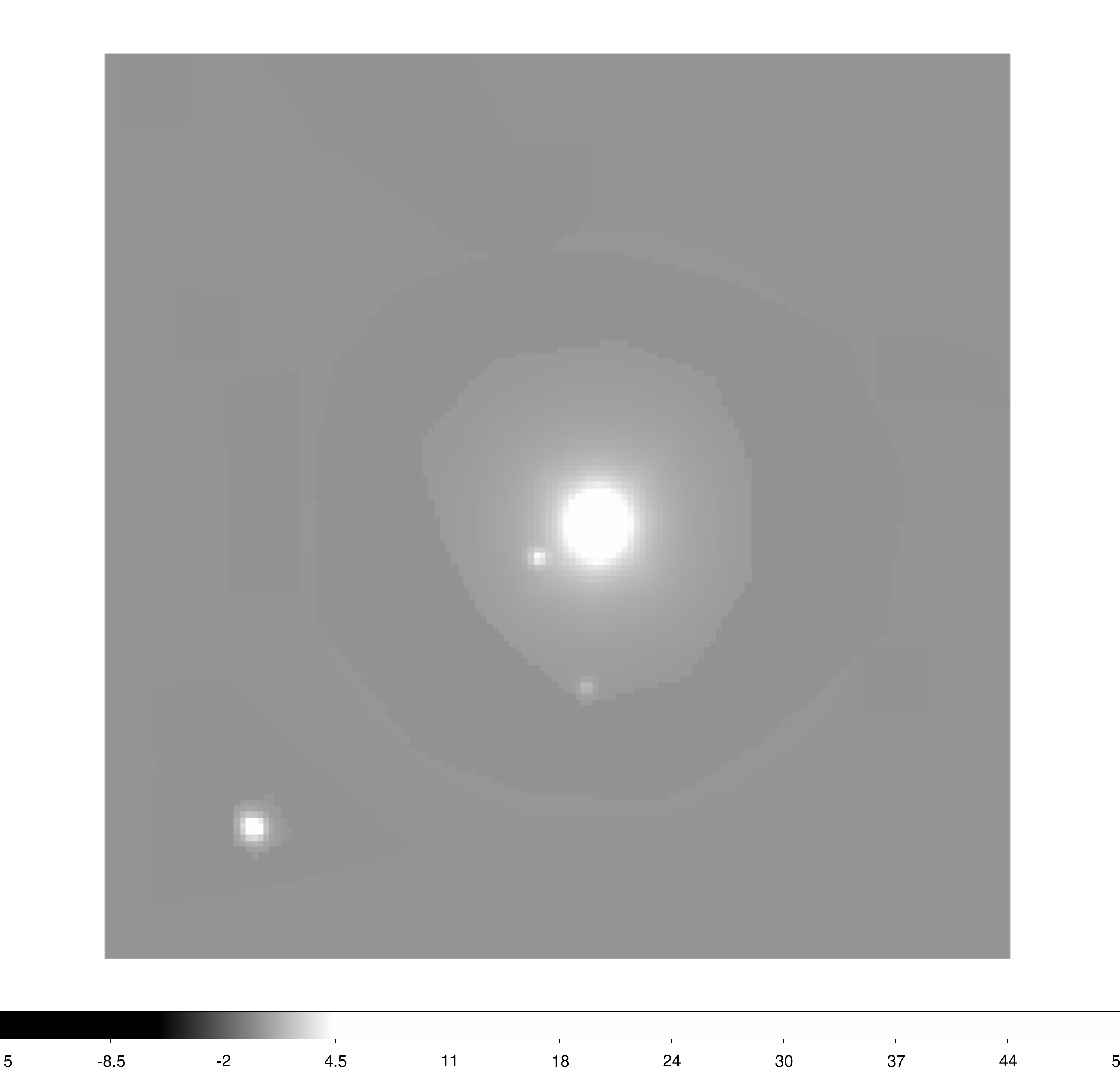} \label{fig:bestfit_cham:pred}}\hfill
\subfigure[normalized residuals]{\includegraphics[trim=50 48 50 0, clip, width=0.65\columnwidth]{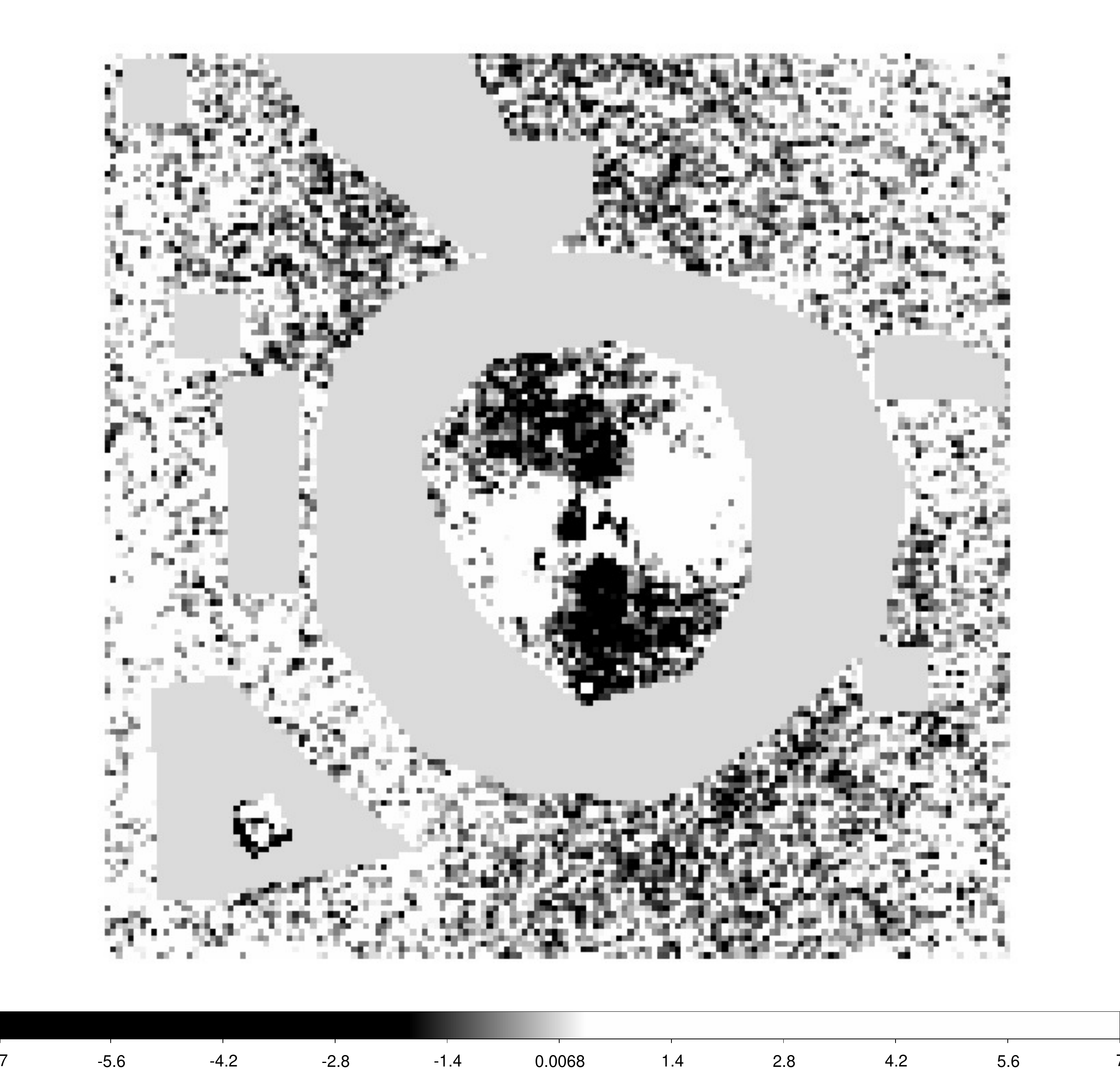} \label{fig:bestfit_cham:normres}}\hfill
\caption{The best-fit model for the lens light distribution. The left image shows the observation of the \CH\ in the F160W-band, whereas the central panel shows the predicted light distribution. This model includes three chameleon profiles (see Eq. \ref{eq:isothermal}) and two PSF and one de~Vaucouleurs profiles for the three objects. 
The right image shows, in a range between $-7 \ \sigma$ and $+7 \ \sigma $, the normalised residuals of this model. The constant gray regions are the masked-out areas (containing lensed arcs and neighboring galaxies) in order to fit only to the flux of the lens. The figures are oriented such that North is up and East is left.}
\label{fig:bestfit_cham}
\end{figure*}

We will use both filters in the extended source modeling (see
Sec.~\ref{sec:compositeModel:esr}) while in the models using identified image positions we only use the
F160W band for the lens light fitting. The parameter values of this best-fit model are
used for the mass modeling (given in Table \ref{tab:bestfit_esr_light}) and the corresponding image is shown in
Fig.~\ref{fig:bestfit_cham}. The left image shows the observed
intensity and the middle the modeled intensity. In the right panel one
can see the normalized residuals of this model in a range ($-7 \sigma,
+7 \sigma$). The constant gray regions are the masked-out areas
(containing lensed arcs and neighbouring galaxies) in order to fit
only to the flux of the lens. Although there are significant image 
residuals visible in the right panel, the typical baryonic mass
residuals (corresponding to the light residuals scaled by $M/L$) would
lead to a change in the deflection angle that is smaller than the
image pixel size of $0.\arcsec13$ at the locations of the radial arc.

\begin{figure}[ht!]
\centering
\includegraphics[angle = 0, trim=10 5 20 40, clip, width=1.0\columnwidth]{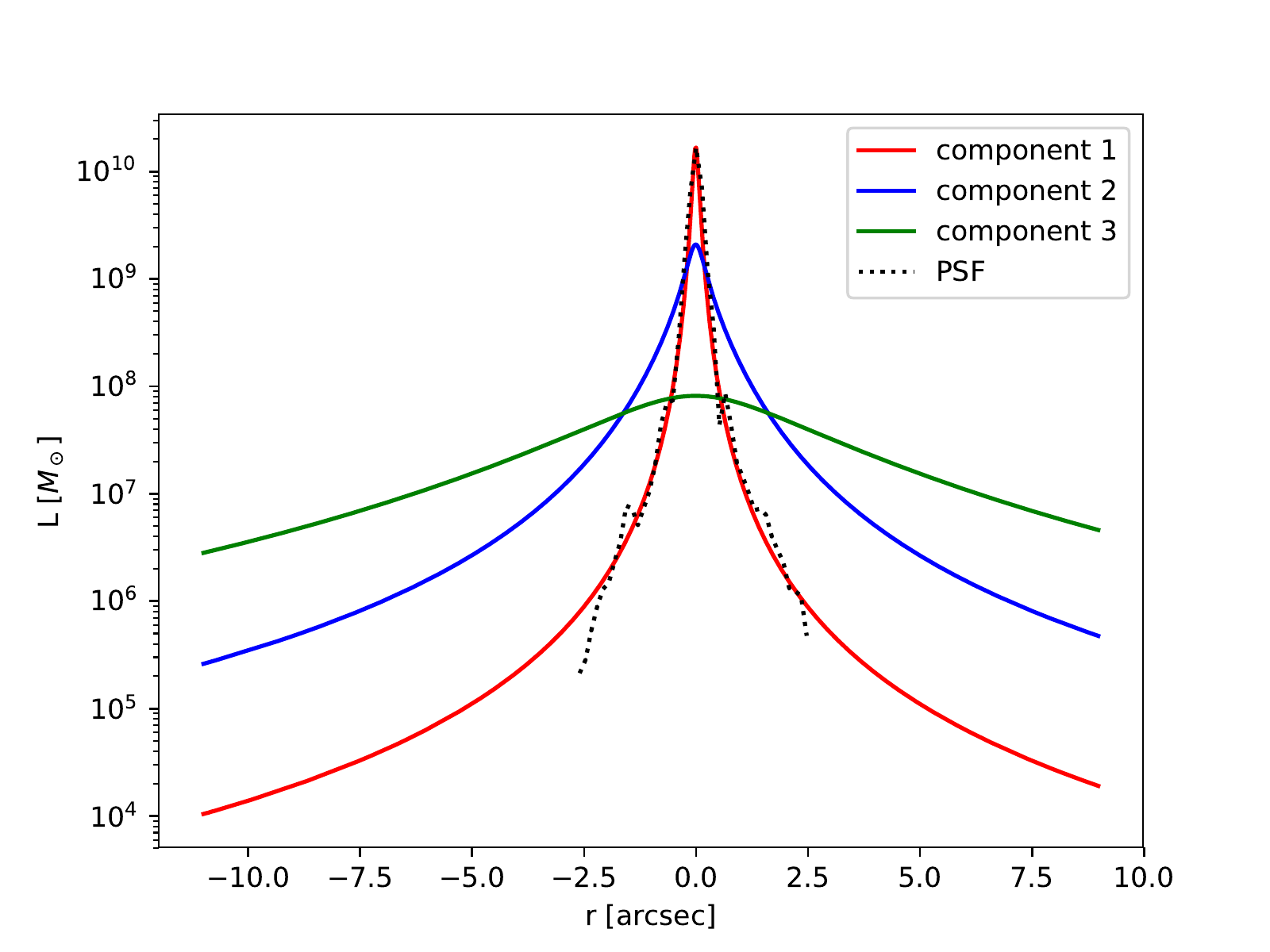}
\caption{Different components of the chameleon profiles shown in units of solar luminosity, respectively in red (``inner''), blue (``medium''), and green (``outer'' component). The total light observed from the \CH\ lens galaxy in the \HST\ filter F160W is described by the sum of all three components. For comparison of the width of the components the scaled PSF is plotted with a black dotted line.}
\label{fig:lenslight_with_PSF}
\end{figure}

In Fig.~\ref{fig:lenslight_with_PSF} we show the contributions of the
different components, plotted along the $x$-axis of the cutout in
units of solar luminosities for comparison of the contribution of the
different light profiles. To compare those components' widths to that
of the PSF, in the same figure we show the latter (black dotted line)
scaled to the lens light of the central component (plotted in red). 

To convert the fitted light distribution into the baryonic mass, we assume at first a
constant mass-to-light ratio. This means we scale all three light
components by the same $M/L$ value. Additionally, we explore models
with different $M/L$ values for the different components, either two
ratios with $M/L_\text{central} = M/L_\text{medium}$ or
$M/L_\text{medium} = M/L_\text{outer}$ and the remaining different, or
with three different $M/L$ values one for each component. 
These baryonic mass models are considered in the Sections
\ref{sec:compositeModel:esr} and \ref{sec:compositeModel:redef:threecham}.   
Furthermore,
since the width of the central component, shown in red in
Fig~\ref{fig:lenslight_with_PSF}, is comparably to the PSF's width,
and based on our modeling results in Section
\ref{sec:compositeModel:redef:threecham}, we model in Section
\ref{sec:LensingDynamics} this central component
by a point mass with Einstein radius described by
\be
\theta_{E, \text{point}} = \frac{4 G M}{c^2 D_\text{d}} 
\ee
(where the Einstein radius is defined here for a source at redshift
infinity), superseding the model that scales the central component
with an $M/L$.  Here $G$ is the gravitational constant, $M$ the point
mass, $c$ the speed of light, and $D_\text{d}$ the distance to the
deflector.  For the remaining two components (blue and green in
Fig.~\ref{fig:lenslight_with_PSF}) we assume either one or two
different mass-to-light ratios to scale the light to a mass.

\subsubsection{Dark matter halo mass distribution}
\label{sec:compositeModel:DMmass}

In the previous section we have derived the baryonic component by
modeling the light distribution. To disentangle the baryonic mass from
the dark component, we model the dark matter distribution using three
different profiles. At first we use a NFW \citep{navarro97} profile
but, since newer simulations predict deviations from this simple
profile, we present in addition the best-fit mass model obtained
assuming a power-law profile \cite[Singular Power-Law Elliptical Mass
Distribution]{barkana98} (with parameters $q$ as axis ratio, 
$\theta_\text{E}$ as Einstein radius, and $r_\text{c} $ as core
radius) and a generalized 
version of the NFW profile, given by
\begin{equation}
\rho (r) =  \frac{\rho_\text{s}}{\left(\frac{r}{r_\text{s}} \right)^{\gamma_\text{g}} \times \left( 1 + \frac{r}{r_\text{s}} \right)^{3-\gamma_\text{g}}} \ ,
\label{eq:gnfw}
\end{equation}
where $\gamma_\text{g}$ is the inner dark matter slope. The generalized NFW profile reduces to the standard NFW profile in the case $\gamma_\text{g}=1$.

We assume an axisymmetric lens mass distribution (axisymmetric in 3
dimensions), and impose the projected orientation of the dark matter
profile to be $0^\circ$ or $90^\circ$ rotated with respect to that of
the projected light distribution. We find that the $90^\circ$
orientation gives a better $\chi^2$, and thus the dark matter halo
seems to be prolate, for an axisymmetric system that has its rotation
axis along the minor axis of the projected light distribution. Since
strong lensing is only sensitive on scales of the Einstein radius, we
assume four different values for the scale radius in the NFW and gNFW
profile, namely $r_\text{s} \equiv 18.11 \arcsec$, $36.22\arcsec$,
$90.54\arcsec$, and $181.08\arcsec$. These values correspond to 100
kpc, 200 kpc, 500 kpc, and 1000 kpc, respectively, for the lens
redshift in the considered cosmology. We include the mass of the
radial arc source in the model, using a singular isothermal sphere
(SIS) profile, as this source galaxy's mass will deflect the light
coming from the background tangential arc source. The center of this
profile is set to the coordinates for the radial arc source which we
obtained from the multiplane lensing, calculated by the weighted mean of the mapped positions of the radial arc and its counter image on the redshift plane of the radial arc. 

\subsection{Extended source modeling}
\label{sec:compositeModel:esr}

In the next stage of our composite mass model, we reconstruct the
source surface brightness (SB) distribution and fit to the observed
lensed source light, i.e. the main arc and the radial arc with its
counter image. This will help us to refine our image positions
afterwards. For this, we start with the mass model obtained
in Sec.~\ref{sec:compositeModel:DMmass}, which includes the lens light
distribution described by the three chameleon profiles scaled with a constant mass-to-light ratio as baryonic
mass and a power-law profile for the dark matter halo. We then allow the mass parameters to
vary and, for a given set of mass parameter values, \GLEE\
reconstructs the source SB on a grid of pixels \citep{suyu06}. This
source is then mapped back to the image plane to get the predicted
arc. To infer the best-fit parameters, one optimizes with \GLEE\ the
posterior probability distribution which is proportional to the
product of the likelihood and the prior of the lens mass parameters
(we refer to \citet{suyu06} and \citet{suyu10} for more
details).
The fitting of the SB distribution has
\begin{equation}
\chi_\text{SB}^2 = ( \vec{d} -\vec{d}^\text{pred})^T C_\text{D}^{-1} (\vec{d} -\vec{d}^\text{pred})~,
\end{equation}
where $\vec{d} = \vec{d}^\text{lens} + \vec{d}^\text{arc}$ is the
intensity values $d_j$ of pixel $j$ written as a vector with length
$N_\text{d}$, the number of image pixels, and $C_\text{D}$ is the
image covariance matrix. 
In the pixellated source SB reconstruction, we impose
curvature form of regularization on the source SB 
pixels \citep{suyu06}.

\begin{figure*}[ht!]
\centering
\subfigure[observed]{\includegraphics[trim=70 50 70 50, clip, width=0.49\columnwidth]{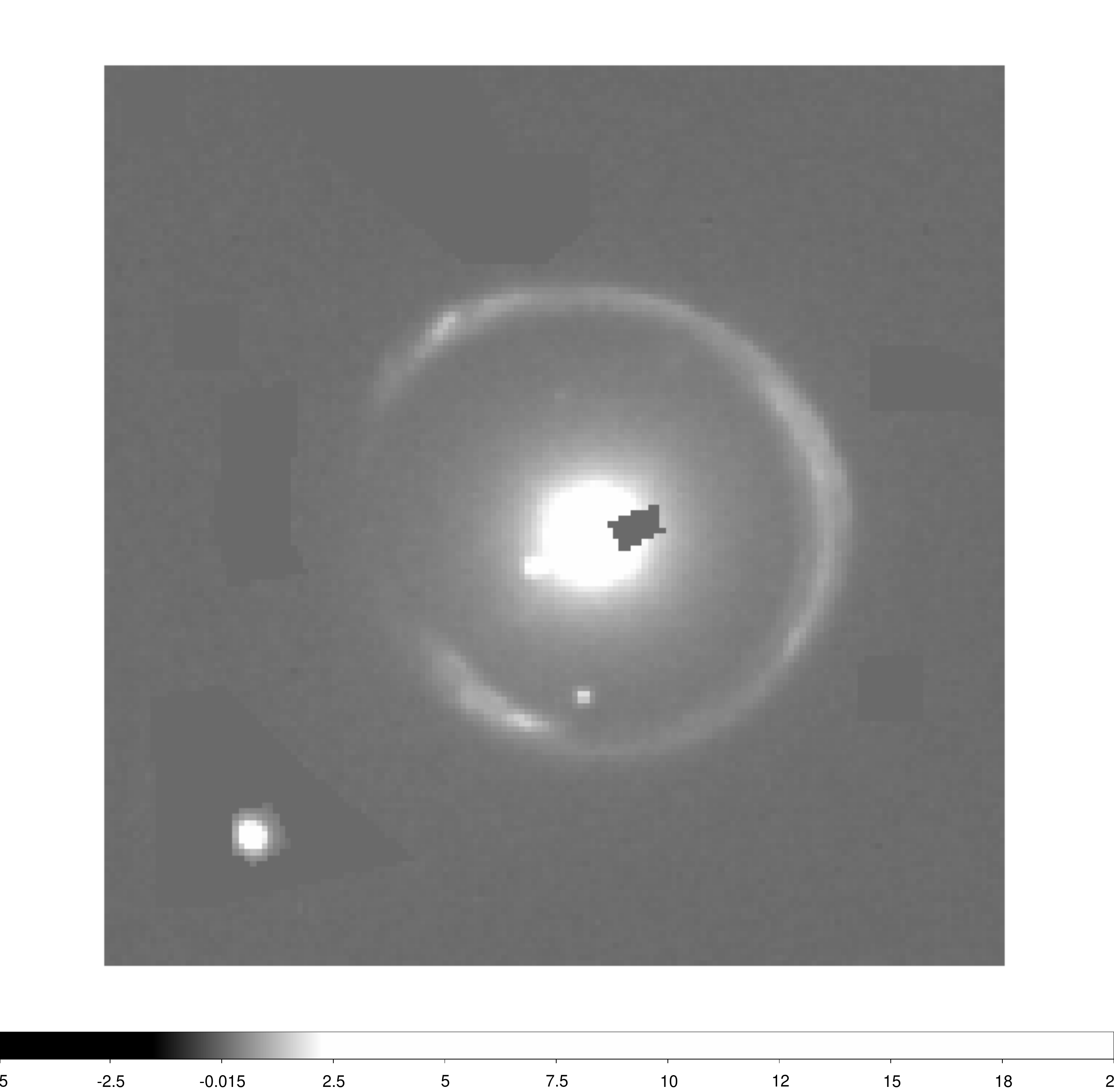} \label{fig:bestfit_esr_160Wobs}}\hfill
\subfigure[predicted]{\includegraphics[trim=70 50 70 50, clip, width=0.49\columnwidth]{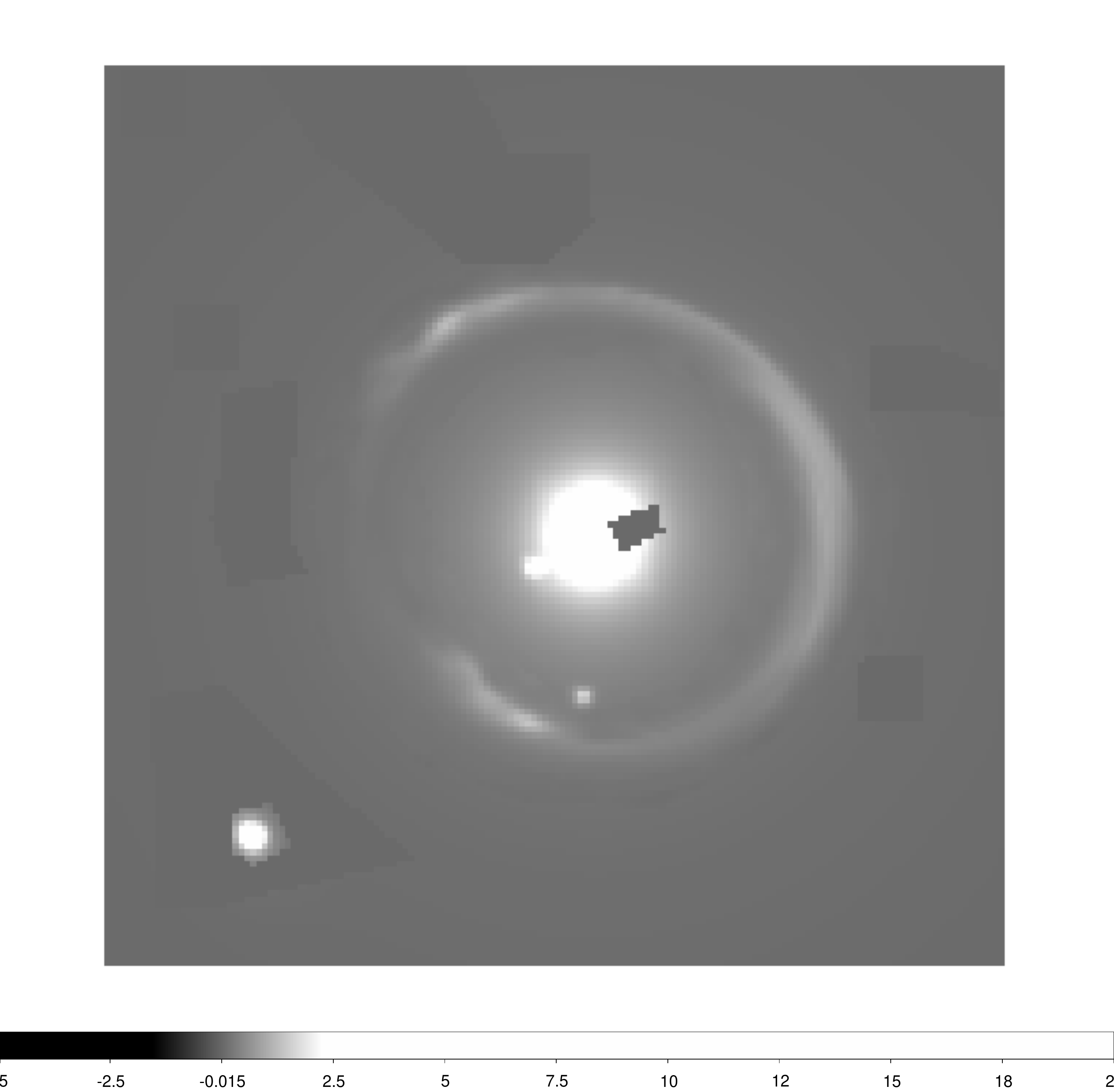} \label{fig:bestfit_esr_160Wpred}}\hfill
\subfigure[normalized residuals]{\includegraphics[trim=70 50 70 50, clip, width=0.49\columnwidth]{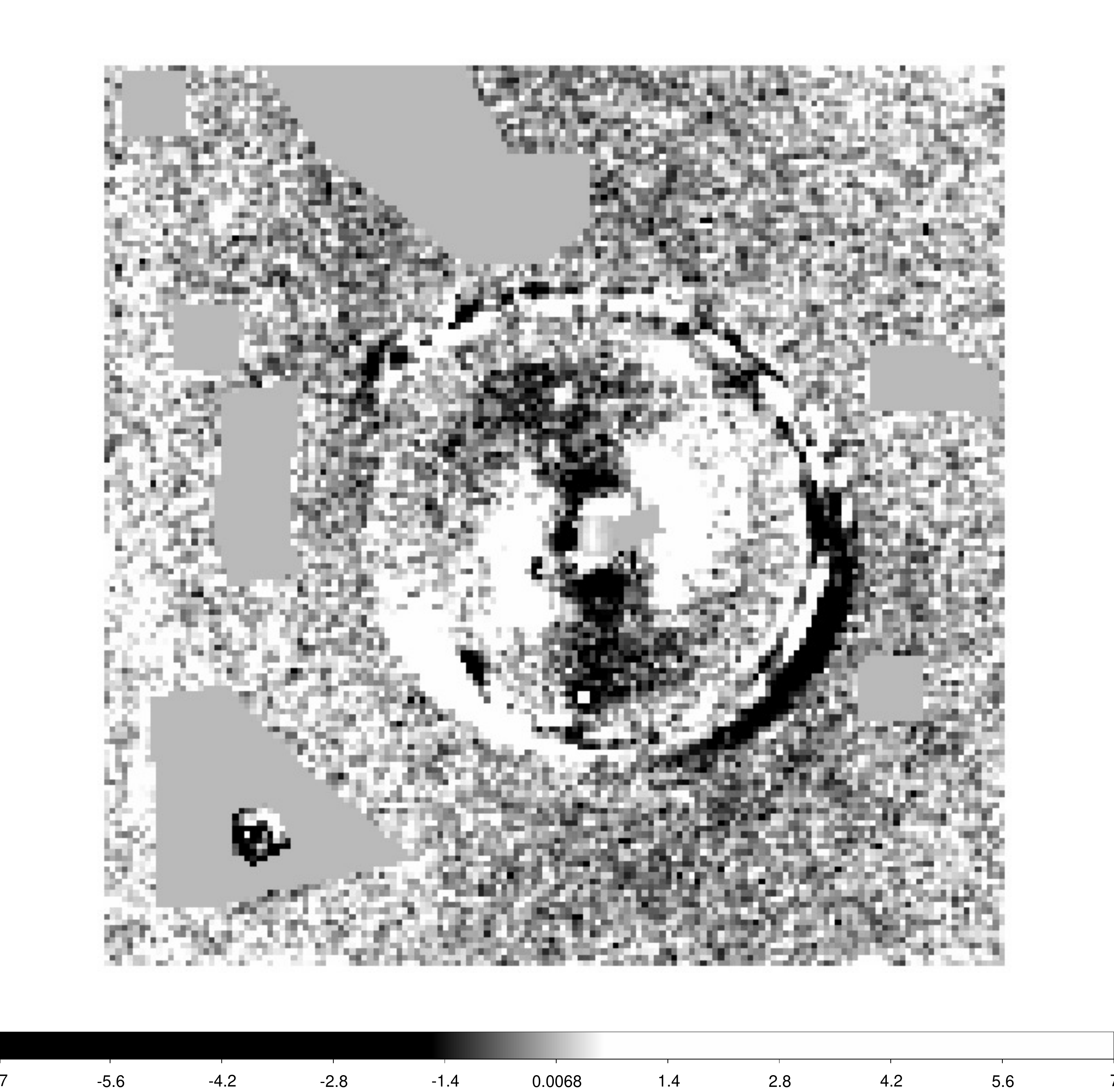} \label{fig:bestfit_esr_160Wsigma}}\hfill
\subfigure[source reconstruction]{\includegraphics[trim=70 50 70 50, clip, width=0.49\columnwidth]{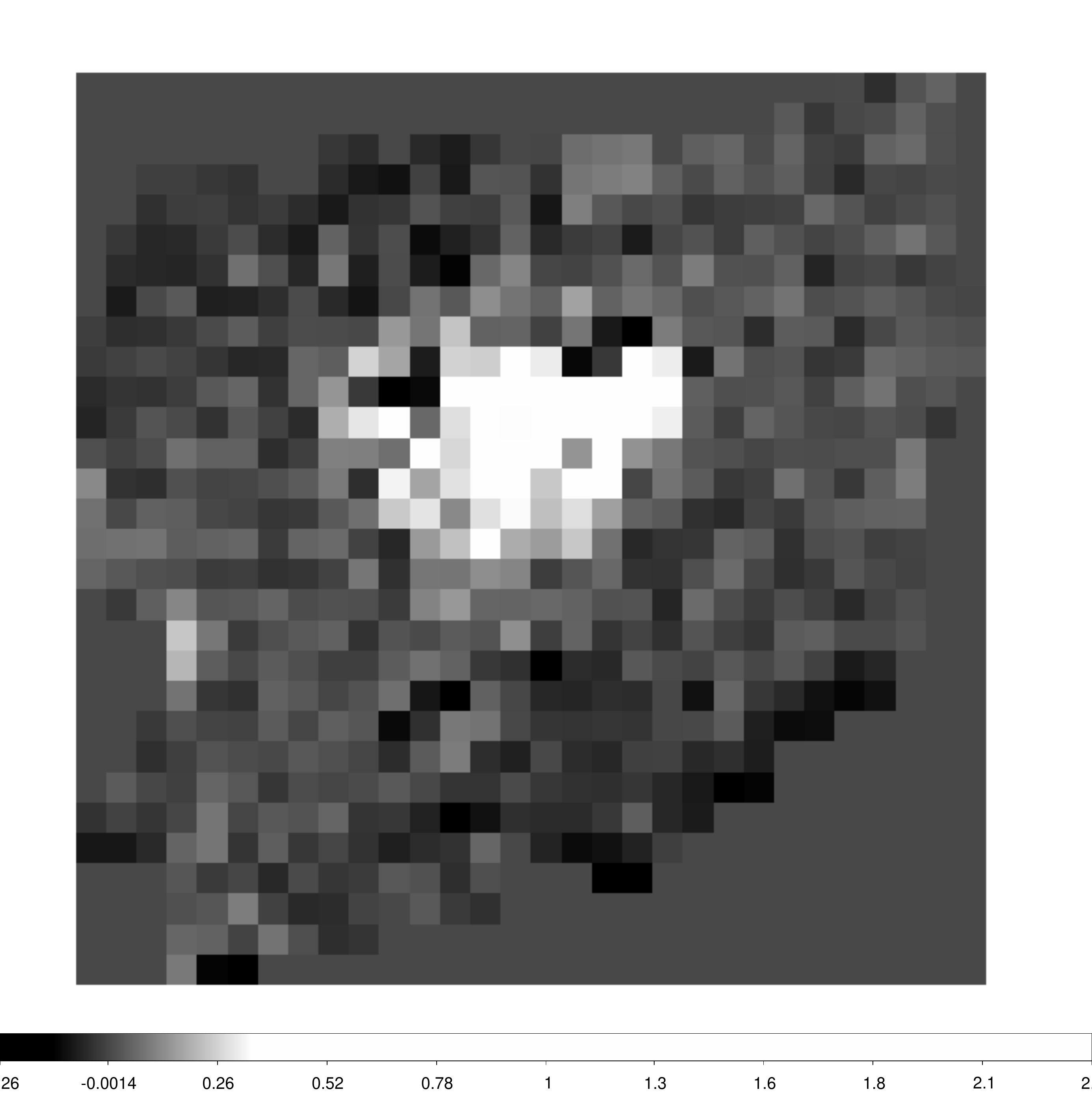} \label{fig:bestfit_esr_160Wsr}}\hfill
\subfigure[observed]{\includegraphics[trim=70 50 70 50, clip, width=0.49\columnwidth]{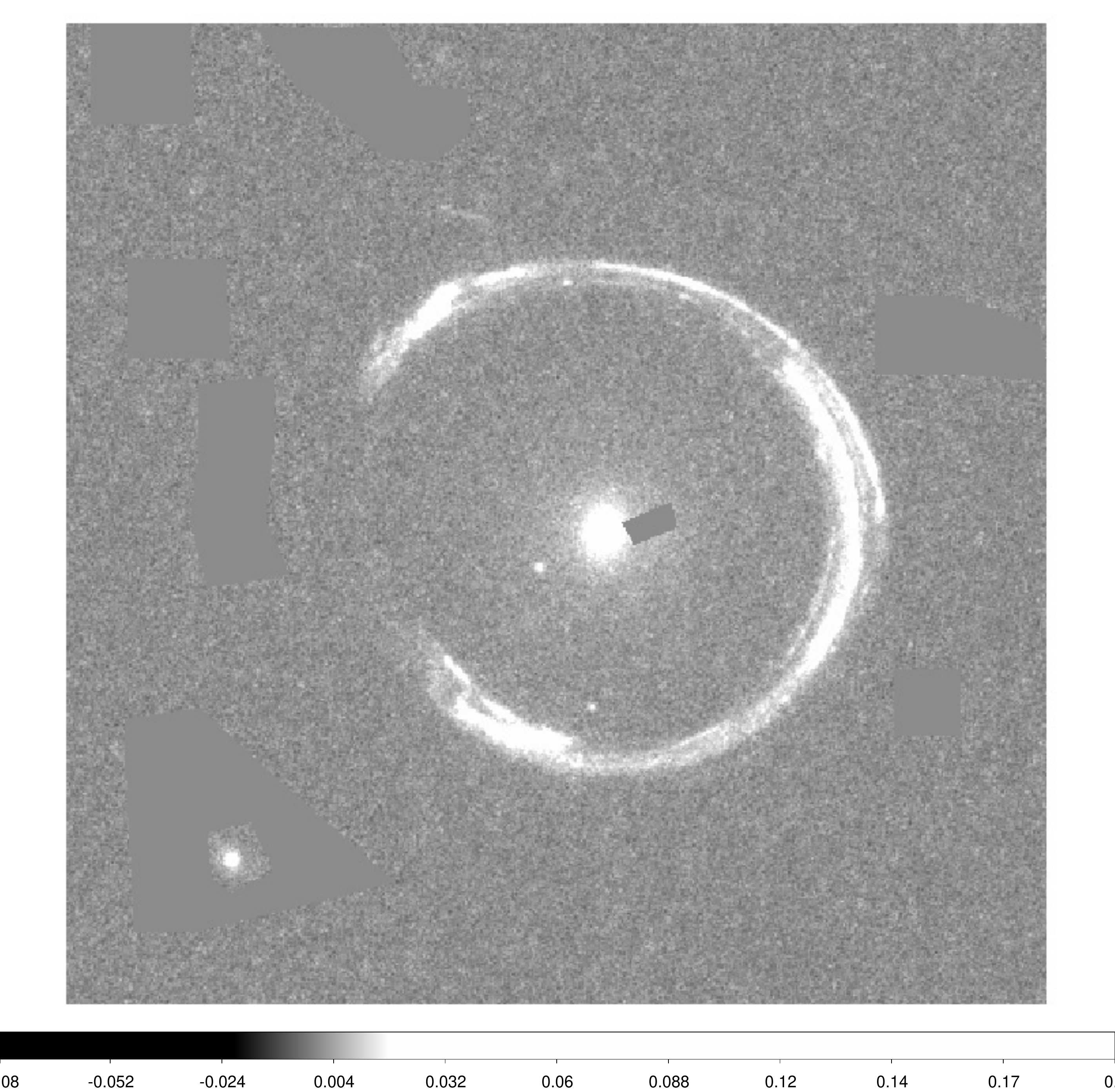} \label{fig:bestfit_esr_475Wobs}}\hfill
\subfigure[predicted]{\includegraphics[trim=70 50 70 50, clip, width=0.49\columnwidth]{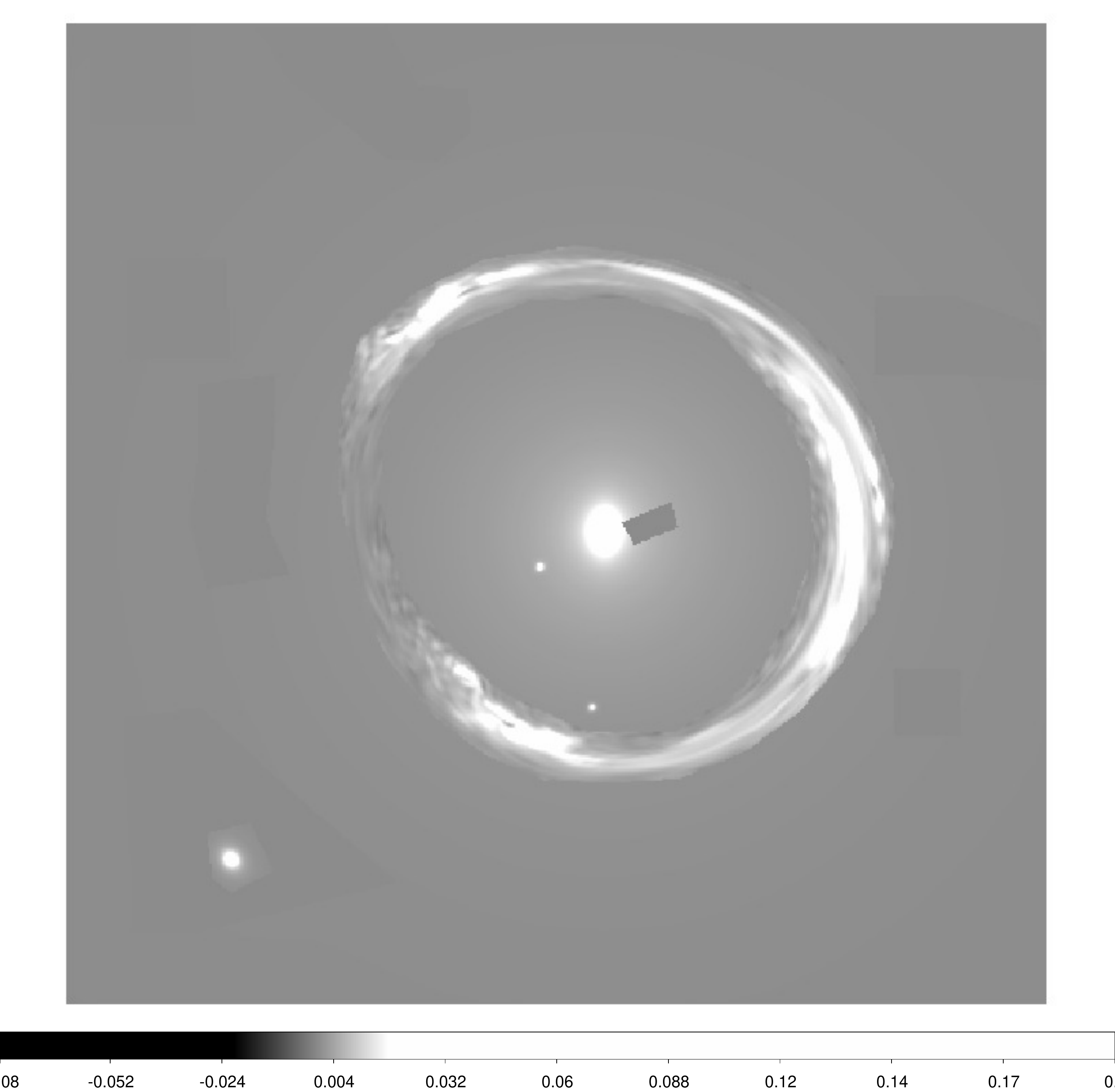} \label{fig:bestfit_esr_475Wpred}}\hfill
\subfigure[normalized residuals]{\includegraphics[trim=70 50 70 50, clip, width=0.49\columnwidth]{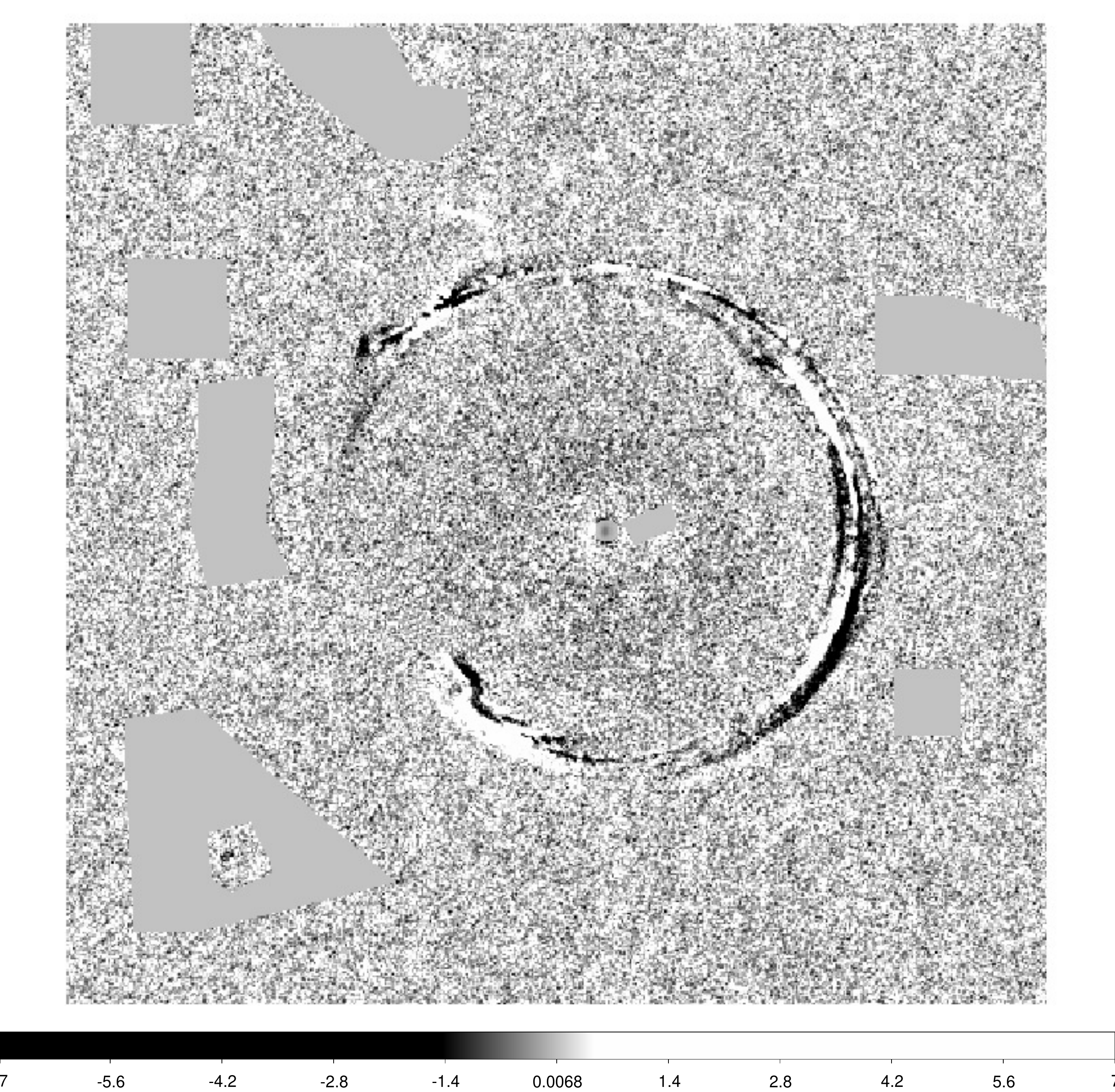} \label{fig:bestfit_esr_475Wsigma}}\hfill
\subfigure[source reconstruction]{\includegraphics[trim=70 50 70 50, clip, width=0.49\columnwidth]{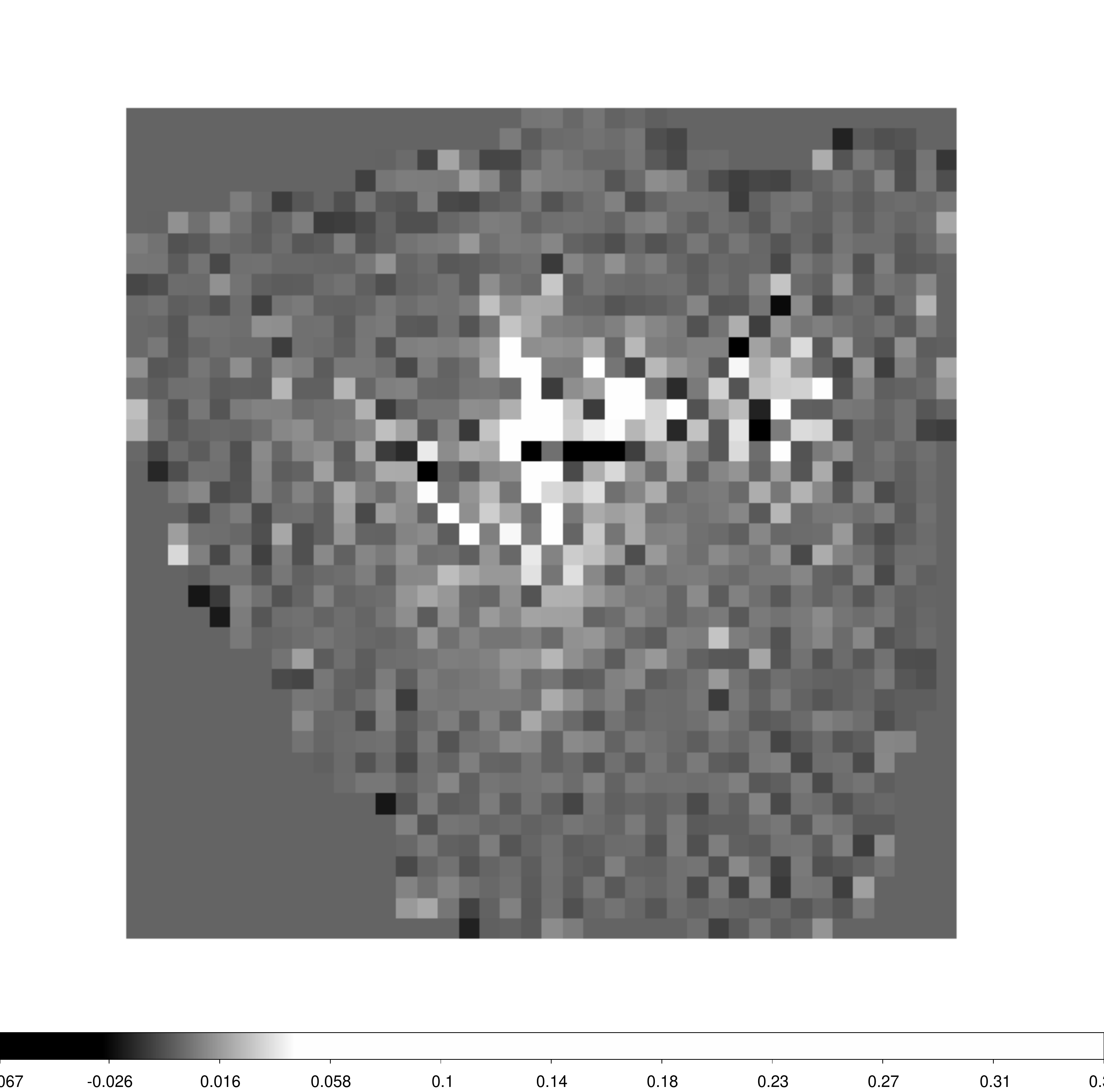} \label{fig:bestfit_esr_475Wsr}}\hfill
\subfigure[observed]{\includegraphics[trim=70 50 70 50, clip, width=0.49\columnwidth]{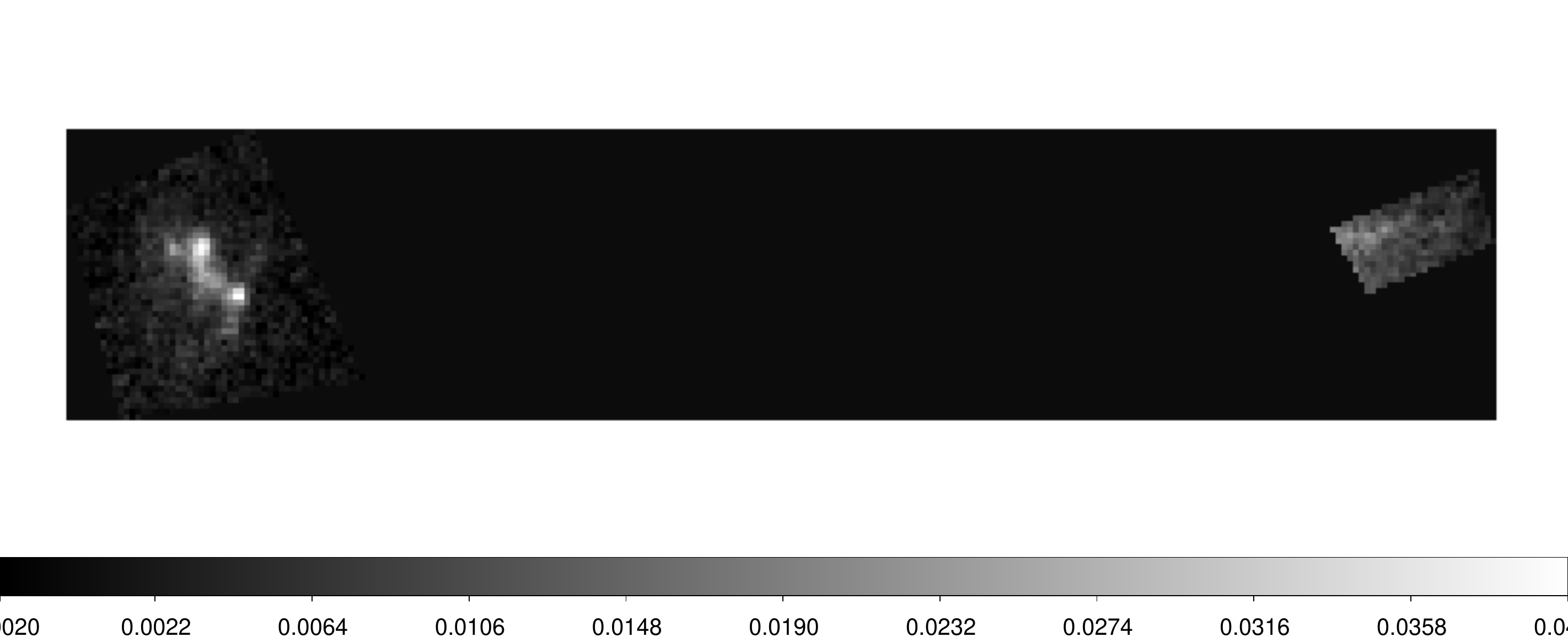} \label{fig:bestfit_esr_475Wobsrad}}\hfill
\subfigure[predicted]{\includegraphics[trim=70 50 70 50, clip, width=0.49\columnwidth]{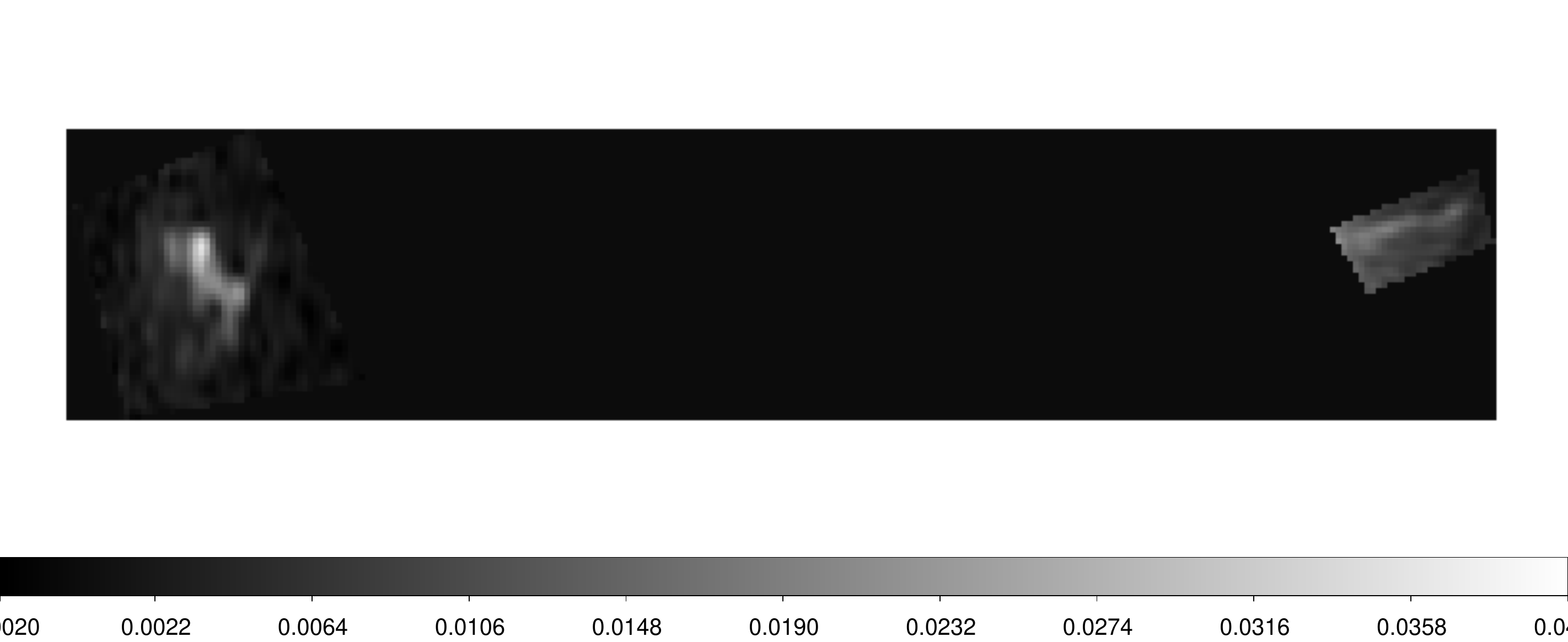} \label{fig:bestfit_esr_475Wpredrad}}\hfill
\subfigure[normalized residuals]{\includegraphics[trim=70 50 70 50, clip, width=0.49\columnwidth]{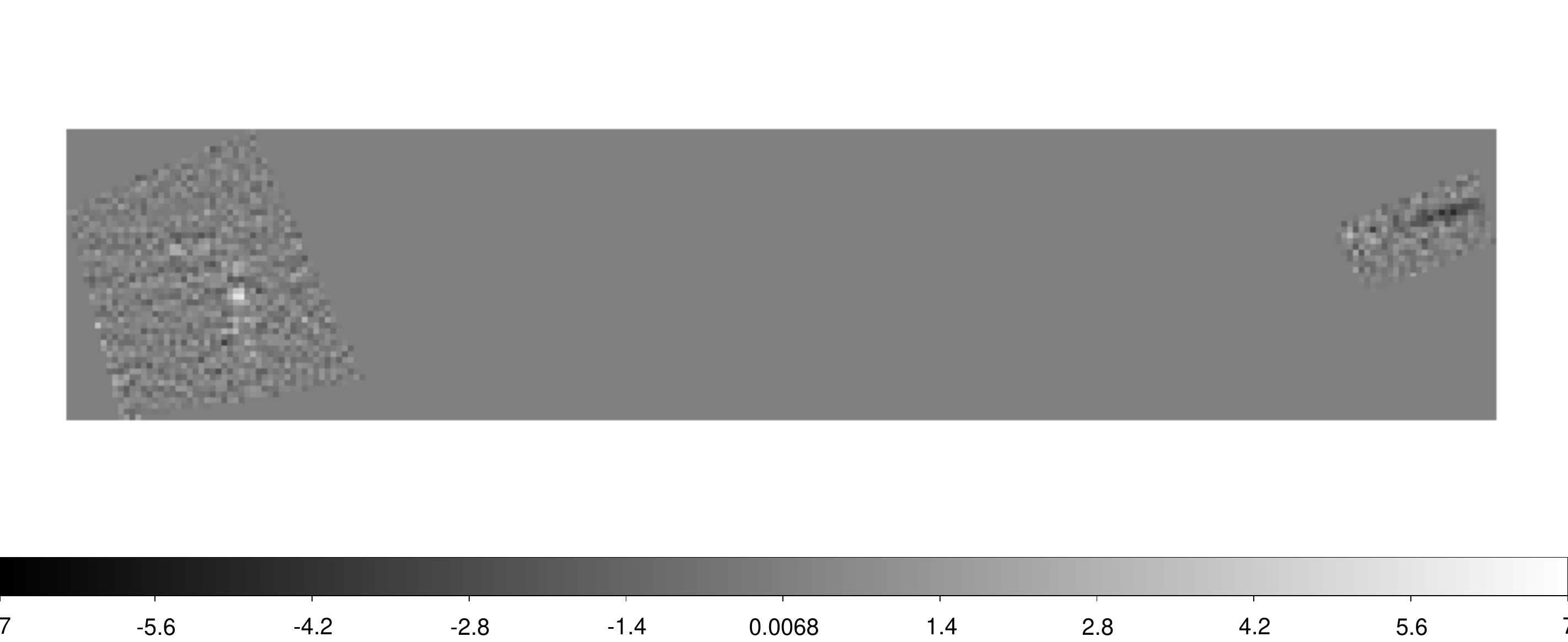} \label{fig:bestfit_esr_475Wsigmarad}}\hfill
\subfigure[source reconstruction]{\includegraphics[trim=70 50 70 50, clip, width=0.49\columnwidth]{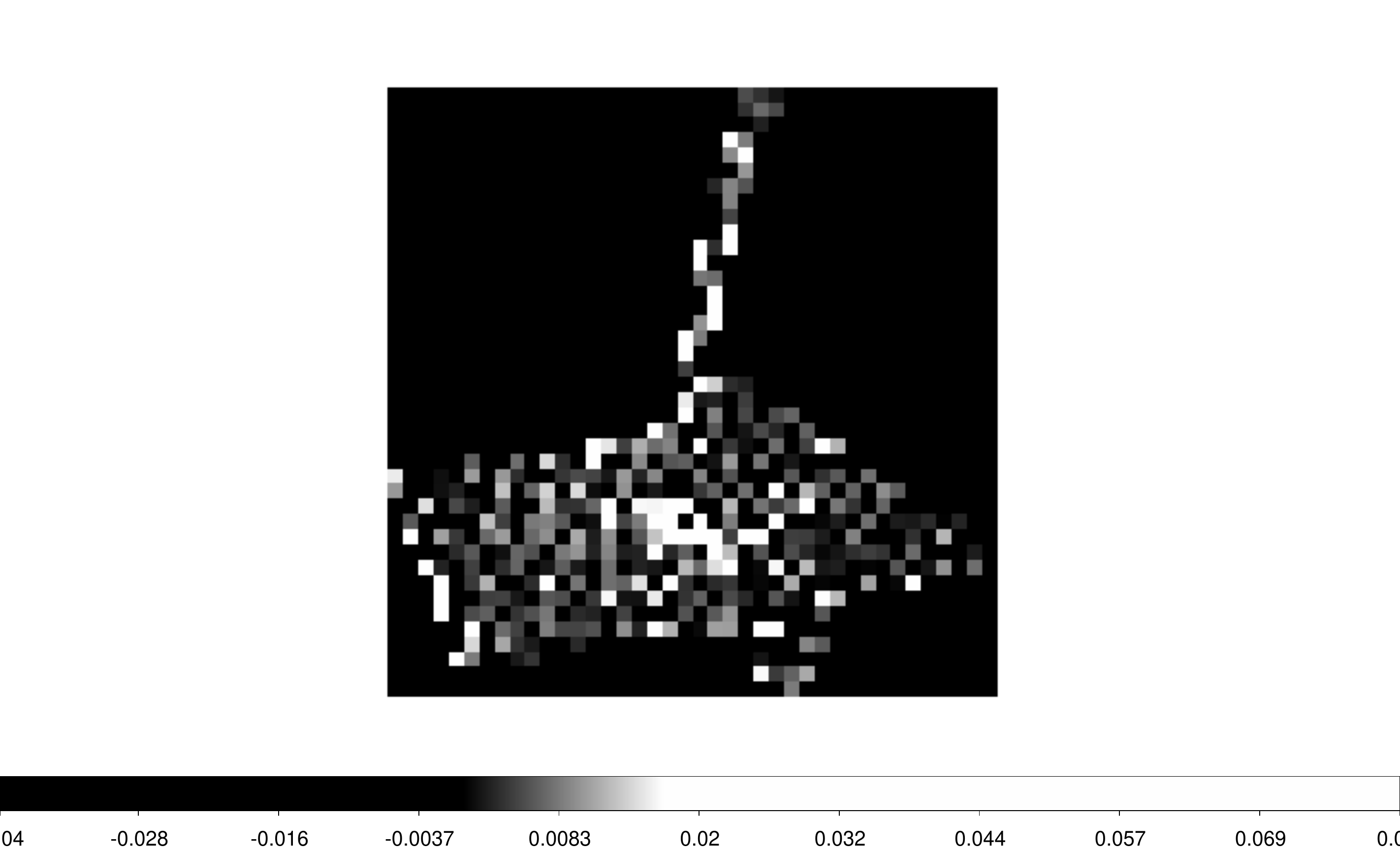} \label{fig:bestfit_esr_475Wsrrad}}\hfill
\caption{Images for the best-fit model which includes the source
  surface brightness reconstruction. In the top row one sees the
  images of the F160W band, and in the middle (tangential arc with
  lens) and bottom (radial arc) rows the images of the F475W band,
  respectively. To separate the radial arc and the tangential arc is
  needed since they lie at a different redshift. The images are
  ordered from left to right as follows: observed data, predicted
  model, normalized residuals in a range from $-7 \sigma$ to $+7
  \sigma$ and the reconstructed source SB on a grid of pixels.}
\label{fig:bestfit_sourcreconstruction}
\end{figure*}

Since we use the observed intensity of the arc to constrain our mass
model and the F475W band has the brightest arc relative to the lens
light, we include the F475W band in addition to the F160W which is
used for the lens light model. For simplicity we assume the same
structural parameters of the lens light profiles in the two bands
(such as axis ratio $q$, center, and orientation $\theta$) and model
only the amplitude of the three chameleon profiles and of the three
objects included. Explicitly, we model the lens galaxy's light in both filters
and reconstruct the observed intensity of the Einstein ring in
both. We also need to specify and model the radial arc and its counter
image separately due to their different redshift from the tangential
arc. This is done only in the F475W filter. The light component parameter values of
this model, with a $\chi^2_{\rm SB}$ of $7.2 \times 10^4$ for the F160W
filter and $3.1 \times 10^5$ for the F475W filter (the corresponding
reduced $\chi^2_{\rm SB}$ for the total model is 1.37), are presented in Table
\ref{tab:bestfit_esr_light}. In the same table we also give the median values
with 1-$\sigma$ uncertainty. The corresponding images of the best-fit model are presented
in Fig. \ref{fig:bestfit_sourcreconstruction}. In the top row one sees
the images of the F160W band, in the middle row the images of the
tangential arc and lens light in the F475W band, and in the bottom row
the images of the radial arc in the F475W band, respectively. The
images are ordered, for each row from left to right, as follows: the
first image shows the observed data, the second the predicted, the
third image shows the normalized residuals and the fourth image
displays the reconstructed source. Despite visible residuals in the
reconstruction, some of which are due to finite source pixel size, we
are reproducing the global features of the tangential arcs (compare
panels a to b, and e to f), to allow us to refine our multiple image
positions.

\begin{table*}[ht!]
 \renewcommand{\arraystretch}{1.4} 
 \caption{Best-fit and marginalized parameter values for the lens light component of the mass model obtained by reconstructing the source surface brightness.
 }
 \begin{center}
 \begin{tabular}[width=\textwidth]{c|c|c|c|c|c|c}
            & \multicolumn{2}{c|}{Chameleon 1 (lens)} & \multicolumn{2}{c|}{Chameleon 2 (lens)} & \multicolumn{2}{c}{Chameleon 3 (lens)}\\
  parameter & best-fit value & marginalized value & best-fit value & marginalized & best-fit value & marginalized \\ \hline 
  $x [\arcsec]$       & 11.00 &$ - $ &11.00  & $ - $ & 11.00 & $ - $ \\ 
  $y [\arcsec]$       & 9.67  &$ - $ &9.67   & $ - $ & 9.67  & $ - $  \\
  $q_\text{L}$ & 0.62 &$0.64^{+0.02}_{-0.03} $ & 1.00 & $1.00^{+0.00}_{-0.01} $  & 1.00 &$1.00^{+0.00}_{-0.01} $ \\
  $\theta$[rad]  & 1.52  &$- $ &1.52  & $- $ & 1.52 & $-$  \\
  $L_0$ (F160W)  & 46.67  &$- $ &3.50 & $- $ & 8.56 & $-$\\
  $w_\text{c}$ & 0.08 &$0.07^{+0.01}_{-0.01} $ & 1.95 & $2.04^{+0.06}_{-0.07} $ & 0.18 & $0.20^{+0.03}_{-0.02} $\\
  $w_\text{t}$ & 0.18 &$0.18^{+0.01}_{-0.01} $ & 6.99 & $7.01^{+0.03}_{-0.06} $ & 1.24 & $1.31^{+0.02}_{-0.03} $      \\ \hline
  $L_0$ (F475W)& 0.11 &$0.11^{+0.01}_{-0.01} $ &0.027& $0.029^{+0.001}_{-0.001} $ & 0.010 &$0.010^{+0.001}_{-0.002} $\\
  \end{tabular}
 \end{center}
 \label{tab:bestfit_esr_light}
 Note. This model includes three chameleon profiles (see Eq. (\ref{eq:isothermal})) for the F160W filter and additionally the same profiles with the same structural parameters for the F475W band. We fix the amplitudes of the F160W band since we are multiplying them with the mass-to-light ratio (variable parameter) in constructing the baryonic mass component.
\end{table*}

We also model the \CH\ observation with source SB reconstruction
assuming the NFW or gNFW for the dark matter halo mass. The fits give
for the NFW based model a $\chi^2_{\rm SB}$ of $3.76 \times 10^5$
(corresponding to a reduced $\chi^2_{\rm SB} = 1.37$) and very similar values for the gNFW model. 
From this, it seems that the gNFW fits almost as well as the NFW
profile. Compared with the power-law extended source model, the $\chi^2$ is slightly
higher, but still comparable. The images reproduce the observations
comparably well assuming the power-law profile, as shown in
Fig. \ref{fig:bestfit_sourcreconstruction}.

\subsection{Image position modeling}
\label{sec:compositeModel:redef}

Finally, we refine multiple image systems using the extended surface brightness
modeling results of the last section. This time we find, similarly to
what was done in Sec \ref{sec:compositeModel:LensMass}, eight sets of
multiple images systems, in addition to the radial arc and its counter
image.

\subsubsection{Three chameleon profiles}
\label{sec:compositeModel:redef:threecham}
\begin{figure}
     \begin{picture}(250,250)
     \put(0,0){\includegraphics[trim=0 5 0 5, clip, width=1.0\columnwidth, angle=0]{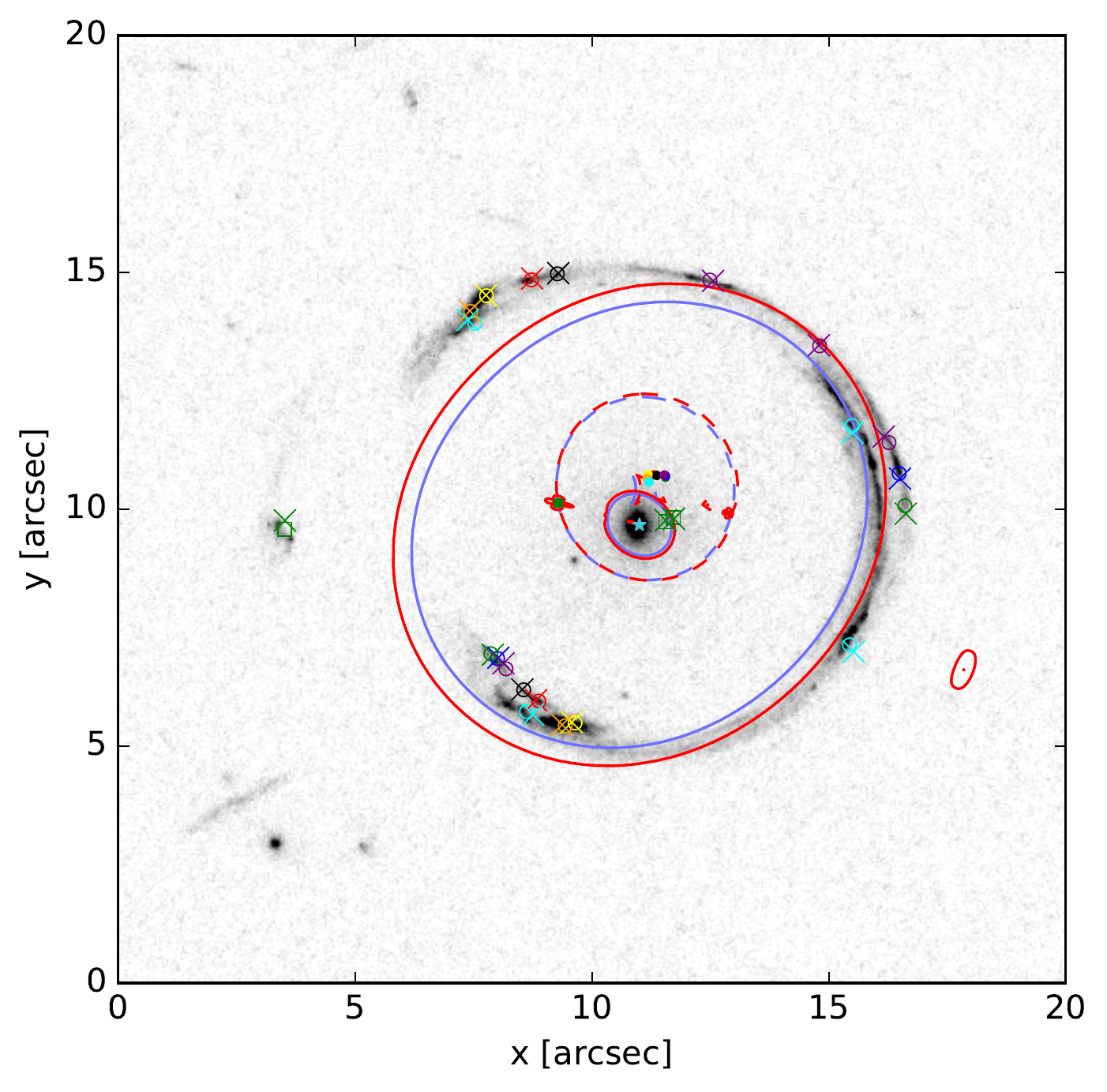}}
     \put(29,175){\includegraphics[trim=190 120 190 150, clip, width=0.27\columnwidth, angle=0]{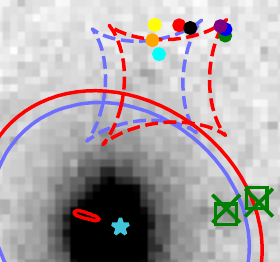}}
     \put(29,23){\includegraphics[trim=105 120 270 150, clip, width=0.27\columnwidth, angle=0]{HorseshoePlots/lensing_lensonly_with_PL_ML1_zoom2.pdf}}
     \end{picture}
    \caption{Best-fit model of the lensed source positions of the \CH, which are identified using our best-fit mass model with source SB reconstruction. This model assumes a power-law profile for the dark matter distribution. It is obtained using, as constraints, eight multiple image systems for the Einstein ring (circles) and the radial arc and its counter image (squares). We mark the predicted image positions with a cross. One can see that all predicted images are very close to the selected ones. The blue lines correspond to the critical curves (solid) and caustics (dashed) computed for the redshift of the radial arc, i.e. $z_\text{s,r} = 1.961$, and the red line to the critical curves (solid) and caustics (dashed) computed for the redshift of the tangential arc, i.e. $z_\text{s,t}=2.381$. The lens position is marked with a blue star. The small additional red features near the radial arc source position, shown in the lower left corner in detail, and on the right hand side are probably due the presence of radial arc source, i.e. as a result of multi-plane lensing. Indeed, we can see that these features do not appear in the single-plane case (blue line). The filled squares and circles correspond to the weighted mean positions of the predicted source position, which are shown in more detail in the zoom in the upper/lower left corner. The figure is oriented such that North is up and East is left.}
   \label{fig:sourcePositionModel_redef}
\end{figure}

If we assume a constant $M/L$ for all three chameleon profiles to scale the light to the baryonic mass, our model predicts the positions very well, with a $\chi^2$ of 20.23,
which corresponds to a reduced $\chi^2$ of 1.07 (in equation \ref{eq:chi2pt}) . Here we use the
best-fit model obtained in Sec.~\ref{sec:compositeModel:esr}, which
adopts the power-law profile, now with core radius set to $10^{-4}$,
for the dark matter distribution. This is done since the value is
always very small and we want to focus on constraining the
slope. Another reason is that we need to fix one parameter for our
dynamics-only model which is explained more in
Sec.~\ref{sec:LensingDynamics}. The model with the selected multiple
image systems is shown in
Fig. \ref{fig:sourcePositionModel_redef}. The figure shows also the
critical curves and caustics for both redshifts, $z_\text{s,r} =
1.961$ and $z_\text{s,t}=2.381$, as well as the predicted image
positions from \GLEE. The filled squares and circles correspond to the
model source position (which is the magnification-weighted mean of the
mapped source position of each image).

To compare how much constraints we get from the radial arc, we
treat also a model based on these image positions excluding the radial
arc and its counter image. Here we have to remove the SIS profile
which we adopt for the radial arc source mass. With this model we get
a best-fit $\chi^2$ of 18.87 which corresponds to a reduced $\chi^2$
of 1.18.

Similarly as before, we test how well we can fit the same multiple
image systems, i.e. these eight sets for the tangential arc and the
radial arc with its counter image as shown in
Fig. \ref{fig:sourcePositionModel_redef}, with our model by assuming a
NFW or gNFW dark matter distribution. It turns out that our model
based on the NFW profile gives a $\chi^2$ of 35.48 (reduced $\chi^2 =
1.87$) whereas the model based on the gNFW profile gives a $\chi^2$ of
35.19 (reduced $\chi^2 = 1.96$). This means that we do not fit the
refined multiple image systems with the NFW or gNFW dark matter
distribution as well as with the power-law. We see a big difference in
$\chi^2$ compared to the models where we exclude the radial arc and
its counter image. Explicitly, without radial arc are the $\chi^2$
values 25.44 (reduced $\chi^2=1.59$) and 25.40 (reduced $\chi^2=1.70$)
for the NFW and gNFW profile, respectively.

While the power-law halo model fits well to the image positions, it yields a $M/L$ of around 0.4 $M_\odot/L_\odot$ that is unphysically low.  On the other hand, the NFW and gNFW with a common $M/L$ for all three light components cannot fit well to the image positions, particularly those of the radial arc. Since
newer publications \citep[e.g.,][]{samurovic16, sonnenfeld18,
  bernardi18} predict variations in the stellar mass-to-light ratio of
massive galaxies, we treat our model of the refined image
position models with different mass-to-light ratios for each chameleon
profile. Different ratios result in a similar effect as a
radial-varying ratio. We treat this variation of different $M/L$ for
all our models, that means both with and without radial arc as well as
for all three different dark matter profiles NFW, gNFW and
power-law. This will be considered further in
Sec.~\ref{sec:LensingDynamics}.  

\subsubsection{Central point mass with constant $M/L$ of extended chameleon profiles}
\label{sec:compositeModel:redef:ptmass}
Since (1) we get a very small $M/L$ for the central component
(compare red line in Fig. \ref{fig:lenslight_with_PSF}) in the
previous model, (2) this 
component is very peaky that the width is smaller as the PSF width, and
(3) the \CH\ galaxy is known to be radio active, we infer that the
central component is a luminous point component like an AGN. Thus we
cannot assume an $M/L$ 
for it to scale to the baryonic matter. Therefore we treat also
models where we
assume a point mass instead of the central light component. The mass
range is restricted to be between $10^8 M_\odot$ and $10^{10} M_\odot$
as these are the known limits of black hole masses
\citep[e.g.,][]{thomas16, rantala18}. For the two other, extended
chameleon profiles, we assume a $M/L$ to scale them to the baryonic
mass. Under this assumption we are able to reproduce a good, physical
meaningful model for all three adopted dark matter profiles. Since our
final model will also include the kinematic information of the lens
galaxy, we will discuss details only for this model in Section
\ref{sec:LensingDynamics}. 

\section{Kinematics \& Dynamics}
\label{sec:Dynamics}

In Sec.~\ref{sec:compositeModel} we construct a composite mass model
of the \CH\ lens galaxy using lensing alone. In this section we
present the kinematic data of the \CH\ lens galaxy taken from
\citet{spiniello11} and a model based on dynamics-only
\citep[e.g.,][]{yildirim16, nguyen17, yildirim17, wang18}.

For the dynamical modeling we use a software which was further
developed by Ak{\i}n Y{\i}ld{\i}r{\i}m \citep{yildirim18} and which is
based on the code from Michele Cappellari \citep{ cappellari03,
  cappellari08}. For an overview of the Jeans ansatz and the
considered parameterization, the Multi-Gaussian-Expansion (MGE)
method, see Appendix \ref{sec:dynamicstheory}.  We infer the best fit
parameters again using {\sc Emcee} as already done for the lensing
part.

\subsection{Lens stellar kinematic data}
\label{sec:Dynamics:kinematic}

Following the discovery of the famous \CH\ by
\citet{belokurov07}, several follow-up observations were done. In
particular, \citet{spiniello11} obtained long slit kinematic data for
the lens galaxy G in March 2010. This was part of their X-Shooter
program (PI: Koopmans). The observations covered a wavelength range
from 300~\AA\ to 25000~\AA\ simultaneously with a slit centered on the
galaxy, a length of 11$\arcsec$ and a width of 0.\arcsec 7.

To spatially resolve the kinematic data, they defined seven apertures
along the slit and summed up the signal within each aperture. The size
of each aperture was chosen to be bigger than the seeing of $\approx
0.\arcsec 6$, such that independent kinematic measurement for each
aperture were obtained. These data are listed in
Table~\ref{tab:kindata}, together with the uncertainties. The obtained
weighted average value of the velocity dispersion is $344 \pm 25 \
\text{km} \ \text{s}^{-1}$. This is within the uncertainty of the
measurements. Due to the small number of available data and the huge
errors we will consider the symmetrized values and uncertainties as
given in Table \ref{tab:kindata}.

For further details on the measurement process or the data of the
stellar lens kinematics see \citet{spiniello11}.

\begin{table*}[ht!]
 \renewcommand{\arraystretch}{1.4} 
 \caption{Stellar kinematic data of the \CH\ lens galaxy.
 }
 \begin{center}
  \begin{tabular}[width=\textwidth]{c|c|c|c|c}
  Aperture distance $[\arcsec]$ & $v ~[\text{kms}^{-1}]$          & $\sigma~[\text{km~s}^{-1}]$& $v_\text{rm~s}~[\text{km~s}^{-1}]$& $v_\text{rms, sym}~[\text{km~s}^{-1}]$ \\ \hline 
  $-2.16$      & $-100 \pm 100$& $350 \pm 100$ & $364 \pm 101$ & $ 406 \pm 101$\\
  $-1.36$      & $-80 \pm 100 $& $311 \pm 76$  & $321 \pm 78$  & $ 340 \pm  89$\\
  $-0.64$      & $-9 \pm 25$   & $341 \pm 26$  & $341 \pm 27$  & $ 353 \pm  26$\\
  $ 0.00$      & $0 \pm 12$    & $332 \pm 16$  & $332 \pm 16$  & $ 332 \pm  16$\\
  $+0.64$      & $62 \pm 18$   & $360 \pm 25$  & $365 \pm 25$  & $ 353 \pm  26$\\
  $+1.36$      & $77 \pm 80 $  & $350 \pm 100$ & $358 \pm 100$ & $ 340 \pm  89$\\
  $+2.16$      & $180 \pm 100$ & $410 \pm 100$ & $448 \pm 101$ & $ 406 \pm 101$\\
 \end{tabular}
 \end{center}
 \label{tab:kindata}
 Note. We give the distance along the slit measured with respect to the center, the corresponding rotation $v$ \citep{spiniello11}, the velocity dispersion $\sigma$ \citep{spiniello11}, the second velocity moments $v_\text{rms}$ obtained through Eq.~(\ref{eq:2moment_normal}), and the symmetrized values $v_\text{rms, sym}$. The uncertainties $\delta v_\text{rms}$ is calculated through the formula $\delta v_\text{rms} = \sqrt{v^2 \delta v^2 + \sigma^2 \delta \sigma^2}/v_\text{rms}$. The last row are the considered values in this section. 
\end{table*}

\subsection{Dynamics-only modeling}
\label{sec:Dynamics:dynamics}

Before we combine all available data to constrain maximally the mass
of the \CH\ lens galaxy, we model the stellar kinematic data alone. We
start from the best-fit model from lensing, and include the parameters
anisotropy $\beta$ and inclination $i$. Since we have only seven data
points (see Table \ref{tab:kindata}), we can vary at most six
parameters. Thus we set the core radius $r_\text{c}$ of the power-law,
which turned out to be very small in our lensing models, to
$10^{-4}$. For a correct comparison to the refined lensing models (see
Sec.~\ref{sec:compositeModel:redef}) we fix the core radius there
too. For dynamics we will only adopt power-law and NFW dark matter
distribution, i.e. no longer the generalization of the NFW
profile. The reason is the small improvement compared to the NFW
profile. One further reason is that otherwise we have to fix one
parameter to vary fewer parameters than the available data points. In
other words, for considering the generalized NFW profile we have to
fix one parameter such that the number of free parameters is smaller
than the number of data points. In analogy to the case of the
power-law profile where we fix the core radius, we would set for the
generalized NFW profile the slope $\gamma_\text{g} \equiv 1$. This
would result in the NFW profile.

\begin{figure}[ht!]
 \centering
   \includegraphics[angle = 0, trim=15 8 30 30, clip, width=1.0\columnwidth]{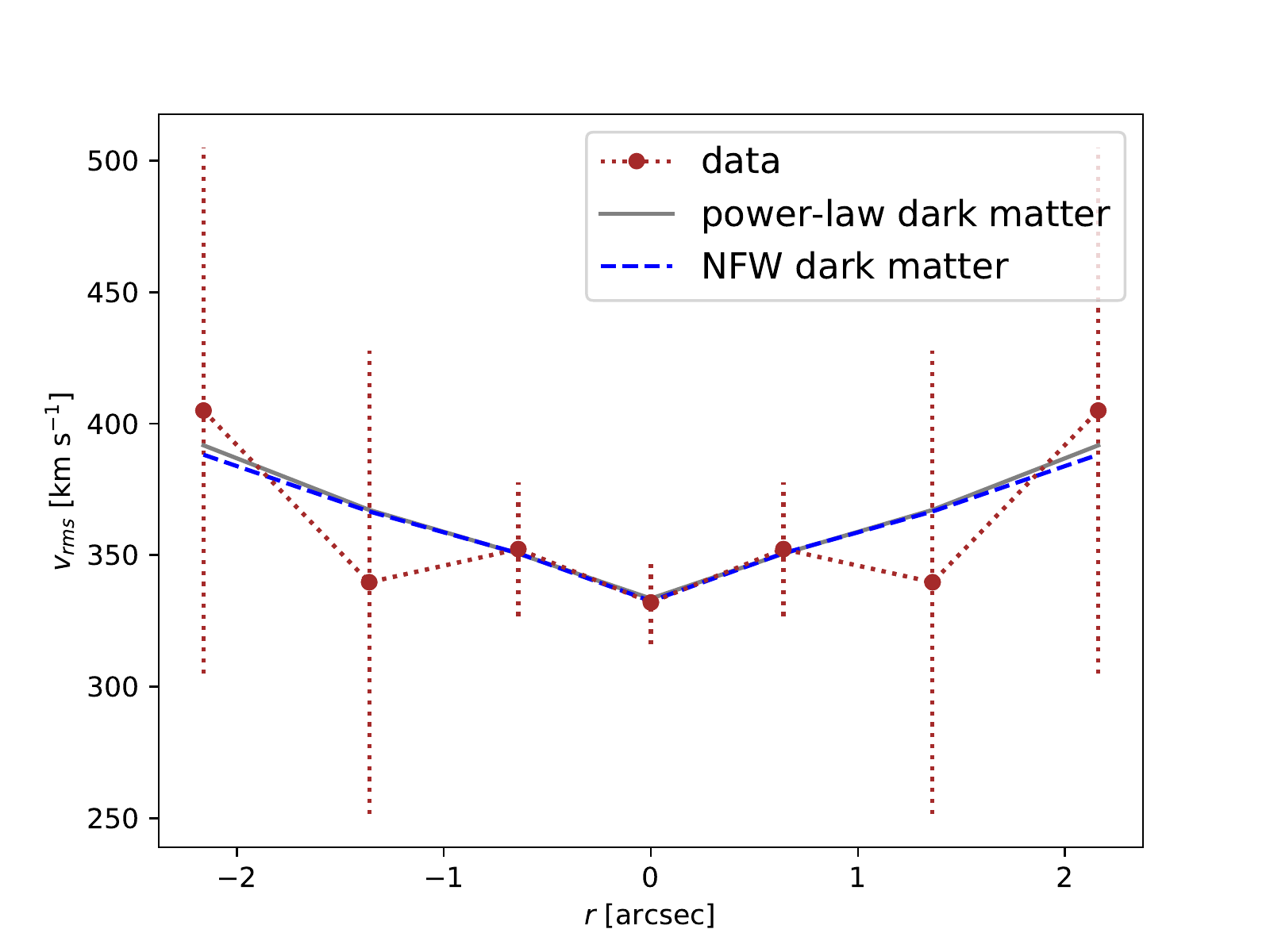}
   \caption{Values for the second velocity moments $v_\text{rms}$ obtained by adopting the power-law dark matter distribution (solid gray) or NFW (dashed blue) for dynamics-only. In brown are shown the measured data points with the full error bars.
     }
   \label{fig:bestfit_dynonly}
\end{figure}

The power-law dark matter distribution gives a dynamics-only best-fit
model with $\chi^2 = 0.25$. The reason why our model has a $\chi^2$
much smaller than 1 is due to the big uncertainties. The data points
are shown in Fig. \ref{fig:bestfit_dynonly} (blue) with our
dynamics-only model assuming power-law (solid) or NFW (dashed) dark
matter distribution. Since we can easily fit to these seven data
points in the given range, we treat the same model also with
forecasted 5\% uncertainties for every measurement. The obtained
best-fit dynamics-only model has a $\chi^2$ of 4.95, which is clearly
much higher than for the full error. The best-fit parameter and median
values with 1-$\sigma$ uncertainty are given in Table
\ref{tab:bestfit_dynonly} for the model assuming the actual measured
errors. As expected, most parameters are within the 1-$\sigma$ range
and the mass-to-light ratio is in a good range. The relatively large
errors on the parameters are due to the small number of data points we
use as constraints.

\begin{table}[ht!]
 \renewcommand{\arraystretch}{1.4} 
 \caption{Best-fit parameter values for our model based on the power-law dark matter distribution with dynamics-only.
 }
 \begin{center}
  \begin{tabular}{c|c|c|c}
 component  & parameter     & best-fit value & marginalized \\ \hline 
 kinematics & $\beta$       & 0.10 & $0.01^{+0.2}_{-0.3}  $\\
            & $i$           & 0.1  & $0.1^{+0.2}_{-0.2} $\\ \hline
 dark matter& $q$		    & 0.82 & $0.83^{+0.2}_{-0.09} $\\
 (power-law)& $\theta_\text{E}$ $[\arcsec]$& 2.3& $2.4^{+0.5}_{-0.4}  $	\\
            & $r_\text{c}$ $[\arcsec]$	& $ \equiv 10^{-4}$& $-$	\\
            & $\gamma'$	    & 1.20 & $1.36^{+0.4}_{-0.2} $	\\ \hline
 baryonic matter& $M/L~[M_\odot/L_\odot]$		& 1.8 & $1.6^{+0.4}_{-0.6}
 $\\ 
  \end{tabular}
  \end{center}
 \label{tab:bestfit_dynonly}
 Note. The parameters are the anisotropy $\beta$, the inclination $i$, the axis ratio $q$, the strength $\theta_\text{E}$, the core radius $r_\text{c}$, and the slope $\gamma'$. In the last row we give the mass-to-light ratio for the baryonic component. Since we have only seven data points with huge uncertainties and vary six parameters in this model, we get also a large range of parameter values within 1-$\sigma$. The corresponding $\chi^2$ is 0.25. Note that we do not obtain any constraints on the anisotropy or inclination, given the assumption of a prior range of $\beta \in [-0.3,+0.3]$ and $i \in [0, +0.3]$.
\end{table}

For the NFW dark matter distribution we fit comparably well as with
the power-law model ($\chi^2 = 0.25$ compared to $\chi^2 = 0.26$),
when using the full kinematic uncertainty, while the $\chi^2$ is
slightly higher for the reduced (forecasted 5\%) uncertainty on the
kinematic data 
($\chi^2 = 4.95$ compared to $\chi^2=
5.61$). Comparing power-law and the NFW, we do not find a remarkable
difference, apart for the radius, which appears to be lower in the NFW
forecasted case. This, however, is in agreement with the higher
$\chi^2$ of the NFW since the predicted $v_\text{rms}$ values are in
both versions, power-law and NFW profile, lower than the
measurement. For a further detailed analysis based on dynamics-only
spatially, resolved kinematic measurements would be helpful.

\section{Dynamical and lensing modeling}
\label{sec:LensingDynamics}

After modeling the inner mass distribution of the \CH\ lens galaxy
based on lensing-only (Sec.~\ref{sec:compositeModel}) and
dynamics-only (Sec.~\ref{sec:Dynamics}), we now combine both
approaches. In the last years huge effort has been spent to combine
lensing and dynamics for strongly lensed observations to get a more
robust mass model \citep[e.g.,][]{treu02, treu04, mortlock00,
  gavazzi07, barnabe09, auger10, barnabe11, sonnenfeld12, grillo13,
  lyskova18}. Since strong lensing has normally the constraints at the
Einstein radius $r \approx \theta_\text{E}$, which is in our case
$\approx 5 \arcsec$, and kinematic measurements are normally in the
central region around the effective radius (here $r \lesssim 2
\arcsec$), one combine information at different radii with these two
approaches. This will result in a better constrained model and one
might break parameter degeneracies thanks to the complementary of
these two approaches. However, in our particular lens system, we also
use the radial arc as lensing constraints in the inner regions.

Although using the \HST\ surface brightness observations would provide
more lensing constraints, we consider here only the refined image
positions presented in Sec.~\ref{sec:compositeModel:redef}. The reason
for this choice is that we would otherwise overwhelm the 7 data points
from dynamics with more than $10^5$ surface brightness pixel from the
images. The data points coming from the identified image positions are
still higher, but at the same order of magnitude. Moreover, with this
method we are able to weight the contribution of the radial arc and
its counter image
more.

When we combine dynamics and lensing, we consider again models with
and without radial arc, each adopting power-law or NFW dark matter
distribution, and all four versions with the full uncertainty of the
kinematic data as well as with 5\% as a forecast. 
Additionally, we
treat all models with one single $M/L$ ratio as well as with
different $M/L$ ratios as already done for lensing-only (see Sec.~\ref{sec:compositeModel:lensLight:cham} for details). Based on the same arguments as for the
lensing-only, we treat also models by replacing the PSF-like central
component (shown in red in Fig. \ref{fig:lenslight_with_PSF}) by a
point mass.

\subsection{Three chameleon mass profiles}
\label{sec:LensingDynamics:3cham}

By combining lensing and dynamics we consider different composite mass
model. As first, we use the lens light, which is composed by three
chameleon profiles as obtained in
Sec.~\ref{sec:compositeModel:lensLight}, scaled by a constant
mass-to-light ratio as baryonic component. Under this assumption, the
best-fit has, when using a power-law dark matter mass distribution, a
$\chi^2$ of 25.08, and, when using a NFW dark matter distribution, a
$\chi^2$ of 71.54. The $\chi^2$ values reveal that the NFW is not as
good at describing the observation as the power-law profile. However,
assuming a power-law dark matter distribution, the $M/L$ value for
scaling all three light components is around $0.1 M_\odot/L_{\odot}$.  This is
unphysically low and results in a very high the dark matter
fraction. 

The next step to model the baryonic component is to allow different
mass-to-light ratios for the different light components shown in
Fig.~\ref{fig:lenslight_with_PSF}. This allows us to fit remarkably
better with the NFW profile, while we do not get much improvement
adopting a power-law dark matter distribution. However, this method
does not allow us to obtain meaningful models, as the central
component needs an unphysically low $M/L$. Therefore, we infer that we cannot
assume a mass-to-light ratio for the central component, irrespective of
the dark matter distribution.

\subsection{Point mass and two chameleon mass profiles}
\label{sec:LensingDynamics:point}

As noted in Sec.~\ref{sec:compositeModel:redef:ptmass}, the central component is probably associated with an AGN, since its light profile width is similar to the width of the PSF (see Fig.~\ref{fig:lenslight_with_PSF}) and its $M/L$ was very low from the previous model in Sec.~\ref{sec:LensingDynamics:3cham}. Thus, assuming a mass-to-light ratio for this component would not
be physically meaningful and we supersede it by a point mass in the
range of a black hole mass.
From our previous models and from the fact that the lens galaxy is very massive, we expect this point mass to be comparable to that of a supermassive black hole. For the two other light components we still assume the two fitted chameleon profiles scaled by a $M/L$, either the same $M/L$ for both components, or a
different $M/L$ for each component. Moreover, we test the effect of relaxing the scale parameter $r_\text{s}$ of the NFW profile. It turn out to be very similar to the model by assuming a fixed value, as expected, such that we present only plots of the model with free $r_\text{s}$. 

\begin{figure}[ht!]
 \centering
 \begin{picture}(100,445)
\put(-80,0){\includegraphics[angle = 0, trim=0 0 40 10, clip, width=1.0\columnwidth]{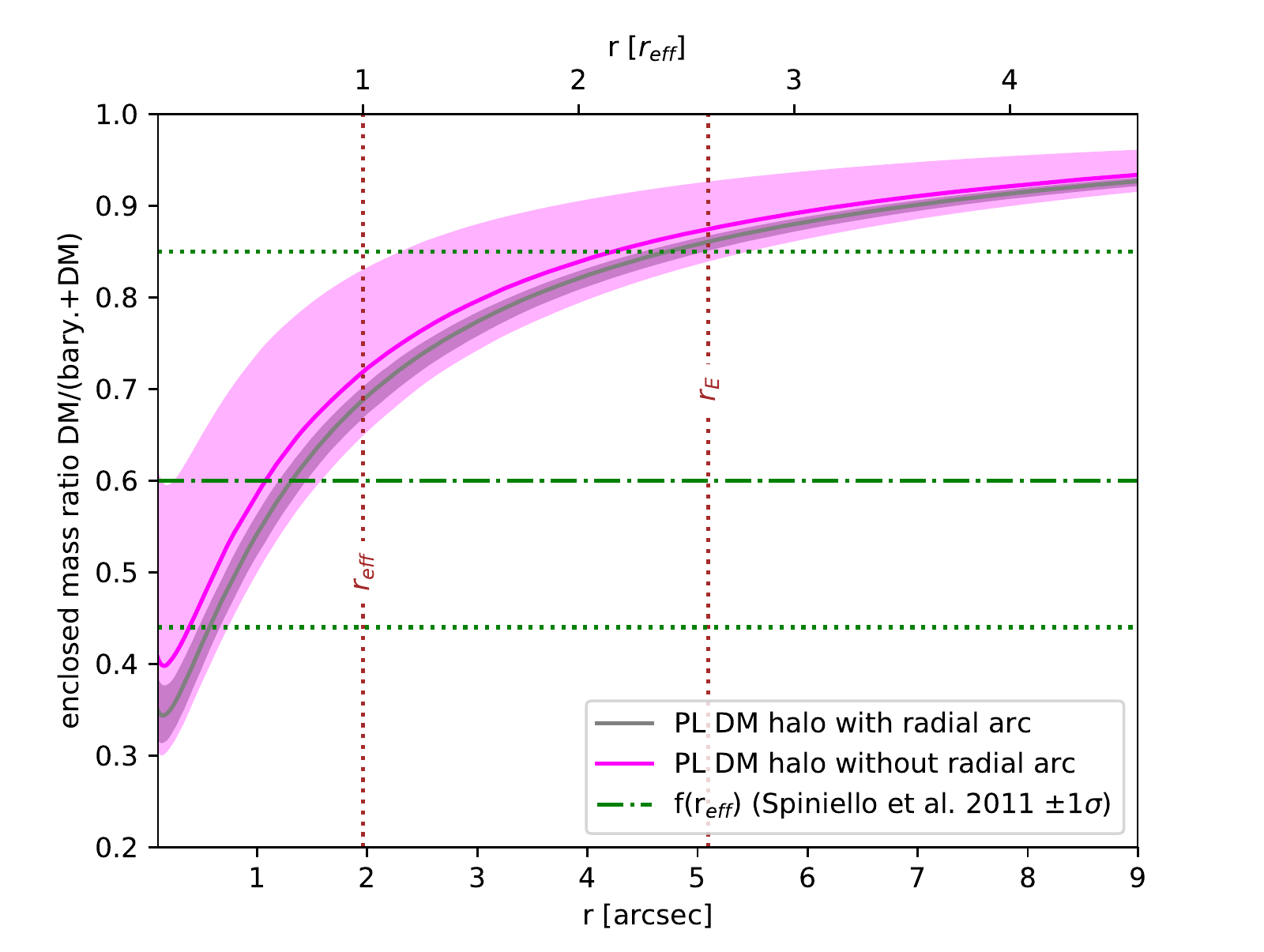}}
\put(-80,220){\includegraphics[angle = 0, trim=0 0 40 10, clip, width=1.0\columnwidth]{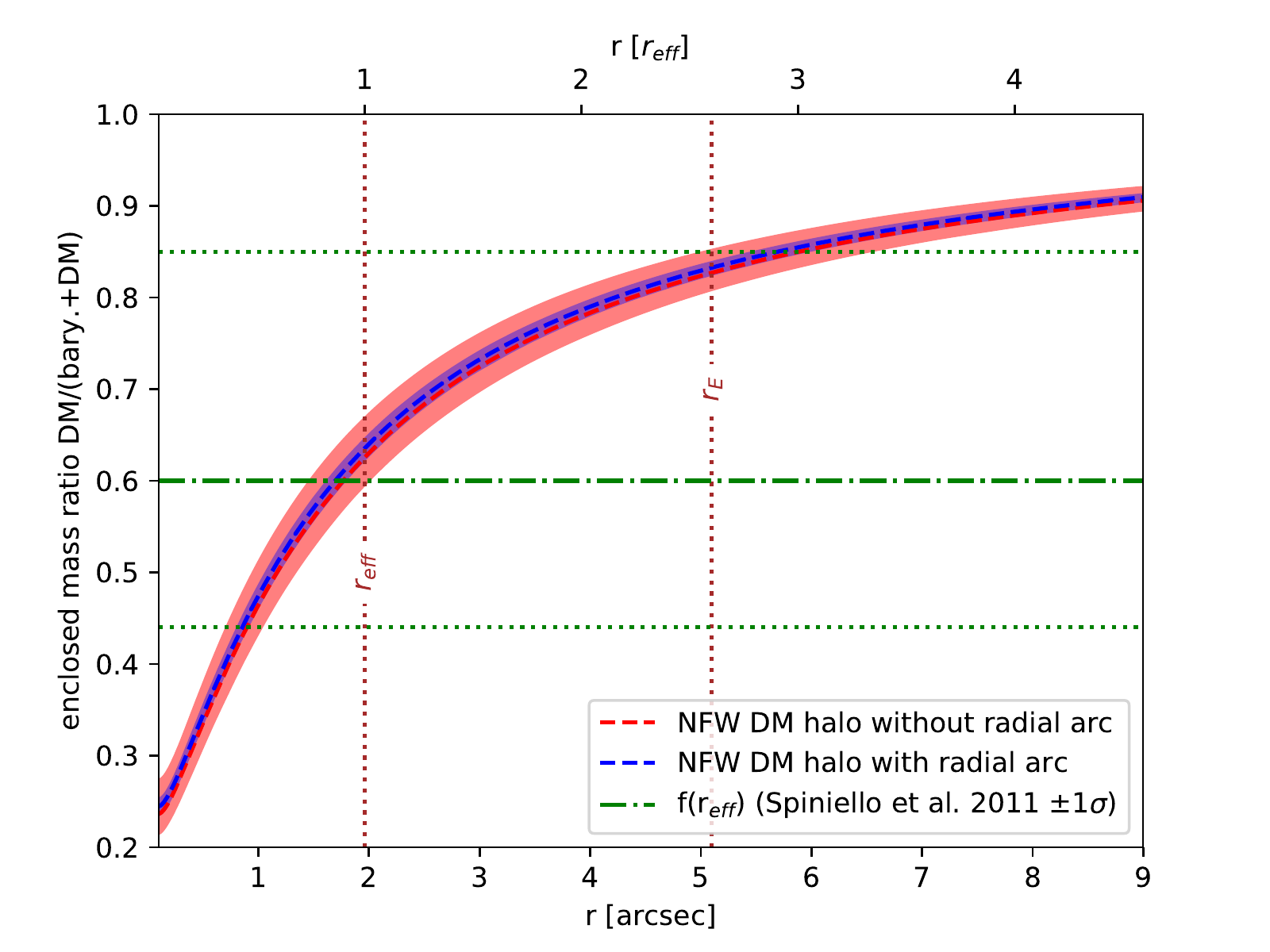}}
\put(-20,210){\color{black} \Large{power-law dark matter halo}} 
\put(-10,430){\color{black} \Large{NFW dark matter halo}} 
\end{picture}
   \caption{Enclosed dark matter fraction, i.e. dark matter divided by
     the sum of baryonic matter from the scaled lens light and dark
     matter enclosed in radius $r$, of the models adopting a power-law
     (bottom) or NFW (top) dark matter distribution. All models
     include the point mass in the lens center, which is not
     considered among the baryonic mass and thus not shown in this
     plot. For comparison, the value of the dark matter fraction within one
effective radius of \citet{spiniello11} is shown by the horizontal
solid line (for the value) and dashed lines (for 1$\sigma$
uncertainties).
     }
   \label{fig:bestfit_dynlensPoint_massfraction}
\end{figure}

We see by comparison of the different models with the point mass that
both dark matter profiles result in a similar $\chi^2$ value (see
Table \ref{tab:overview_pointcomp}). Both dark matter distributions
seem to fit the observation with an acceptable dark matter fraction
between 60\% and 70\%. The corresponding plot is shown in
Fig. \ref{fig:bestfit_dynlensPoint_massfraction} for the final models:
\begin{itemize}
\item[{\color{magenta}{\textbullet}}] lensing \& dynamics, power-law dark matter, without radial arc
\item[{\color{gray}{\textbullet}}] lensing \& dynamics, power-law dark matter, with radial arc
\item[{\color{red}{\textbullet}}] lensing \& dynamics, NFW dark matter, without radial arc
\item[\textcolor{blue}{\textbullet}] lensing \& dynamics, NFW dark matter, with radial arc
\end{itemize}

The dark matter fraction is defined here as
the dark matter divided by the sum of baryonic matter from the scaled
lens light and dark matter enclosed in the radius $r$. To be noted is
that the point mass is not assumed to be pure baryonic matter, and
thus not included in the baryonic component in the calculation. This
results in the profile of dark matter fraction having a concave curve in the very central region. Including the point mass with less than $10^{10} M_\odot$ would shift the fraction insignificantly to lower values. The
best-fit parameter values for these four models are given with the
corresponding median values with $1\sigma$ uncertainties in Table
\ref{tab:bestfit_pointPL} (adopting power-law dark matter
distribution) and Table \ref{tab:bestfit_pointNFW} (adopting NFW with free scale radius $r_\text{s}$).

\begin{table}[ht!]
 \renewcommand{\arraystretch}{1.4} 
 \caption{Overview of the different final best-fit models with the point-mass component representing the innermost light component (red profile in Fig. \ref{fig:lenslight_with_PSF}).
 }
 \center
  \begin{tabular}[width=\textwidth]{c|cc|cc||cc}
   DM profile   & \multicolumn{2}{c|}{radial arc}&  \multicolumn{2}{c||}{one $M/L$} & \multicolumn{2}{c}{two $M/L$} \\ \hline
                & with       & without    & $\chi^2$& $\chi^2_\text{red}$ & $\chi^2$& $\chi^2_\text{red}$  \\\hline \hline
  power-law     &$\checkmark$&            & 21.65 & 0.84 & 20.71 & 0.83 \\   
                &            &$\checkmark$& 19.90 & 0.91 & 19.58 & 0.94 \\ \hline
  NFW           &$\checkmark$&            & 20.14 & 0.78 & 19.95 & 0.80 \\
                &            &$\checkmark$& 19.87 & 0.91 & 19.53 & 0.93 
 \end{tabular}
 \label{tab:overview_pointcomp}
\end{table}

\begin{table*}[ht!]  
 \renewcommand{\arraystretch}{1.4} 
 \caption{Power-law dark matter halo model: Best-fit and marginalized
   parameter values for the mass model based on our image positions
   shown in Fig. \ref{fig:sourcePositionModel_redef} and the stellar
   kinematic data $v_\text{rms, sym}$ given in Table
   \ref{tab:kindata}.
 }
 \begin{center}
  \begin{tabular}[width=\textwidth]{c|c|c|c||c|c}
   \multicolumn{2}{c|}{}& \multicolumn{2}{c||}{with radial arc} & \multicolumn{2}{c}{without rad. arc} \\
  component & parameter & best-fit & marginalized & best-fit & marginalized\\ \hline 
  kinematics & $\beta$    &0.00    & $-0.04^{+0.2}_{-0.2} $     & $-0.02$  & $-0.06^{+0.4}_{-0.3} $\\
             & $i$        &0.15    & $0.11^{+0.07}_{-0.08} $    & 0.3  & $0.2^{+0.2}_{-0.2} $\\ \hline
             & $q$        &0.91   & $0.91 ^{+0.02}_{-0.02} $   & 0.90 & $0.91^{+0.04}_{-0.05} $\\
  dark matter& $\theta_\text{E} \ [\arcsec]$  &1.69  &$1.66^{+0.07}_{-0.07} $ & 1.7 & $1.7^{+0.3}_{-0.4} $ \\
  (power-law)& $r_\text{c} \ [\arcsec] $ & $\equiv 10^{-4}$& $-$& $\equiv 10^{-4}$ & $-$ \\
             & $\gamma' $  &1.28    & $1.26^{+0.04}_{-0.04} $    & 1.26 & $1.3^{+0.1}_{-0.1} $ \\ \hline
  shear      & $\gamma_\text{ext}$  &0.077 & $0.076^{+0.007}_{-0.006} $ & 0.08  & $0.07^{+0.03}_{-0.02} $\\
             & $\phi_\text{ext}$    &2.81  & $2.80^{+0.04}_{-0.04} $    & 2.8  & $2.8^{+0.1}_{-0.1} $\\ \hline
  baryonic matter& $M/L ~[M_\odot/L_\odot]$    &1.8    & $1.9^{+0.2}_{-0.1} $    & 1.7  & $1.7^{+0.6}_{-0.7} $\\ 
             & $\log(\theta_\text{E, point})$ & $-1.01$ &$-1.09^{+0.08}_{0.3}$ & $-2.0$  & $-2.4^{+1.3}_{-1.6}$  
 \end{tabular}
 \end{center}
 \label{tab:bestfit_pointPL}
 Note. The parameters are the anisotropy $\beta$, the inclination $i$, the axis ratio $q$, the strength $\theta_\text{E}$, the core radius $r_\text{c}$, the slope $\gamma'$, the shear magnitude $\gamma_\text{ext}$, and the shear orientation $\phi_\text{ext}$. Additionally, we give the mass-to-light ratio $M/L$, and the strength of the point mass $\theta_\text{E, point}$ in logarithmic scale (i.e. $-1$ corresponds to around $10^{10} M_\odot$).\\
\end{table*}

\begin{table*}[ht!]
 \renewcommand{\arraystretch}{1.4} 
 \caption{NFW dark matter halo model: Best-fit and marginalized
   parameter values for the mass model based on our image positions
   shown in Fig. \ref{fig:sourcePositionModel_redef} and the stellar
   kinematic data $v_\text{rms, sym}$ given in Table
   \ref{tab:kindata}.
 }
 \begin{center}
  \begin{tabular}[width=\textwidth]{c|c|c|c||c|c}
   \multicolumn{2}{c|}{}& \multicolumn{2}{c||}{with radial arc} & \multicolumn{2}{c}{without rad. arc} \\
  component & parameter & best-fit & marginalized & best-fit &
  marginalized \\ \hline 
  kinematics & $\beta$    &0.06    & $0.0^{+0.2}_{-0.2} $      & $-0.1$  & $-0.08^{+0.2}_{-0.2} $\\
             & $i$        &0.10   & $0.10^{+0.07}_{-0.07} $   & 0.2  & $0.1^{+0.1}_{-0.1} $\\ \hline
             & $q$        &0.95 & $0.95^{+0.01}_{-0.01} $& 0.95 & $0.95^{+0.01}_{-0.01} $\\
  dark matter& $\theta_\text{E} \ [\arcsec]$  &0.63  &$0.64^{+0.03}_{-0.02} $ & 0.62 & $0.64^{+0.04}_{-0.03} $ \\
  (NFW)      & $r_\text{s} \ [\arcsec] $ & $185$  & $170^{+22}_{-28}$ & $ 177$ & $180^{+18}_{-23}$ \\ \hline
  shear      & $\gamma_\text{ext}$  &0.08 & $0.08^{+0.01}_{-0.01} $   & 0.08  & $0.08^{+0.01}_{-0.01} $\\
             & $\phi_\text{ext}$    &2.81  & $2.80^{+0.03}_{-0.04} $      & 2.82  & $2.80^{+0.08}_{-0.05} $\\ \hline
  baryonic matter& $M/L~[M_\odot/L_\odot]$ &2.2 & $2.3^{+0.2}_{-0.1} $       & 2.5  & $2.4^{+0.3}_{-0.4} $\\
             & $\log(\theta_\text{E, point})$ &$-1.01$ & $-1.10^{+0.08}_{-0.2}$ & $-1.02$ & $-2.0^{+0.9}_{-1.3}$
 \end{tabular}
 \end{center}
 \label{tab:bestfit_pointNFW}
 Note. The parameters are the anisotropy $\beta$, the inclination $i$, the axis ratio $q$, the strength $\theta_\text{E}$, the scale radius $r_\text{s}$, the shear magnitude $\gamma_\text{ext}$, and the shear orientation $\phi_\text{ext}$. Additionally, we give the mass-to-light ratio $M/L$, and the strength of the point mass $\theta_\text{E, point}$ in logarithmic scale (i.e. $-1$ corresponds to around $10^{10} M_\odot$).\\
\end{table*}

Allowing two independent $M/L$ seems not to decrease the $\chi^2$
notably, and we see from the sampling that the outer $M/L$ is not well
constrained and highly degenerate with the other mass components. This
might come from the flatness of the profile (see green line in
Fig.~\ref{fig:lenslight_with_PSF}) and less constraints on the outer
part where the profile is dominant.

\begin{figure*}[ht!]
 \centering
\begin{picture}(200,210)
\put(-170,0){\includegraphics[angle = 0, trim=10 0 40 30, clip, width=1.0\columnwidth]{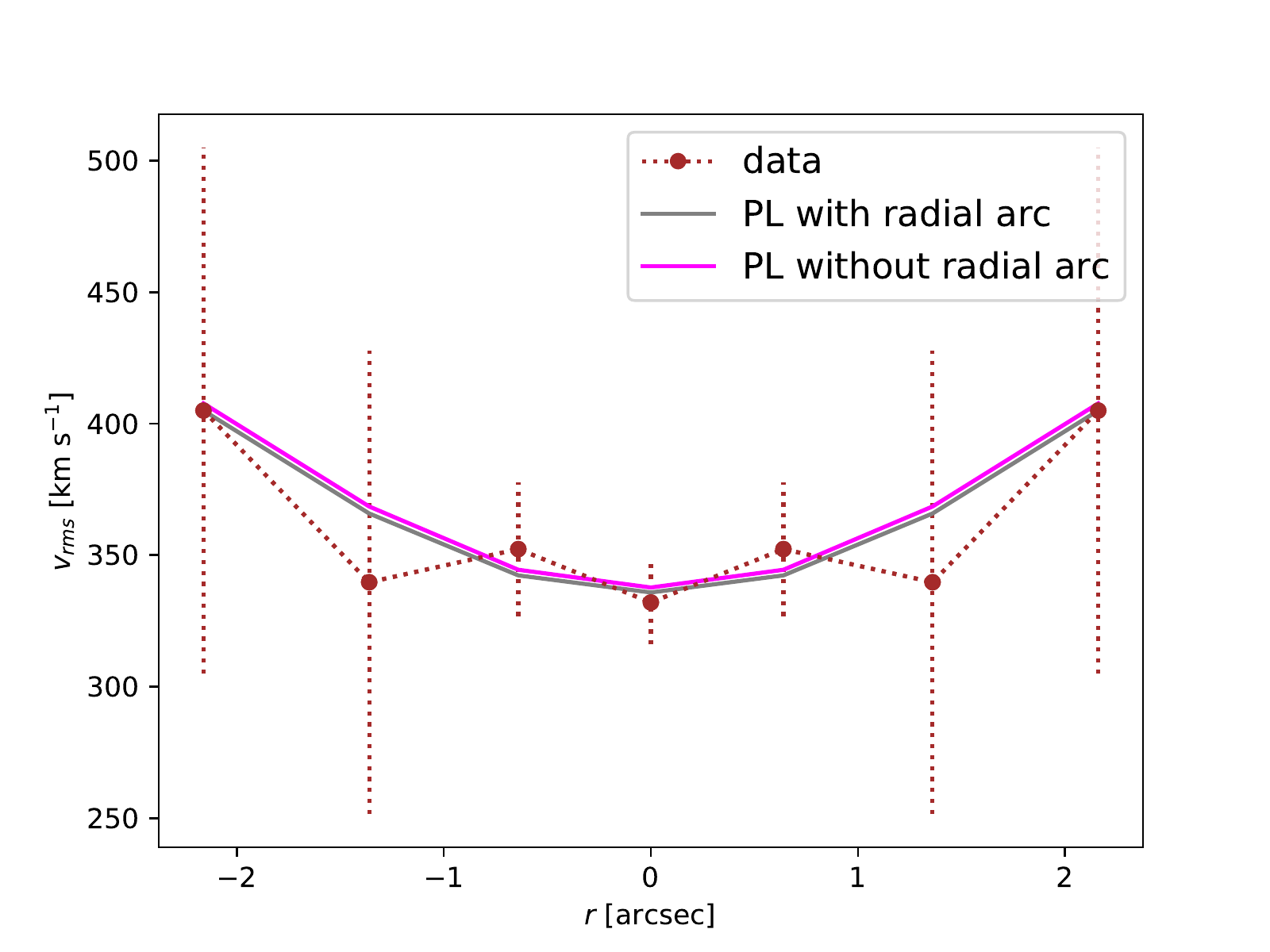}}
\put(100,0){\includegraphics[angle = 0, trim=10 0 40 30, clip, width=1.0\columnwidth]{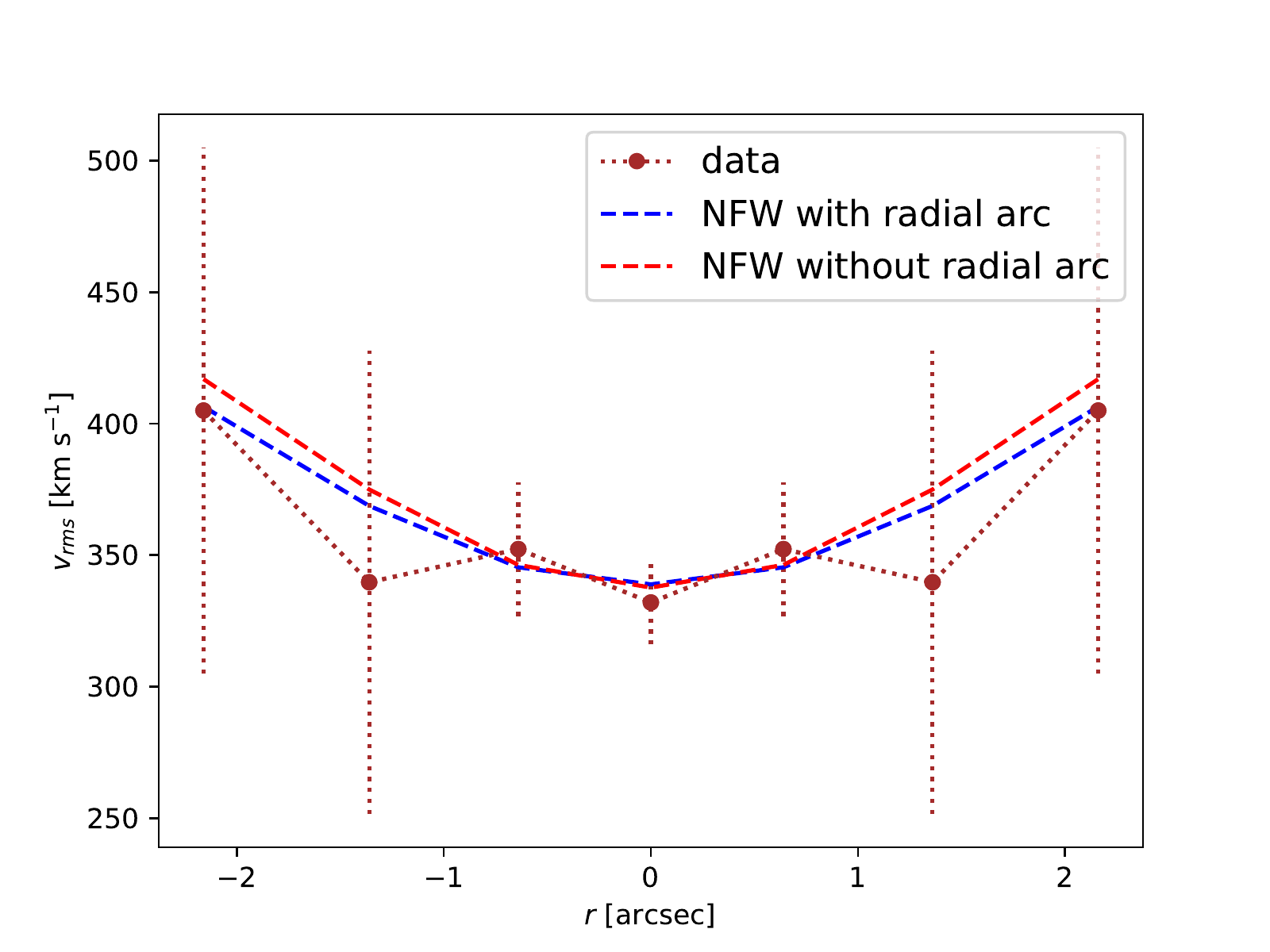}}
\put(-120,200){\color{black} \Large{power-law dark matter halo}} 
\put(170,200){\color{black} \Large{NFW dark matter halo}} 
\end{picture}
   \caption{Values for the second velocity moments $v_\text{rms}$ obtained by adopting the power-law dark matter distribution (left) or NFW (right) for dynamics and lensing. We use here the full uncertainties on the kinematic measurements and the point mass instead of the central component of the fitted light. In brown are shown the measured data points with the error bars.
     }
   \label{fig:bestfit_dynlensPoint_full}
\end{figure*}

In all models, we can fit very well to the kinematic data with a
dynamics-$\chi^2$ of around 0.5. This can be seen in
Fig.~\ref{fig:bestfit_dynlensPoint_full} and is expected because of
the large uncertainties and small number of data points
available.  
According to that, we see from Table
\ref{tab:bestfit_pointPL} and Table \ref{tab:bestfit_pointNFW} that we
cannot well constrain the anisotropy $\beta$ and inclination $i$ given
a prior range of $\beta \in [-0.3,0.3]$ and $i \in [0,
0.3]$. Moreover, from those two tables we see that the radial arc
definitely helps to constrain the model better based on the $1\sigma$
values. Especially parameters which are associated with the central region
(e.g., the point mass) are much better constrained using the radial
arc.

\begin{figure*}[ht!]
 \centering
   \includegraphics[angle = 0, trim=0 0 40 0, clip, width=1.0\columnwidth]{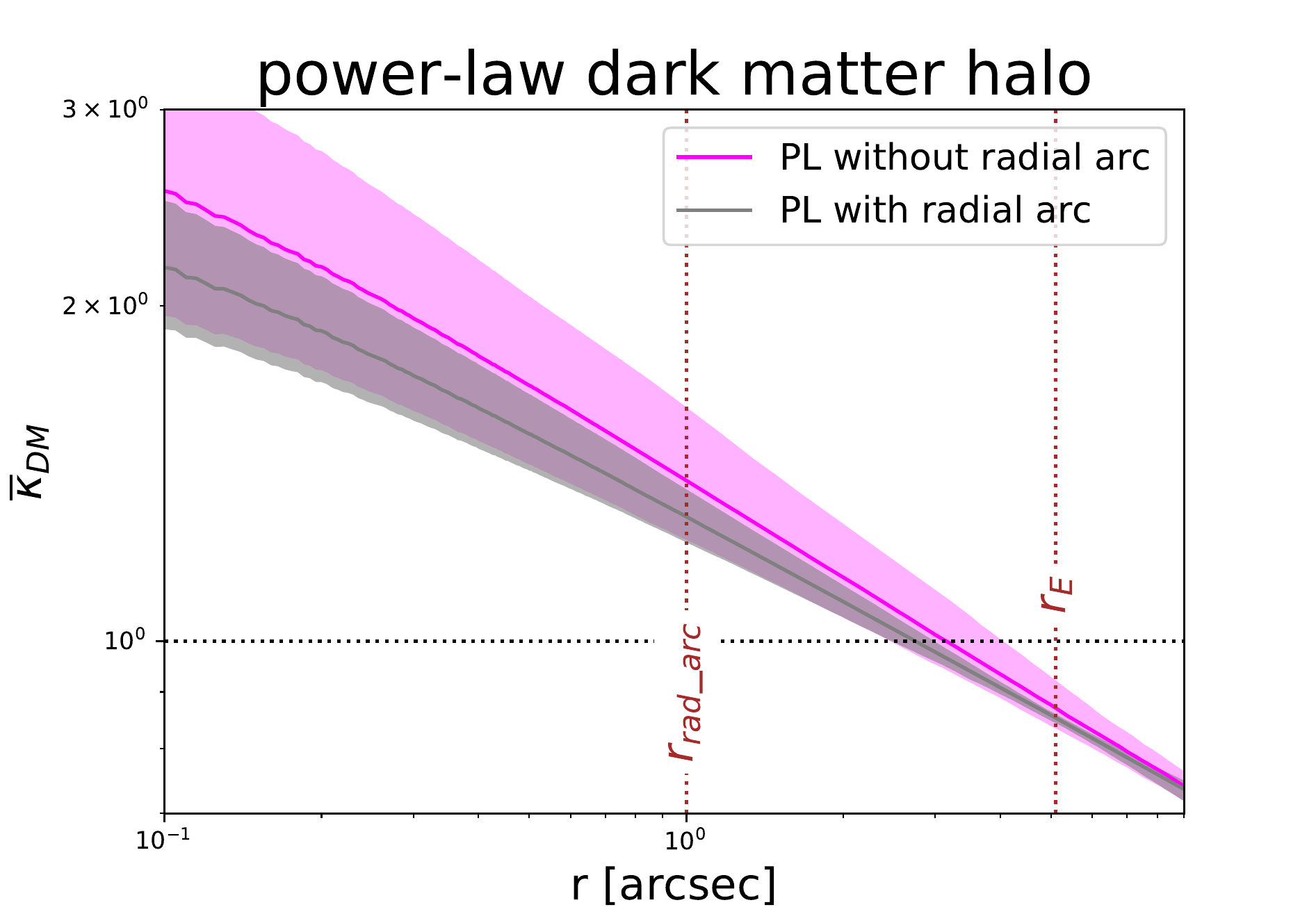}
   \includegraphics[angle = 0, trim=0 0 40 0, clip, width=1.0\columnwidth]{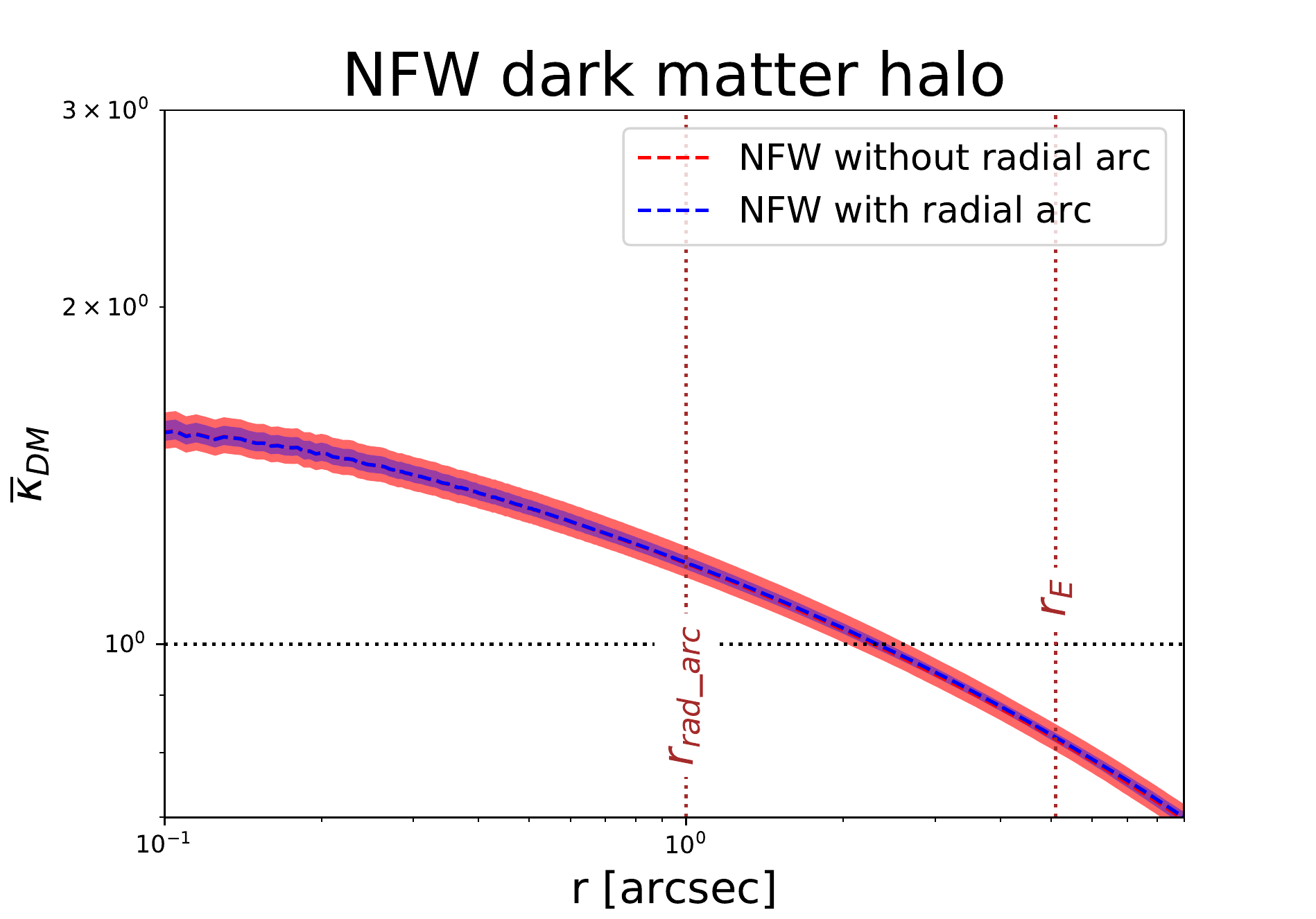}
   \caption{Mean convergence $\overline{\kappa}$ of the model with power-law dark matter component (left) or NFW dark matter component (right), with radial arc and without radial arc as constraint. We show the mean corresponding to the 1-$\sigma$ uncertainty for the dark matter component. We see directly that the radial arc helps to constrain the $\overline{\kappa}(r)$ curve. The brown line indicates the Einstein radius $r_\text{E}$ and radial arc radius $r_\text{rad\_arc}$, respectively, and the black line represents the line $\overline{\kappa} =1 $.
     }
   \label{fig:kappa_bar}
\end{figure*}

\begin{figure*}[ht!]
 \centering
   \includegraphics[angle = 0, trim=0 0 50 15, clip, width=1.0\columnwidth]{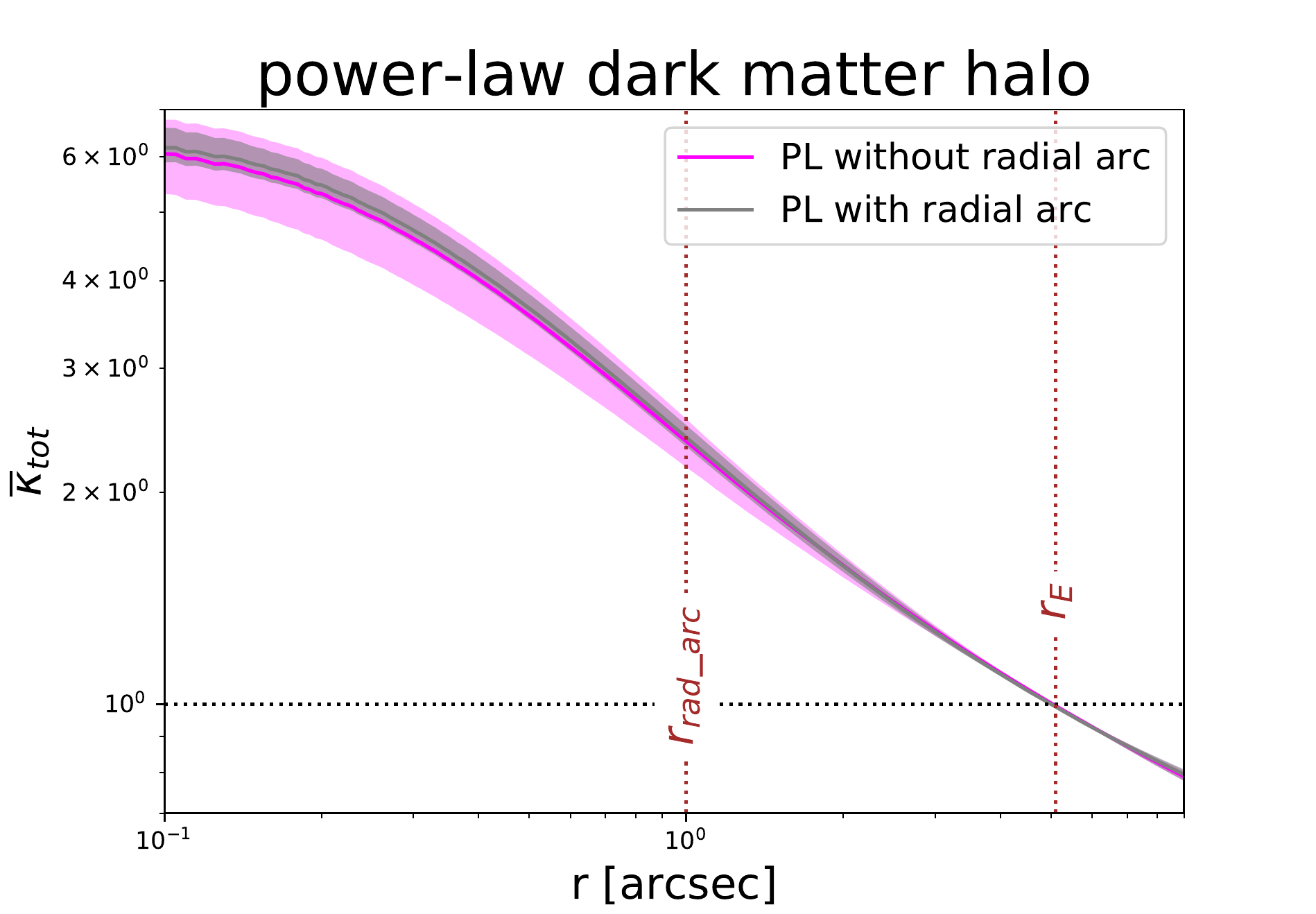}
    \includegraphics[angle = 0, trim=0 0 50 15, clip, width=1.0\columnwidth]{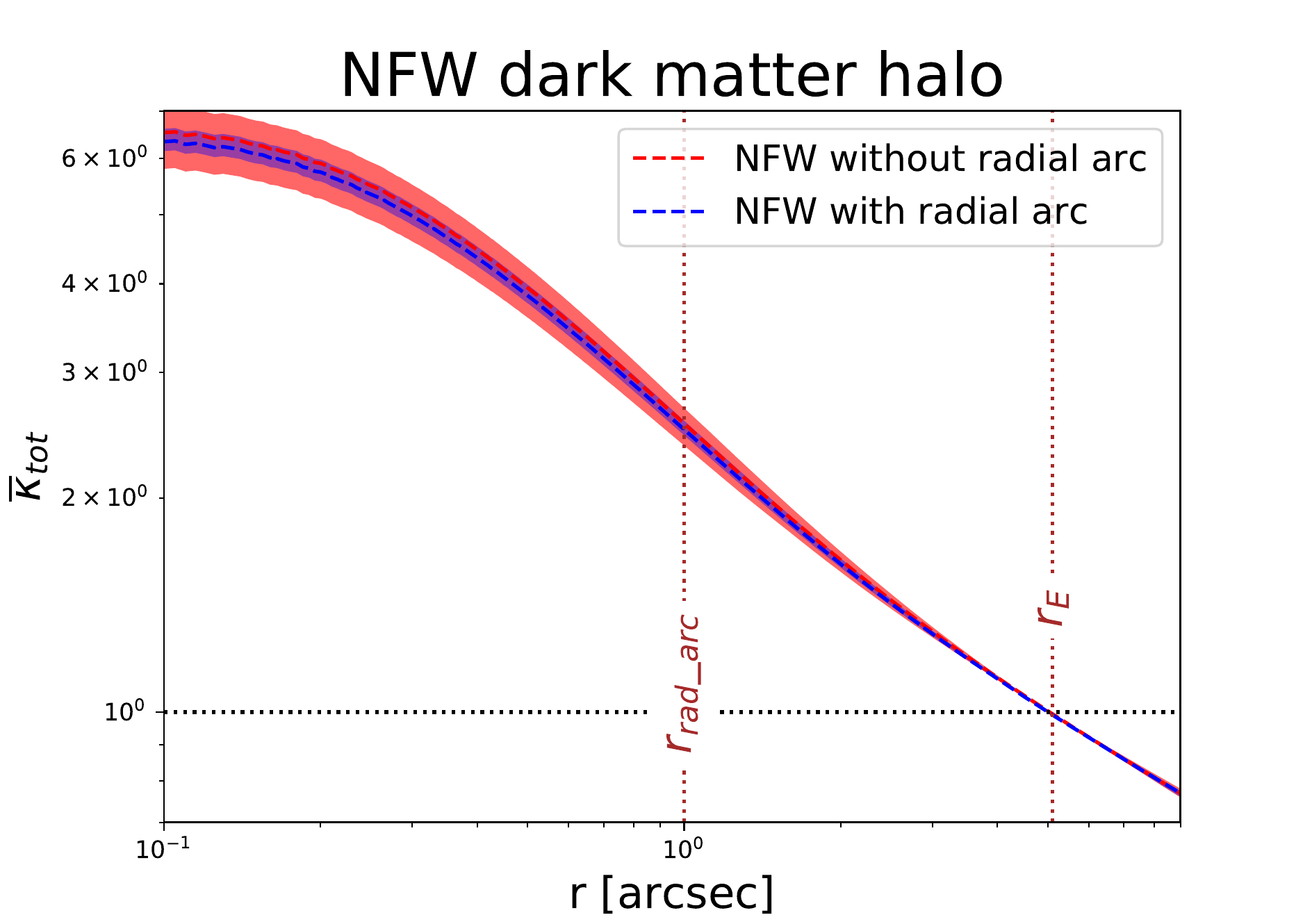}
   \caption{Mean convergence $\overline{\kappa}_\text{tot}$ of the model with power-law dark matter component (left) or NFW dark matter component (right), with radial arc and without radial arc as constraint. We show the mean corresponding to the 1-$\sigma$ uncertanty for baryonic and dark matter component. We see directly that the radial arc helps to constrain the $\overline{\kappa}_\text{tot}(r)$ curve. The brown line indicates the Einstein radius $r_\text{E}$ and radial arc radius $r_\text{rad\_arc}$, respectively, and the black line represents the line $\overline{\kappa}_\text{tot} =1 $.
     }
   \label{fig:kappa_bar_total}
\end{figure*}

Additionally we compare the mean convergence $\overline{\kappa}$ curves of our different models. In Fig.~\ref{fig:kappa_bar} we show
the effect of including the radial arc among the constraints in the
case of the power-law dark matter distribution (left) or NFW dark matter distribution (right) for the dark matter component and in Fig.~\ref{fig:kappa_bar_total} for both baryonic and dark matter component. We can see that the
$\overline{\kappa}_\text{tot}$ are very similar in both cases. In both plots we see the good improvement if we include the radial arc. However, if we compare the NFW and the power-law for the dark matter component, we can see a slight difference in the inner regions. The NFW profile looks less steep.

\begin{figure*}[ht!]
 \centering
 \begin{picture}(200,200)
   \put(-170,0){\includegraphics[angle = 0, trim=0 0 50 35, clip, width=1.0\columnwidth]{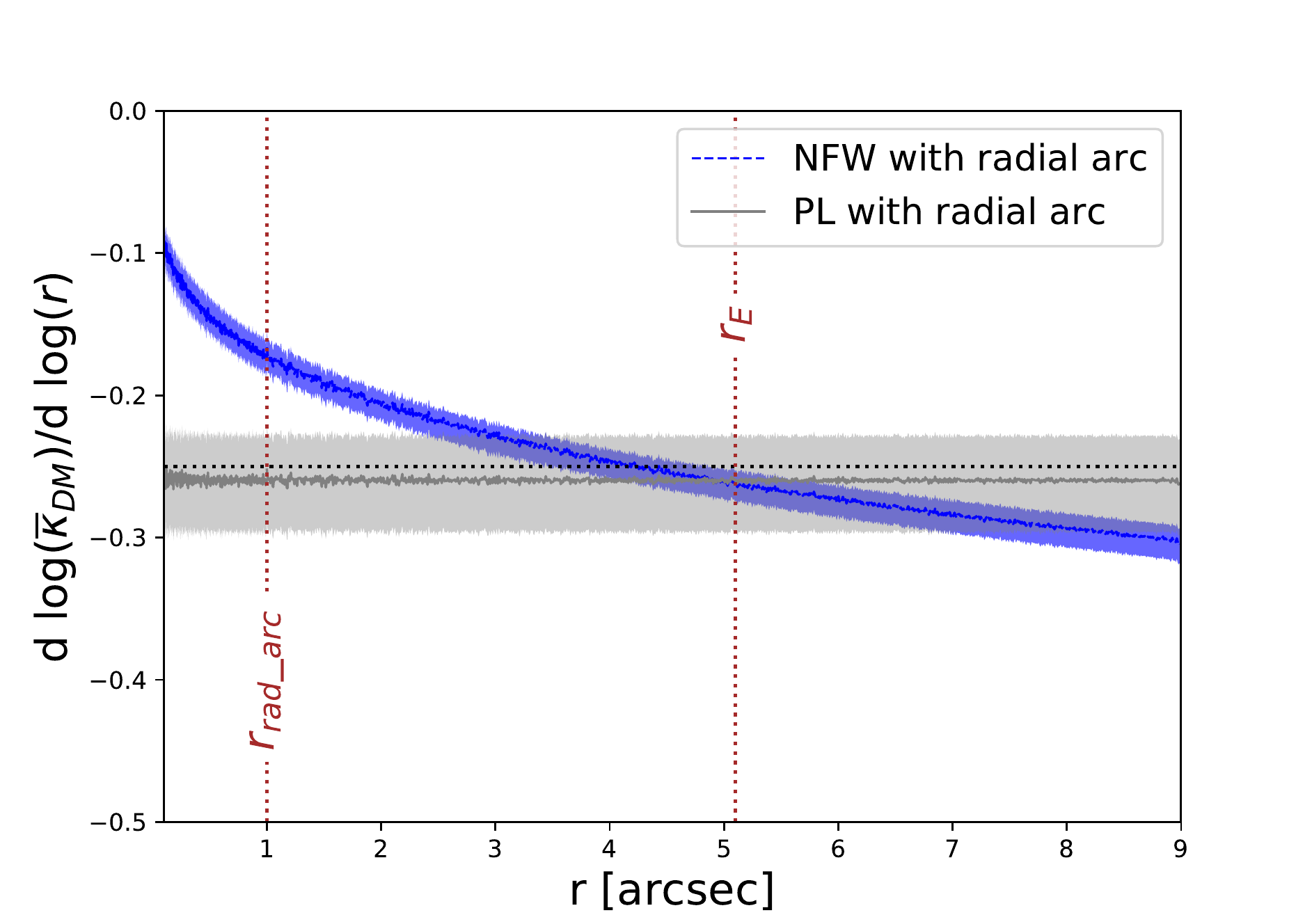}}
    \put(100,0){\includegraphics[angle = 0, trim=0 0 50 35, clip, width=1.0\columnwidth]{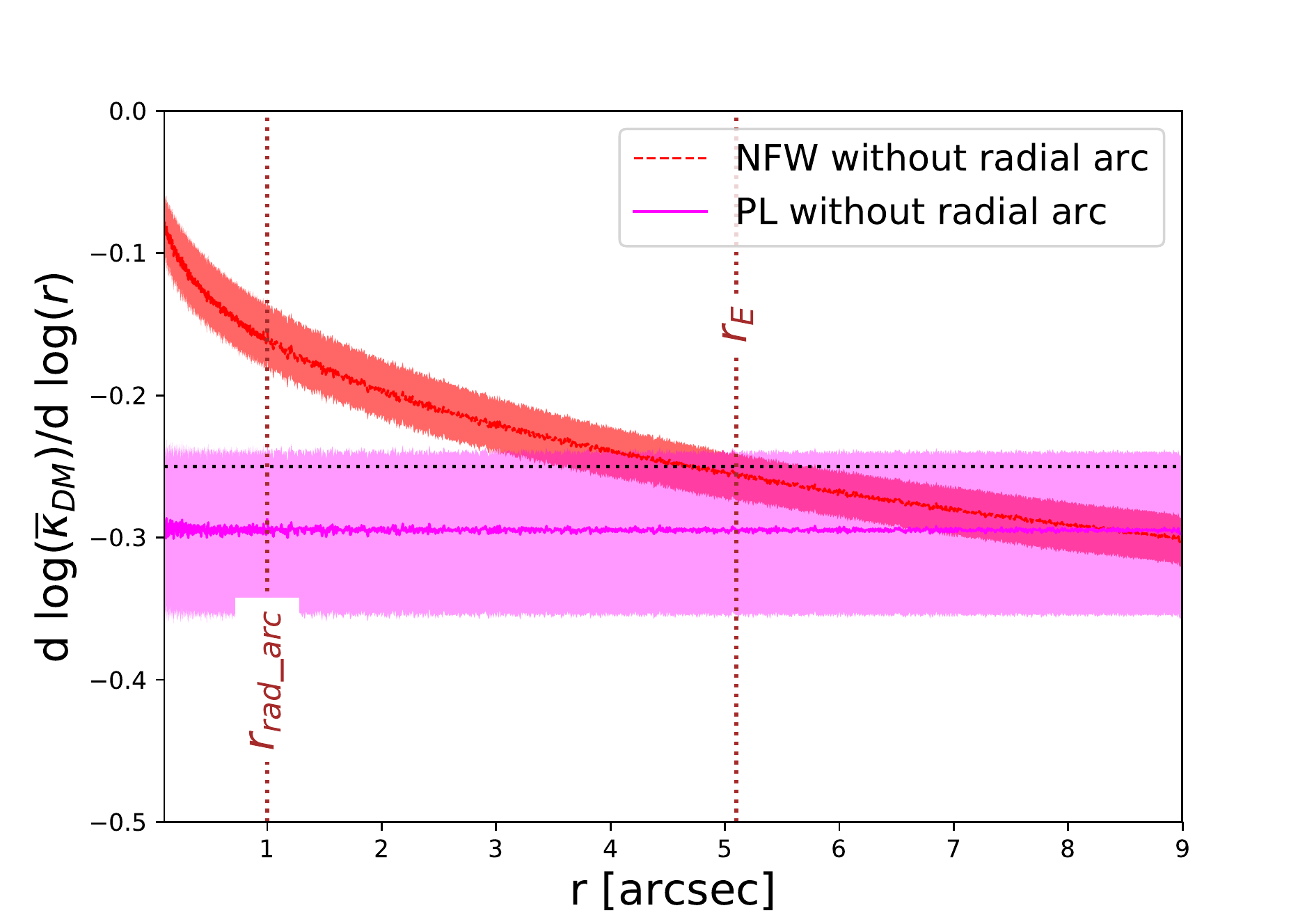}}
    \put(-80,180){\color{black} \Large{with radial arc}} 
    \put(190,180){\color{black} \Large{without radial arc}} 
    \end{picture}
   \caption{Radial slope of dark matter profile, $\text{d} \log(\overline{\kappa}_\text{DM}) / \text{d} \log(r)$, for the models with radial arc (left) and without radial arc (right) as constraint. We show the mean corresponding to the 1-$\sigma$ uncertainty. We see directly that lensing tightly constrains the slope at the Einstein radius, and also the improvement coming from the radial arc. The brown vertical lines indicate the Einstein radius $r_\text{E}$ and radial arc radius $r_\text{rad\_arc}$, respectively, and the black horizontal line represents the line $\text{d} \log(\overline{\kappa}_\text{DM}) / \text{d} \log(r) = 0.25$ for comparison.
     }
   \label{fig:kappa_bar_derivative}
\end{figure*}

Indeed, Fig. \ref{fig:kappa_bar_derivative} shows the logarithm of the slope of the dark matter profile $\text{d} \log(\overline{\kappa}_\text{DM}) / \text{d} \log(r)$. On the left hand panel, we compare the NFW and power-law models including the radial arc while the right hand side shows those excluding the radial arc as constraint. We see that the inferred slope at the Einstein ring is well constrained and independent of the adopted profile. Including the radial arc, we are able to constrain the slope near the radial arc better to a range of $\sim-0.3$ to $\sim-0.15$ at the radial arc radius, covering the spread between the two models.  More information in the central region (of $\sim 1''$), such as spatially resolved kinematics, would be required to break further the model degeneracies in measuring the dark matter profile slope in this region.

\begin{figure*}[ht!]
\centering
\begin{picture}(200,380)
\put(-150,0){\includegraphics[angle = 0, trim=0 0 0 0, clip, width=0.9\textwidth]{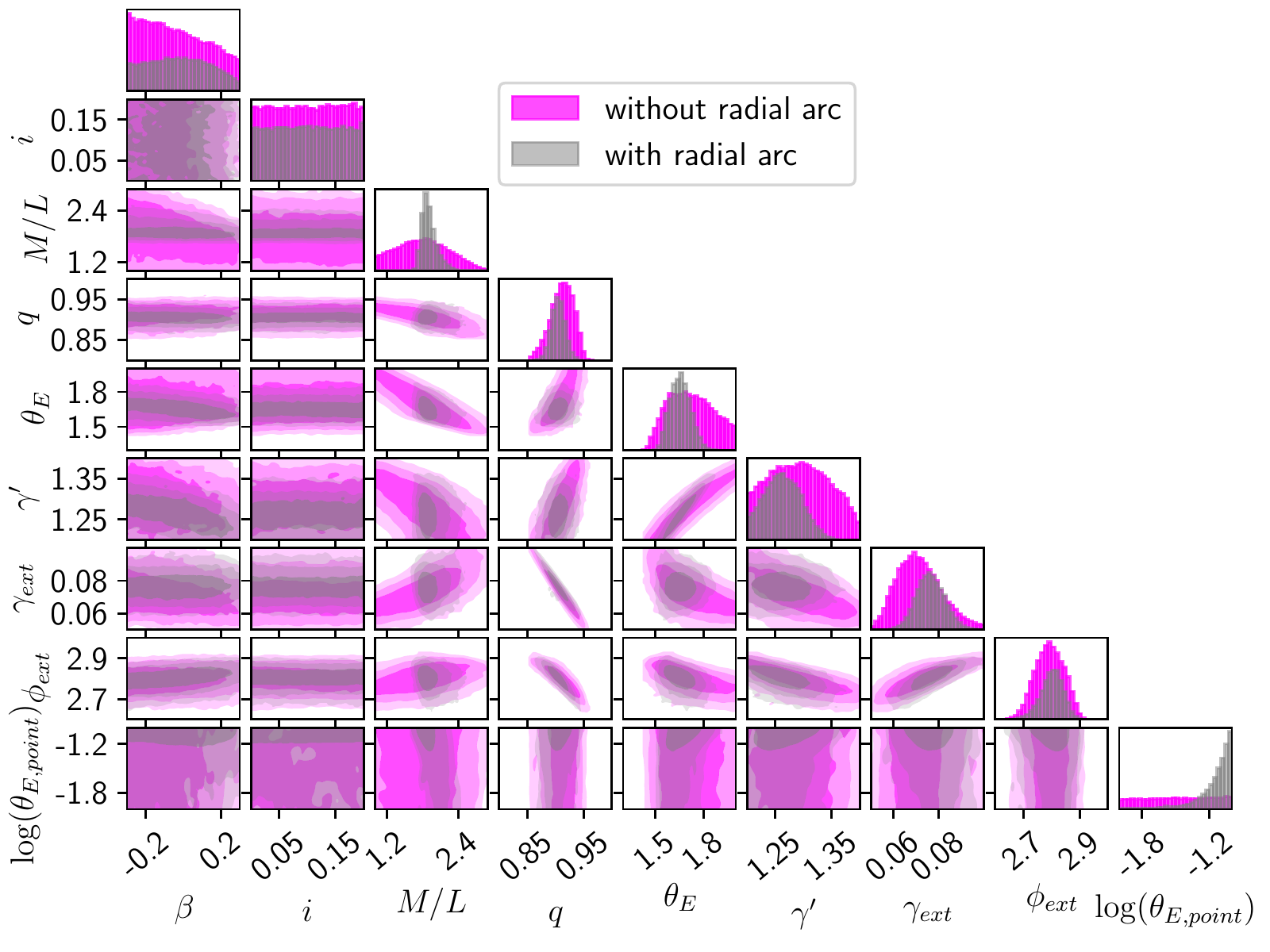}}
\put(0,340){\color{black} \Large{power-law dark matter halo}} 
\end{picture}
\caption{Probability density distribution for our best-fit models, adopting a power-law dark matter profile. In the diagonal one sees the 1-D histograms for the corresponding parameter given on the $x$-axis (and independent of the label in the $y$-axis), while below in the triangle the joint 2-D probability distributions corresponding to the parameters given on $x$- and $y$-axis are shown. The different opacities in the 2-D plots indicate the different sigma ranges. In general, one sees that the model with radial arc is much better constrained than without. 
The parameters are the anisotropy $\beta$, the inclination $i$, the shear magnitude, and its orientation counter clockwise to the x-axis, for the dark matter profile the axis ratio $q$, Einstein angle $\theta_E$, and the slope $\gamma'$. Additionally, we show the mass-to-light ratio $M/L$, which is used to scale the two light components, and the logarithm of the strength of the point mass $\theta_{E, \text{point}}$. The prior range for the point mass is set to $10^{8}-10^{10} M_\odot $ as the known limits of black holes, corresponding to $\log(\theta_{\text{E,point}})$ between $-4$ and $-1$.
}
\label{fig:probdens_PL}
\end{figure*}

\begin{figure*}[ht!]
\centering
\begin{picture}(200,380)
\put(-150,0){\includegraphics[angle = 0, trim=0 0 0 0, clip, width=0.9\textwidth]{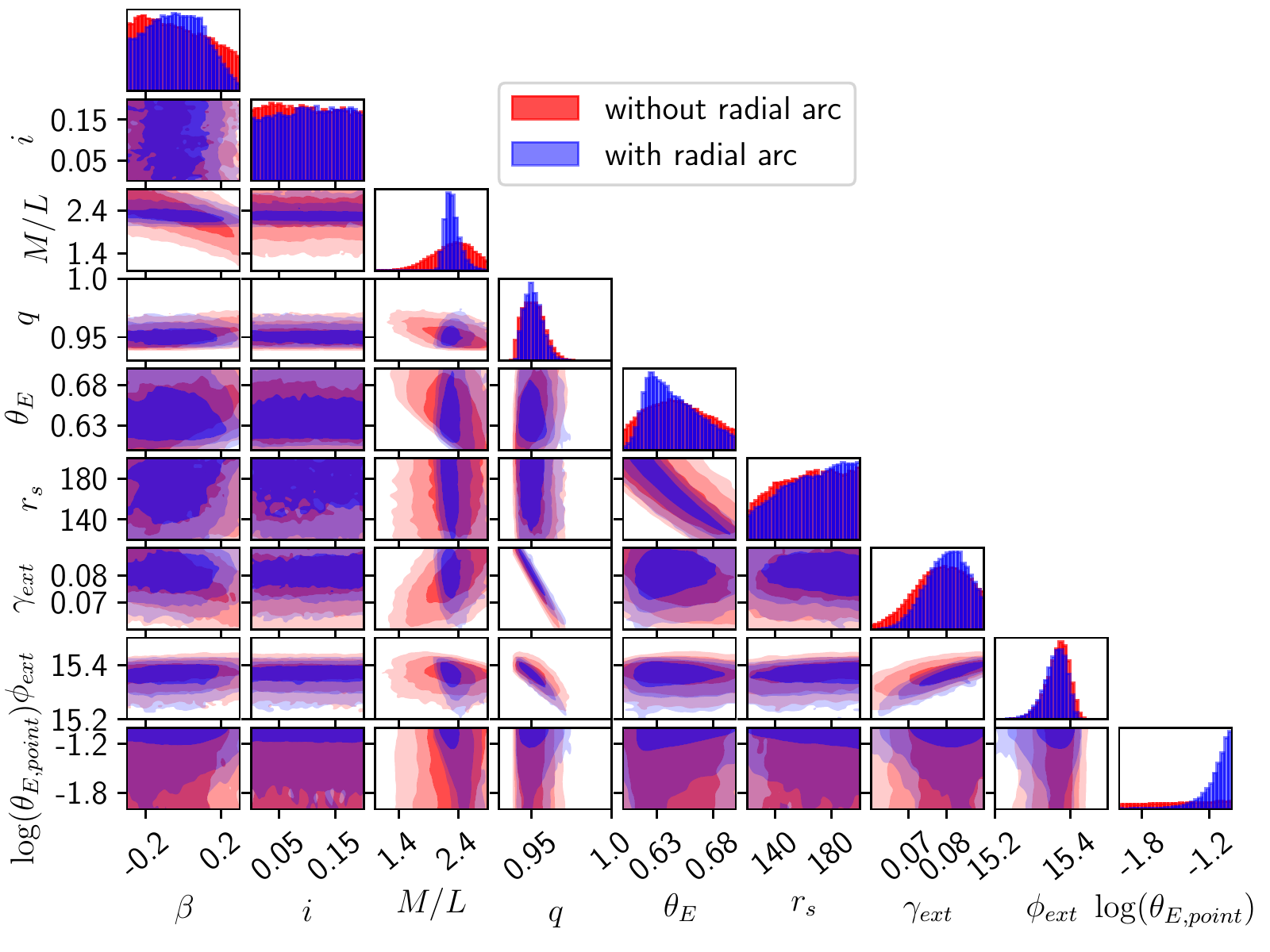} }
\put(0,340){\color{black} \Large{NFW dark matter halo}} 
\end{picture}
\caption{Probability density distribution for our best-fit models, adopting a NFW dark matter profile. In the diagonal one sees the 1-D histograms for the corresponding parameter given on the $x$-axis (and independent of the label in the $y$-axis), while below in the triangle the joint 2-D probability distributions corresponding to the parameters given on $x$- and $y$-axis are shown. The different opacities in the 2-D plots indicate the different sigma ranges. In general, one sees that the model with radial arc is much better constrained than without. 
The parameters are the anisotropy $\beta$, the inclination $i$, the shear magnitude, and its orientation counter clockwise to the x-axis, for the dark matter profile the axis ratio $q$, strength (right) $\theta_E$, and the scale radius $r_\text{s}$. Additionally, we show the mass-to-light ratio $M/L$, which is used to scale the two light components, and the logarithm of the strength of the point mass $\theta_{E, \text{point}}$. The prior range for the point mass is set to $10^{8}-10^{10} M_\odot $ as the known limits of black holes, corresponding to $\log(\theta_{\text{E,point}})$ between $-4$ and $-1$.
}
\label{fig:probdens_NFW}
\end{figure*}

Finally, to further see the contribution of the radial arc we show the probability density distribution of our final best-fit models. In particular, Fig.~\ref{fig:probdens_PL} shows the power-law models, while Fig.~\ref{fig:probdens_NFW} shows the NFW models. From those figures we also see that parameters are 
much better constrained when radial arc is included, especially the point mass parameter, which
is understandable as it is only present in the central region where
the radial arc is observed. The prior range of the point mass strength is for all models the same as we restrict it to be inbetween $10^8-10^{10} M_\odot$ as known mass range of black holes. This corresponds to $\log(\theta_\text{E,point})$ between $-4$ and $-1$s. We see from this distribution that the radial arc forces the point mass to its upper limit. Since the lens galaxy is very massive, a supermassive black hole is realistic. Interestingly, also the mass-to-light
ratio is better constrained by including the radial arc. This confirms
the importance of including the radial arc as constraint. We also see
that the contribution of the constraints coming from dynamics is quite
small, probably due to the small amount of data and the large
uncertainties. 

We also consider all models under the assumption of 5\% uncertainty as the current errors are very huge. Comparing to the figure obtained with the real uncertainties, we do not see a remarkable difference. The uncertainties do not seem to reduce the parameter space substantially, even though the $\chi^2$ is higher. Therefore, to further improve the mass modeling through dynamics, spatially resolved
kinematic measurements would likely be needed in the future.

\section{Summary and Conclusion}
\label{sec:conclusion}

While in the standard CDM model the structure of dark matter is well understood through large numerical dark matter only simulations
\citep[e.g.][]{dubinski91, navarro96b, navarro96a}, one has to include the baryonic component to reach more complex, but realistic models. Since the deflection of light depends on the total matter, strong gravitational lensing provides a good opportunity to obtain the distribution of the lens' mass. In this paper we study the matter distribution of a unique strong lensing observation, known as the Cosmic Horseshoe (\ourlens). This observation shows a radial arc at a redshift of $z_\text{s,r} = 1.961$ inside the huge Einstein ring, whose redshift we measured based on spectroscopic observations presented in this paper. Including that radial arc in our models helps to improve our model as it gives lensing constraints in the central region. For obtaining a self-consistent mass model, we include kinematic measurements of the lens galaxy in our final model.

Before disentangling dark matter and baryonic mass, we first
construct a model of the total lens mass. Based on this model, we
create a composite model with baryonic and dark matter components
separately. We adopt different dark matter profiles, a power-law
profile, a NFW profile, or a generalization of the NFW profile. For
the baryonic component we adopt the lens light distribution, which is
described by three components, scaled by a mass-to-light ratio. As one
component is very peaky and thus AGN-like, we suggest in our final
model to supersede this component by a point mass as we cannot assume
a physical meaningful $M/L$ value. The other two components are still
scaled by a mass-to-light ratio. We then include stellar kinematic
information of the lens, thanks to which we are able to construct a
self-consistent mass model. As we are also interested to see the
improvement coming from the radial arc, we always model with and
without radial arc and compare those models. From our study of the
matter distribution we obtain the following key results:
\begin{itemize}
\item Since the width of the central component of the fitted lens
  light is comparable with the PSF width (compare
  Fig.~\ref{fig:lenslight_with_PSF}) and the lens galaxy emits in the
  radio wavelengths, our modeling results support a mass model for the
  \CH\ lens galaxy with a point component in the center instead of a
  luminous component scaled by a mass-to-light ratio.  The two outer
  components are scaled with a mass-to-light ratio to account for the
  baryonic mass.  The dark matter component could follow either a
  power-law or an NFW profile, since both profiles could adequately
  fit to the current data.
\item We can construct a better mass model thanks to the contribution
  of radial arc and its counter image. Thus we infer the radial arc is
  part of the full \CH\ system. It turns out that the radial arc
  improves the dark matter halo parameter constraints independently of
  the adopted dark matter distribution.
\item When adding the contribution of dynamical modeling, we find that
  actually this method is not able to constrain significantly better
  the possible parameter range. We suspect this might be due to the
  lack of data points and large uncertainties. When using the
  forecasted 5\% uncertainties on the kinematic measurements, we
  constrain the parameter ranges slightly better.
\item When trying to model the baryonic matter independently of the
  dark matter, we consider two scenarios: the matter aligned with the
  lens light or to be 90$^\circ$ offset. Here we find that the model with the 90$^\circ$ rotated orientation fits better. Thus,
  the major axis of the projected halo mass distribution seem to be
  perpendicular to the major axis of the baryonic mass
  distribution.
\item For all the tested models, we obtain a mass of around $5.2
  \times 10^{12} \text{M}_\odot$ enclosed in the Einstein ring. This
  is in agreement with previous studies of the \CH, e.g.,
  \citet{dye08}. 
  We predict the dark matter fraction at one effective radius to be
  $~0.65$, whose value is slightly higher as predicted by
  \citet{spiniello11}, but within their $1\sigma$
  range. The exact fraction depends also on the specific model.
\end{itemize}

From this work, we demonstrate the utility of having a radial arc in
constraining the dark matter profile, particularly in the inner
regions.  This is important for the future, when we might discover more
lenses in current and future surveys. New single-galaxy lens systems
with radial arcs would provide great opportunities to model the inner
dark matter distributions and probe galaxy formation scenarios. This
would also give a more general statement rather than from one anecdotal
example.

Moreover, we see that combining lensing and kinematic data helps to
constrain the model better, even though current kinematic data of the
\CH\ are limited. Thus further kinematic measurements, particularly if
spatially resolved, of such strong lens observations
would help to construct a better mass
model.

\FloatBarrier
\begin{acknowledgements}
  We thank M.~Auger for useful discussions.\\ 
  SS, GC, SHS and AY thank the Max Planck Society for support through the Max Planck Research Group for SHS.  This research was supported in part by Perimeter Institute for Theoretical Physics. Research at Perimeter Institute is supported by the Government of Canada through the Department of Innovation, Science and Economic Development and by the Province of Ontario through the Ministry of Research, Innovation and Science.
\\
  The analysis is based on: 1) observations made with the NASA/ESA
  \textit{Hubble Space Telescope}, obtained at the Space Telescope
  Science Institute, which is operated by the Association of
  Universities for Research in Astronomy, Inc., under NASA contract
  NAS 5-26555. These observations were done in May 2010 with Proposal ID 11602 and in November 2011 with Proposal ID 12266;
2) Observations obtained at the Gemini Observatory, which is operated by the Association of Universities for Research in Astronomy, Inc., under a cooperative agreement with the NSF on behalf of the Gemini partnership: the National Science Foundation (United States), the National Research Council (Canada), CONICYT (Chile), Ministerio de Ciencia, Tecnolog\'{i}a e Innovaci\'{o}n Productiva (Argentina), and Minist\'{e}rio da Ci\^{e}ncia, Tecnologia e Inova\c{c}\~{a}o (Brazil).
\end{acknowledgements}

\bibliographystyle{aa}
\bibliography{Horseshoe}

\FloatBarrier

\section{Dynamical modeling using Multi-Gaussian-Expansion (MGE) parameterization and Jeans ansatz}
\label{sec:dynamicstheory}

Here we introduce briefly the Jeans formalism which we use for
dynamical modeling in our mass model of the \CH. Since we assume an
axisymmetric model, we only consider this specific case here, and
refer for details and the general case to \citet{cappellari08}.

We start from the general axisymmetric Jeans equations \citep{Jeans22}
\bea
\label{eq:jeans1}
\frac{\mu  \overline{v^2_\text{R}} - \mu \overline{v^2_\phi}}{R} + \frac{\partial (\mu \overline{v^2_\text{R}}) }{\partial R} + \frac{\partial (\mu \overline{v_\text{R} v_\text{z}}) }{\partial z} &=& - \mu \frac{\partial \Phi }{\partial R} \\
\frac{\mu  \overline{v_\text{R}v_\text{z}}}{R} + \frac{\partial (\mu \overline{v^2_\text{z}}) }{\partial z} + \frac{\partial (\mu \overline{v_\text{R} v_\text {z}}) }{\partial R} &=& - \mu \frac{\partial \Phi }{\partial z} \ ,
\label{eq:jeans2}
\eea 
which are only two equations since the third reduces to zero in the axisymmetric case. Here, $\Phi$  is the gravitational potential, $(R,z, \phi)$ standard cylindrical coordinates, $ \mu \overline{v_k v_j} $ an abbreviation for $\int v_k v_j f \text{d} ^3 \vec{v}$ with $f(\vec{x},\vec{v})$ the distribution function (DF) at position $\vec{x}$ and with velocity $\vec{v}$ and $\mu$ the luminosity density (not $\nu$ as in \citet{cappellari08} to distinguish better from the velocity $v$). One can reduce these two equations to
\bea
\label{eq:jeansaxisym1}
\frac{b \mu  \overline{v^2_\text{z}} - \mu \overline{v^2_\phi}}{R} + \frac{\partial (b \mu \overline{v^2_\text{z}})}{\partial R} &=& - \mu \frac{\partial \Phi }{\partial R} \\
\frac{\partial (\mu \overline{v^2_\text{z}})}{\partial z} &=& - \mu \frac{\partial \Phi }{\partial z} 
\label{eq:jeansaxisym2}
\eea
by assuming that the velocity ellipsoid is aligned with the cylindrical coordinate system $(R,z, \phi )$ and that the anisotropy $b$ is constant and given by 
\be
\overline{v^2_\text{R}} = b  \times \overline{v^2_z} \ .
\ee
The situation $b=1$ is the so-called semi-isotropic or two-integral case.

For the stellar density and the total density we adopt the \textit{Multi-Gaussian Expansion} (MGE) parameterization \citep{bendinelli91, monnet92} as described in \citet{cappellari02} because of its accuracy in reproducing the surface brightness and its robustness. By assuming that the $x$-axis is aligned with the photometric major axis, the surface brightness $\Sigma $ is given by
\be
\Sigma (x',y') = \sum_{k=1}^{N} \frac{L_k}{2 \pi \sigma^2_k q'_k} \, \exp \left[ - \frac{1}{2 \sigma_k^2} \, \left(x'^2 + \frac{y'^2}{q'^2_k} \right) \right]
\ee
at the position $(x',y')$ of the plane of sky. Here, $N$ is the number of adopted Gauissians with luminosity $L_k$, observed axis ratio $q'_k$ between 0 and 1, and dispersion $\sigma_k$ along the major axis.

Since the galaxies have an unknown inclination $i$, one needs a deprojection of the surface brightness to get the intrinsic luminosity density. This is not unique unless one considers edge-on ($i = 90 ^\circ$) oriented galaxies \citep{rybicki87, kochanek96}. As described in \citet{cappellari08}, one advantage of the MGE method is that one can relatively well include the roundness of the model to get realistic densities and fulfill the \textit{morphological criterion}, which is described in detail in \citet*{cappellari03}. Thereafter one can write the deprojected MGE oblate axisymmetric luminous density $\mu$ as
\be
\mu (R,z) = \sum_{k=1}^{N} \frac{L_k}{(2 \pi)^{3/2} \sigma_k^3 q_k} \, \exp \left[ - \frac{1}{2 \sigma_k^2} \left( R^2 + \frac{z^2}{q_k^2} \right) \right] \, ,
\label{eq:lumdens}
\ee
with the intrinsic axial ratio of each Gaussian component
\be
q_k = \frac{\sqrt{ q'^2_k - \cos ^2 (i) }}{\sin (i)} \, .
\ee

As we said, we adopt for the total density $\rho$ an MGE parameterization as well, such that one can write it as a sum of $M$ Gaussians:
\be
\rho (R,z) = \sum_{j=1}^M \frac{M_j }{(2 \pi)^{3/2} \sigma_j^3 q_j} \exp \left[ - \frac{1}{2 \sigma_j^2} \left( R^2 + \frac{z^2}{q_j^2} \right) \right] \, .
\label{eq:totdens}
\ee

After applying the MGE formalism to the solution of axisymmetric anisotropic Jeans equations \ref{eq:jeansaxisym1} and \ref{eq:jeansaxisym2}, i.e. one substitutes Eq. \ref{eq:lumdens} and the gravitational potential obtained from Eq. \ref{eq:totdens} into equations \ref{eq:jeansaxisym1} and \ref{eq:jeansaxisym2}, then one can perform the integral analytically. With that, one can integrate along the line-of-sight (LOS) to obtain the observables which we then want to compare to the galaxy kinematics. These are the total observed second moment and the first moment. For the last one, we need additional assumptions because one has to decide how the second moment separates into the contribution of ordered and random motion, which is defined by
\be
\overline{v^2_\phi} = \overline{v_\phi}^2 + \sigma_\phi^2 \, ,
\label{eq:2moment}
\ee
or in simplified, but often used notation
\be
v_\text{rms}^2 = v^2 + \sigma^2 \, .
\label{eq:2moment_normal}
\ee
Here $v_\text{rms}$ is the second velocity moment, $v$ the rotation, and $\sigma$ the velocity dispersion. These necessary additional assumptions are the reason why one considers often the
second velocity moment, which is the more general formula. However, the first moment are very useful to quantify the amount of rotation in galaxies and are thus sometimes used \citep[e.g.,][]{nagai76, satoh80, binney90, marel90}.

In the case that the anisotropy $b_k$ is different for each Gaussian, the total luminosity-weighted anisotropy of an MGE model, under the assumptions noted above, is given by the definition \citep{binney82, cappellari08}
\be
\beta_z (R,z) \equiv 1 - \frac{\overline{v^2_z}}{\overline{v^2_R}} = 1 - \frac{ \sum_{k=1}^N [\mu \overline{v^2_z}]_k}{\sum_{k=1}^{N} b_k [\mu \overline{v^2_z}]_k } \, .
\ee
For further theoretical discussion we refer the reader to the paper \citet{cappellari08}.

\end{document}